\documentclass[twocolumn]{aastex631}
\usepackage{lineno}
\shorttitle{Ten New BH Spins}
\shortauthors{Draghis et al.}

\begin{document}

\title{A Systematic View of Ten New Black Hole Spins}
\correspondingauthor{Paul A. Draghis}
\author[0000-0002-2218-2306]{Paul A. Draghis}
\email{pdraghis@umich.edu}
\affiliation{Department of Astronomy, University of Michigan, 1085 South University Avenue, Ann Arbor, MI 48109, USA}

\author[0000-0003-2869-7682]{Jon M. Miller}
\affiliation{Department of Astronomy, University of Michigan, 1085 South University Avenue, Ann Arbor, MI 48109, USA}

\author[0000-0002-0572-9613]{Abderahmen Zoghbi}
\affiliation{Department of Astronomy, University of Maryland, College Park, MD 20742, USA}
\affiliation{HEASARC, Code 6601, NASA/GSFC, Greenbelt, MD 20771, USA}
\affiliation{CRESST II, NASA Goddard Space Flight Center, Greenbelt, MD 20771, USA}

\author[0000-0003-1621-9392]{Mark Reynolds}
\affiliation{Department of Astronomy, University of Michigan, 1085 South University Avenue, Ann Arbor, MI 48109, USA}
\affiliation{Department of Astronomy, Ohio State University,  140 W 18th Avenue, Columbus, OH 43210, USA}

\author[0000-0001-8470-749X]{Elisa Costantini}
\affiliation{SRON Netherlands Institute for Space Research, Niels Bohrweg 4, 2333 CA Leiden, The Netherlands}
\affiliation{Anton Pannekoek Astronomical Institute, University of Amsterdam, P.O. Box 94249, 1090 GE Amsterdam, the Netherlands}

\author{Luigi C. Gallo}
\affiliation{Department of Astronomy \& Physics, Saint Mary’s University, 923 Robie Street, Halifax, Nova Scotia, B3H 3C3, Canada}

\author[0000-0001-5506-9855]{John A. Tomsick}
\affiliation{Space Sciences Laboratory, 7 Gauss Way, University of California, Berkeley, CA 94720-7450, USA}

\begin{abstract}
The launch of NuSTAR and the increasing number of binary black hole (BBH) mergers detected through gravitational wave (GW) observations have exponentially advanced our understanding of black holes. Despite the simplicity owed to being fully described by their mass and angular momentum, black holes have remained mysterious laboratories that probe the most extreme environments in the Universe. While significant progress has been made in the recent decade, the distribution of spin in black holes has not yet been understood. In this work, we provide a systematic analysis of all known black holes in X-ray binary systems (XB) that have previously been observed by NuSTAR, but have not yet had a spin measurement using the ``relativistic reflection" method obtained from that data. By looking at all the available archival NuSTAR data of these sources, we measure ten new black hole spins: IGR J17454-2919 -- $a=0.97^{+0.03}_{-0.17}$; GRS 1758-258 -- $a=0.991^{+0.007}_{-0.019}$; MAXI J1727-203 -- $a=0.986^{+0.012}_{-0.159}$; MAXI J0637-430 -- $a=0.97\pm0.02$; Swift J1753.5-0127 -- $a=0.997^{+0.001}_{-0.003}$; V4641 Sgr -- $a=0.86^{+0.04}_{-0.06}$; 4U 1543-47 -- $a=0.98^{+0.01}_{-0.02}$; 4U 1957+11 -- $a=0.95^{+0.02}_{-0.04}$; H 1743-322 -- $a=0.98^{+0.01}_{-0.02}$; MAXI J1820+070 -- $a=0.988^{+0.006}_{-0.028}$ (all uncertainties are at the $1\sigma$ confidence level). We discuss the implications of our measurements on the entire distribution of stellar mass black hole spins in XB, and we compare that with the spin distribution in BBH, finding that the two distributions are clearly in disagreement. Additionally, we discuss the implications of this work on our understanding of how the ``relativistic reflection" spin measurement technique works, and discuss possible sources of systematic uncertainty that can bias our measurements.
\end{abstract}

%\keywords{These are the keywords}

\section{Introduction} \label{sec:intro}
It is believed that black holes (BH) were first thought of almost 240 years ago (\citealt{first}), and then rediscovered 132 years later, in 1915, when Karl Schwarzschild found a solution of general relativity that would describe a black hole (\citealt{1999physics...5030S}). In 1972, 57 years later, multiple independent studies concluded that the system Cygnus X-1 must harbor a black hole (\citealt{1972Natur.235...37W, 1972NPhS..240..124B, 1972ApJ...177L...5T}), making this the first BH detected through electromagnetic radiation. In 2016, 44 years later, gravitational wave signals from the merging event of two black holes was first detected (\citealt{2016PhRvL.116f1102A}), and only three years later, the first direct image of the shadow of a Kerr black hole was obtained (\citealt{2019ApJ...875L...1E}). Even though our understanding of black holes has grown exponentially in the past few decades and even though black holes are intrinsically simple objects, described entirely by their mass and angular momentum, many questions remain unanswered. This includes, perhaps most interestingly, a complete understanding of the origin and magnitude of black hole rotation.

The third Gravitational Wave Transient Catalog (GWTC-3, \citealt{2021arXiv211103606T}) contains 90 merging events detected by Advanced LIGO (aLIGO) and Advanced Virgo (AdV). Based on this sample, \cite{2021arXiv211103634T} inferred that the spin distribution of the most rapidly spinning BH of the two in the binary black hole (BBH) system peaks near 0.4\footnote{the dimensionless spin parameter is given by $a=cJ/GM^2$, where, in general, $-0.998 \leq a \leq 0.998$ (\citealt{1974ApJ...191..507T}). However, for the BHs in BBH, the absolute magnitude of the spin and the orientation of the spin vector are reported.}, with the 1st and 99th percentiles at $0.07^{+0.05}_{-0.03}$ and $0.80\pm0.08$, while the spin distribution of the less rapidly spinning BHs in BBH peaks around 0.2, with 99\% of the distribution below $0.54^{+0.09}_{-0.08}$. The spin distribution in these systems is inconsistent with the spin distribution in X-ray binary systems at more than 99.9\% confidence (\citealt{2022ApJ...929L..26F}), hinting at different formation mechanisms between the two populations of BHs. However, it is important to mention that the BH spins obtained based on GW measurements are not directly measured from the observation, rather they are calculated using various models and theoretical assumptions based on the observable quantities in the GW events.

Current estimates (\citealt{2020LRR....23....3A}) predict that with KAGRA joining the aLIGO/AdV collaboration and with the detectors reaching their designed sensitivity,  $79^{+89}_{-44}$ BBH mergers are expected to be detected during O4, which is expected to begin in March 2023. In the near future, the distribution of spin measurements both pre and post merger will become increasingly statistically significant. In contrast, there have been $\sim70$ XB BH candidates discovered since the beginning of the era of X-ray astronomy, of which only $\sim20$ have been confirmed dynamically (see the updated results of \citealt{2016A&A...587A..61C}). In addition to the $\sim1-3$ new BH candidates discovered every year, this number is expected to increase significantly in the future, in the era of 30 meter optical telescopes and X-ray probes such as AXIS (\citealt{2019BAAS...51g.107M}). By looking at the distribution of black hole spins in X-ray binaries (XB) and comparing it to that of the spins in BBH mergers, we begin to look into the final destiny of stars, the origin of black holes, and the nature of BBH systems. Fully describing the two distributions of BH spins is essential to providing a unified understanding of stellar-mass black hole formation and evolution. This will help us to understand the entire BH population, whether or not the two seemingly distinct BH populations have a common origin, and what factors might lead to different median properties. It is therefore important to measure spins in the greatest possible number of stellar-mass black holes in XB. 

%paragraph about the different spin measurement techniques
The spins of BHs in XB are most often measured through either continuum fitting (see e.g., \citealt{2009ApJ...701.1076G, 2012ApJ...745L...7S, 2020MNRAS.499.5891S, 2021ApJ...916..108Z}) or relativistic disk reflection (see e.g., \citealt{2006ApJ...652.1028B, 2009ApJ...697..900M, 2020ApJ...900...78D, 2022MNRAS.514.1422D}). For a detailed review of BH spin measurement techniques, see \cite{2021ARA&A..59..117R}. Continuum fitting models the shape of the emission from the accretion disk around a black hole, but requires prior knowledge of the BH mass, distance to the system, accretion rate, and inclination of the inner accretion disk. Additionally, continuum fitting measurements depend on the assumption of spectral hardening factor and the nature of the hard emission (i.e. direct and reflected coronal emission) in the observations.

%paragraph about relativistic reflection
Relativistic reflection (\citealt{1991MNRAS.249..352G}) models the distortion of spectral features (mainly the Fe K$\alpha$ spectral line, present at 6.4~keV for neutral gas, and the Compton hump present above $\sim20$~keV) in order to infer the proximity of the emitting matter to the BH. The fluorescent Fe K line is produced when one of the two K-shell electrons of an Fe atom is ejected by an ionizing X-ray photon incoming from the compact corona. Following the photoionization, an L-shell electron can decay to the K shell by releasing 6.4 keV of energy either as a photon (with 34$\%$ probability) or as an Auger electron (with 66$\%$ probability). As a Fe atom becomes more ionized, the reduced number of screening electrons in the upper shells increases the energy difference between the K and L shells, therefore causing the fluorescent photons to be emitted with progressively higher energies, up to 6.97~keV for H-like Fe XXVI, where due to the lack of electrons in the K-shell, the Auger effect cannot occur and the Fe K line is produced through recombination with a free electron. Unlike the soft incident X-ray photons which are preferentially absorbed by the accretion disk, the hard X-ray photons are preferentially Compton scattered, leading to a broad spectral shape above $\sim20~\rm keV$, referred to as the ``Compton hump".

By virtue of it being a relative measurement, this method infers distances from the BH in units of gravitational radii ($r_{\rm g}=GM/c^2$), making it an ideal tool for measuring the spin of BHs for which there are no mass or distance estimates. However, while relativistic reflection does not require prior information about the BH in the system, similarly to the continuum fitting method, it also depends on the assumption that the optically thick, geometrically thin accretion disk (\citealt{1973A&A....24..337S}) extends all the way to the Innermost Stable Circular Orbit (ISCO) and that matter within the ISCO is on plunging orbits and has no time to contribute significantly to the observed flux. As the size of the ISCO is determined by gravity (\citealt{1972ApJ...178..347B, 1973blho.conf..343N}), relativistic reflection measures the size of the BH ISCO, directly probing the BH spin. The assumption of an inner disk edge consistent with the ISCO is motivated by numerical simulations (see e.g., \citealt{2008ApJ...675.1048R, 2008ApJ...676..549S, 2016ApJ...819...48S}) and observational results (see e.g., \citealt{2010ApJ...718L.117S, 2013MNRAS.431.3510S, 2015ApJ...813...84G}) suggesting that for Eddington fractions between $10^{-3}\lesssim L_{disk}/L_{Edd}\lesssim 0.3$, there is a sharp inner disk boundary, consistent with the ISCO. It is important to note that works such as \cite{2009ApJ...707L..87T} and \cite{2020ApJ...893...42X} find evidence of disk truncation at Eddington fractions of 0.14\% and 0.18\%, suggesting that while possible in the low Eddington regime, it is not always a requirement for the accretion disk to extend to the ISCO.

Additionally, relativistic reflection depends on assumptions regarding the properties of the compact corona that produces the radiation incident of the disk, that is then reprocessed (``reflected"). Most current models either assume a ``lamp-post" geometry, or parameterize the emissivity of the corona in terms of radial emissivity indices, which remove the need of a geometrical description of the corona. Lastly, the inclination of the inner disk regions also has an effect on determining spectral features, and is usually treated as a free parameter in reflection models. While some studies assume that the inclination of the inner disk is the same as the orbital inclination, that is not necessarily the case, and the two can be different due to disk tearing and the Bardeen-Petterson effect (\citealt{1975ApJ...195L..65B, 2015MNRAS.448.1526N, 2021MNRAS.507..983L}).

\begin{figure}[ht]
    \centering
    \includegraphics[width= 0.55\textwidth]{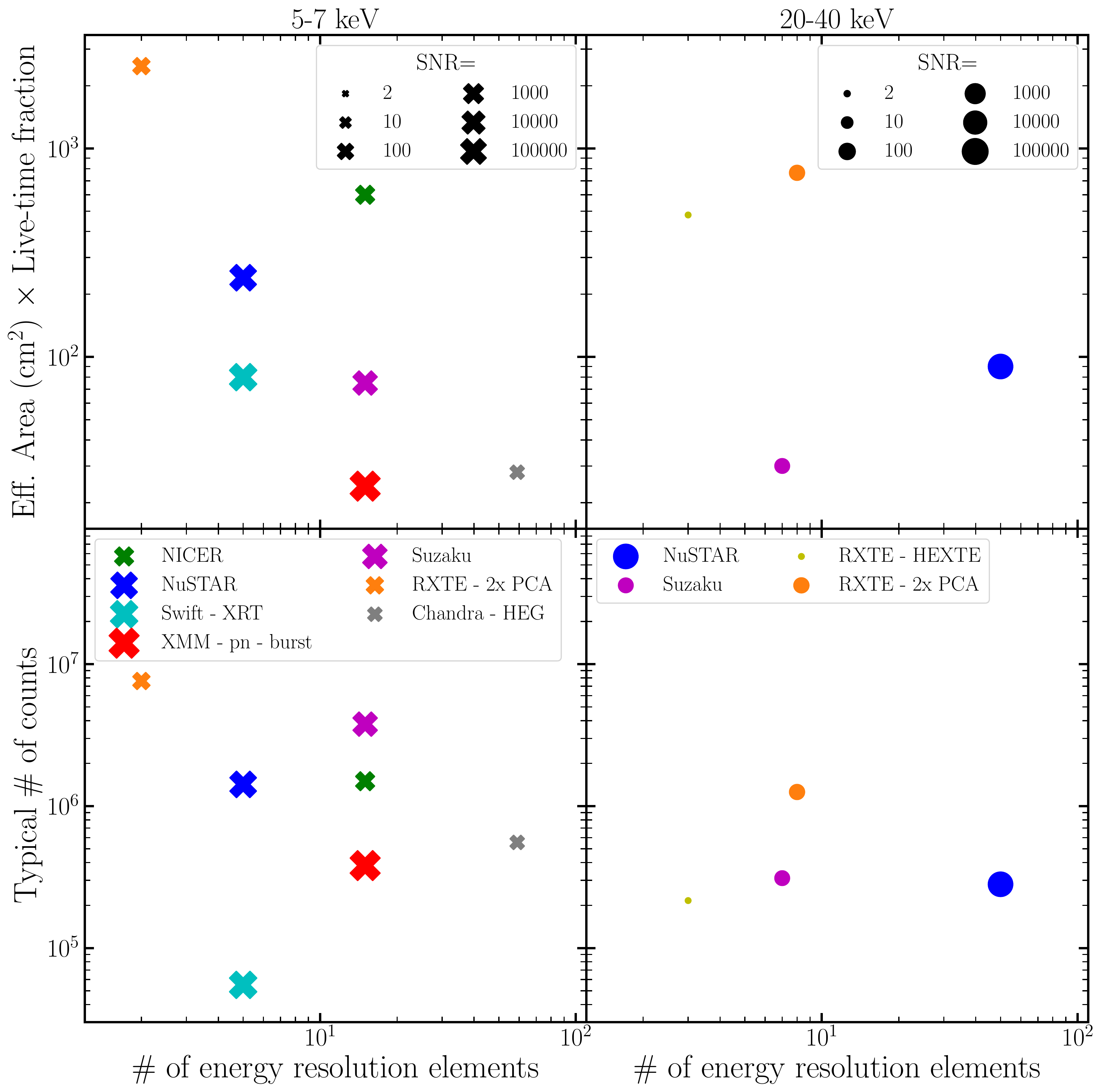}
    \caption{The top panels show the effective area multiplied by the approximate live-time fraction during observations versus the number of energy resolution elements in the 5-7 keV (left) and 20-40 keV (right) for some major X-ray observatories. The bottom panels show the typical number of counts per observation versus the number of energy resolution elements in the same bands. The point sizes are proportional to the typical signal to noise ratio during observations. All values are estimated for a source with a flux of 1 Crab, using typical exposure duration for each observatory: 20ks for NuSTAR, 5ks for NICER, 1ks for Swift, 20ks for XMM-Newton, 80ks for Suzaku, 5ks for RXTE, 30ks for Chandra.}
    \label{fig:instruments}
\end{figure}

%paragraph about NuSTAR
The launch of NuSTAR (\citealt{2013ApJ...770..103H}) in 2012 revolutionized relativistic reflection spin measurements, with the size of the sample of measured stellar-mass BH spins more than tripling through the addition of measurements made using NuSTAR data. This improvement is attributable to the wide band pass that NuSTAR offers, the high sensitivity and spectral resolution at high energies, and the fact that the instrument does not suffer from effects such as pile-up. The top panels of Figure \ref{fig:instruments} show the product of the effective area and the live-time fraction of some major X-ray observatories and the bottom panels show the typical number of counts as a function of the number of resolution elements in the 5-7 keV (left) and 20-40 keV (right) bands, with the size of the points being proportional to the signal to noise ratio (SNR). The number of counts was estimated for a source with a flux of 1 Crab, accounting for the effective area of the instrument, the live-time fraction, and an estimated typical duration of an observation obtained through an analysis of archival observations. In the 5-7 keV band, NuSTAR offers a combination of high SNR, large number of counts, and sufficient number of resolution elements to accurately observe the shape of the Fe K line. In the 20-40 keV, when compared to other instruments, NuSTAR provides unparalleled capabilities to describe the Compton hump feature of reflection spectra. All these advantages make NuSTAR the most appropriate instrument for spin measurements through relativistic reflection. 

There are (on the order of $\sim40$) examples of individual black hole spin measurements in papers adopting different methods. Additionally, most spin measurements report only statistical uncertainties, while the systematic uncertainties of the method have not yet been properly understood, due to the effects mentioned above. This is particularly important for measurements such as that of MAXI J1803-298, where \cite{2021arXiv211202794F} found $a=0.991\pm0.001$, or of EXO 1846-031, where \cite{2020ApJ...900...78D} measured $a=0.997^{+0.001}_{-0.002}$, where even though the accuracy is probably very high, the precision of the measurement is likely overestimated. 

%paragraph about our paper as a whole - nr of sources and description of different parts of the paper.
We aim to make a comprehensive, uniform treatment of black hole spin with entirely consistent methods and systematic errors, which is needed in order to completely understand the events that create black holes, the evolution of black holes in binary systems, accretion onto black holes, the relativistic effects onto the matter surrounding them, and the origins of the population of massive black holes in binary mergers observed through gravitational waves. Therefore, we analyzed the entire sample of archival NuSTAR observations of X-ray binary systems harboring BHs that \textit{did not previously have a spin measurement obtained through relativistic reflection from NuSTAR data}. The sources were selected initially based on the BHs identified in the WATCHDOG (\citealt{2016ApJS..222...15T}) and BackCAT (\citealt{2016A&A...587A..61C}) catalogs, followed by a survey of journal literature indicating BH candidates. We excluded sources that presented evidence or thermonuclear bursts (or burst-like behavior) in order to exclude neutron star candidates.  We analyzed 24 sources across 115 observations, and measured 10 new BH spins. We describe our data analysis methods in Section \ref{sec:analysis}, we present the analysis that led to the 10 new measurements in Section \ref{sec:new_measurements}, and discuss the sources for which a spin measurement was not possible in Appendix \ref{sec:no_measurement}. In Section \ref{sec:distribution}, we present the updated distribution of BH spins in XB. Lastly, we summarize our results and discuss their implications in Section \ref{sec:discussion}. For clarity, we grouped all figures and tables in Appendix \ref{sec:figures_and_tables}.

\section{Data Analysis} \label{sec:analysis}
All the NuSTAR spectra were extracted using the \texttt{nustardas} v2.1.1 routines in Heasoft v6.29c, using the NuSTAR CALDB version 20211103. The spectra from the two NuSTAR FPM detectors were extracted from circular regions with 120'' radius centered at the location of the source, taken from the header of the event file. Background spectra were extracted from annuli with an inner radius of 200'' and an outer radius of 300'', concentric with the source region, and subtracted from the source spectra. Our choice of background region does not significantly change the background spectrum when compared to the spectrum extracted from a circular background region selected on a different location on the detector. This is especially the case since the observed sources are generally bright, with count rates larger than a few counts per second. For additional explanation regarding this choice of source and background regions, together with the impact that this might have on the reported results, see Appendix \ref{sec:regions}. The spectra are then binned using the ``optimal" binning scheme presented by \cite{2016A&A...587A.151K}, using the ``ftgrouppha" ftool, and truncated at the energy where background emission begins to dominate over the source spectrum.

The spectral analysis was performed using \texttt{Xspec} v12.12.0g (\citealt{1996ASPC..101...17A}), using $\chi^2$ statistics with standard weighting, and using the \texttt{leven} minimization method. Throughout the analysis, we used the \texttt{wilm} abundances (\citealt{2000ApJ...542..914W}) and the \texttt{vern} photoelectric absorption cross-sections (\citealt{1996ApJ...465..487V}). 

When fitting NuSTAR spectra from the two FPM detectors, it is customary to link all parameters of the model for the spectra from the two detectors and to allow a multiplicative constant offset between the spectra. However, this sometimes produces different residuals when fitting the two spectra, especially at low energies $\leq5\;\rm keV$ (see e.g., Draghis et al. 2022). Therefore, instead of adopting a multiplicative constant to account for the difference between the spectra from the two FPM NuSTAR detectors, we allowed the normalizations of the model components to be different for the models describing the two spectra, linking all other parameters. This method not only accounts for differences in absolute flux sensitivity between the detectors, but also for possible variations in spectral response. For example, instead of fitting the two spectra with the model \texttt{constant*TBabs*(diskbb+powerlaw)} and linking all parameters except for the \texttt{constant} component, we fit the two spectra using the model \texttt{TBabs*(diskbb+powerlaw)}, but allowing the normalization of the \texttt{diskbb} and \texttt{powerlaw} components to vary between the two detectors and linking all other parameters, therefore introducing one extra free parameter. Despite having one extra free parameter, the quality of the fits is almost always significantly improved by allowing free normalizations when compared to having a multiplicative constant account for the difference between spectra, and the parameters of the models are not generally strongly influenced. By using this method, the fits can better constrain the shape of the continuum without forcing the reflection features to account for instrumental differences between the detectors. This allows more clearly distinguishing reflection features from the underlying continuum.

\subsection{Choice of Models} \label{sec:models}
In order to test the possible systematic effects introduced through our choice of models, throughout our analysis we start by fitting spectra with a baseline array of seven models. First, all models include a \texttt{TBabs} component in order to describe galactic absorption along the line of sight, using the value estimated by the web version of the $\rm N_H$ Column Density FTOOL\footnote{\url{https://heasarc.gsfc.nasa.gov/cgi-bin/Tools/w3nh/w3nh.pl}} for each source as a starting point. Second, in order to describe the emission from the accretion disk, we include a \texttt{diskbb} component to all models, even those describing observations that occurred during particularly hard states. If the \texttt{diskbb} component is not required by the model, the temperature of the disk and/or the normalization of the component will be driven to values indicating that the component does not contribute significantly to the model. Lastly, to describe the coronal emission, we add a \texttt{powerlaw} component. While this does not probe the effects of relativistic reflection, it allows us to quantify statistically if reflection is present: we can compare this to the quality of the fit with models that account for reflection in order to probe if reflection is indeed present. This gives us the first model variation, in \texttt{Xspec} language \texttt{TBabs*(diskbb+powerlaw)}.

In order to model reflection features, we used the \texttt{relxill} v.1.4.3 (\citealt{2014MNRAS.444L.100D, 2014ApJ...782...76G}) family of models. For a complete description of the models, their variations, and of all model parameters, see the website of the model\footnote{\url{http://www.sternwarte.uni-erlangen.de/~dauser/research/relxill/} }, Section 3.1 in \citealt{2021ApJ...920...88D}, or Appendix A in Draghis et al. 2022. We note that since the analysis was performed, an updated version of \texttt{relxill} was launched (v2.0 - \citealt{2022MNRAS.514.3965D}) that includes the effects of returning radiation on the observed spectra. However, since the main focus of the paper is to measure the spins of the compact objects in the systems and since the addition of returning radiation does not influence spin measurements significantly (see Figure 12b in \citealt{2022MNRAS.514.3965D}), we chose to continue using version 1.4.3 in our analysis. %However, given the automated nature of our pipeline, these measurements can be verified in the future using new models.

To probe the extent of the theoretical assumptions of the \texttt{relxill} family of models, we replace the \texttt{powerlaw} component with six variations of the \texttt{relxill} components. In order to test the effect of the assumption regarding the shape of the underlying continuum, we test both the simple \texttt{relxill} variant and \texttt{relxillCp}. To test the effect of the assumed coronal geometry, we also test the \texttt{relxilllp} variant, which assumes a ``lamp-post" geometry, while the previous two models attempt to measure the coronal emissivity directly, without assuming a ``lamp-post" geometry. Lastly, in order to probe the effect of the density of the accretion disk, we also tested the \texttt{relxillD} variant. However, when allowing the disk density to vary, it is almost always unconstrained by fits. Therefore, we test 3 variations of the \texttt{relxillD} component, by fixing the disk density at $n=10^{15}$, $10^{17}$, and $10^{19}\;\rm cm^{-3}$. This gives us the array of six reflection components that we used to describe relativistic reflection, by replacing the \texttt{powerlaw} component in our first model variation. We leave all other parameters in the models free, with the exception of the outer disk radius, which we fix at $r_{\rm out}=990\;r_g$. As the radial emissivity profiles of the coronal emission are generally steep, the outer regions of the disk contribute insignificantly to the total observed emission, making the outer disk radius nearly impossible to constrain for such systems. Therefore, we fixed this parameter at nearly the maximum value allowed in the model. Additionally, unless otherwise noted, we fix the inner disk radius at $r_{\rm in}=r_{\rm ISCO}$ for observations that occur while the source was at an Eddington fraction between $10^{-3}\lesssim L_{disk}/L_{Edd}\lesssim 0.3$, calculated using the flux in the $0.1-10\;\rm keV$ band and the measurements in literature for the BH mass and distance to the systems. Where no mass or distance measurements were available, we analyze the observations that fall within this Eddington fraction range when assuming generic values for these parameters (e.g., $M_{\rm BH}\sim8\pm 4 M_{\odot}$ and $d\sim 10\pm 5 \rm kpc$).

\subsection{MCMC analysis} \label{sec:MCMC}
After fitting the data with our baseline array of models and finding the best-fit solution for each \texttt{relxill} variant tested, we ran a Markov Chain Monte Carlo (MCMC) analysis on the models performing best in terms of the $\chi^2$ statistic. In cases where identifying the best performing models was not trivial, we calculated the Akaike Information Criterion (AIC - \citealt{1974ITAC...19..716A}) and the Bayesian Information Criterion (BIC - \citealt{1978AnSta...6..461S}) using the following formulas:
\begin{equation}
    AIC=Nln(\chi^2/N)+2N_{var}
\end{equation}
\begin{equation}
    BIC=Nln(\chi^2/N)+ln(N)N_{var} \\ 
\end{equation}
where $N$ is the number of data bins and $N_{var}$ is the number of variables in the model. 

We used the ``best-fit" parameter combination to generate a Gaussian proposal distribution for initiating the walkers in the MCMC analysis, and used uniform priors in the range allowed by the parameters in the model. For running the MCMC analysis, we used the \texttt{Xspec} \texttt{emcee} implementation written by A. Zoghbi\footnote{ \url{https://zoghbi-a.github.io/xspec\_emcee/}}. We chose the number of walkers in the analysis as a large integer, on the order of a few (4-5) times the number of free parameters in the analysis. In oder to ensure chain convergence, we ran the chains for $3-4000\times\tau_f$, where $\tau_f$ represents the integrated autocorrelation time (\citealt{Sokal1996MonteCM} recommend running the chains with more steps than $\sim 1000\times \tau_f$ in order to ensure convergence). Therefore, we ran the MCMC chains with 200 walkers for a total of $2.5\times10^6$ steps, and disregarded the first $5\times10^5$ steps of the chains as a ``burn-in" phase. 

When running the MCMC analysis on more than one model, we also computed the Deviance Information Criterion (DIC, \citealt{DIC_text}) and reported the model that performs best in terms of DIC. We computed the DIC using:
\begin{equation}
    DIC=\overline{D(\theta)}+p_D
\end{equation}
where $\theta$ represents the unknown parameters in the model, $D(\theta)$ represents the deviance, computed as $D(\theta)=-2\log(p(y|\theta))+C$ (where $y$ represents the data, $p(y|\theta)$ is the likelihood function, and $C$ is some constant which cancels out when comparing models), and $p_D$ represents the effective number or parameters of the model, computed as $p_D=\frac{1}{2}\overline{{\rm var}(D(\theta))}$ (\citealt{MR2027492}). This information criterion has the advantage of probing not only the ``best-fit" solution, but also the shape of the entire posterior distribution of the parameter space around the best solution. 

The MCMC analysis was run for each individual observation. For sources with multiple observations, this produces a number of measurements for each parameter in the models equal to the number of observations. We report the modes of the posterior distributions and the $1\sigma$ highest posterior density credible intervals of our measurements in the tables at the end of the paper (see Section \ref{sec:figures_and_tables}). We chose to report the modes of the posterior distributions instead of the medians in order to highlight the value that occurs most often in the MCMC analysis for each parameter, suggesting a higher preference for the specific value. 

Additionally, throughout the paper, we chose to report the $1\sigma$ highest posterior density credible intervals (i.e. the shortest interval in the posterior distribution of a parameter that contains $68.3\%$ of the posterior samples) instead of an equal-tailed confidence interval (i.e. having an equal number of samples below the interval as above it) as this is better suited for distributions that are clustered around the edge of a parameter space. As an example, this is often the case for extremely high spin measurements.

\subsection{Combining Measurements} \label{sec:combining}

When there are multiple measurements of the same parameter (i.e. BH spin or inclination), it can be difficult to report a single value. While one could choose to report the parameter combination from a single observation, such as the one with highest total number of counts in the observation, the one happening during the hardest spectral state, or the one where the reflection strength is the highest relative to the continuum, this inevitably leads to a loss of the information from the other observations. An alternative that takes advantage of all available data is to fit all spectra simultaneously, and link parameters that are expected to remain constant throughout the observations, such as the BH spin and the inclination of the inner disk, ensuring that the values measured are constrained by all the spectra at the same time. While this method would be the ideal way forward to ensure that all data is fully used, jointly fitting multiple NuSTAR spectra with complicated models that include multiple components becomes a very slow, computationally intensive process. In practice, fitting more than 2-3 observations, each having two spectra, becomes nearly impossible, with each iteration in the $\chi^2$ minimization algorithm in \texttt{Xspec} lasting on the order of tens of seconds to minutes. With the time required to run each iteration increasing significantly through the addition of extra spectra, fitting $\sim10$ observations simultaneously (for sources such as 4U 1957+11, H 1743-322, or MAXI J1820+070) is currently impossible due to limited computing power. Additionally, even if one were to find the parameter combination for which the "best fit" is achieved, running the MCMC analysis described in Subsection \ref{sec:MCMC} would be an even more computationally demanding process, by many orders of magnitude. 

Therefore, due to these limitations, we ran our analysis by fitting the spectra produced by each observation individually. We did, however, simultaneously fit the two spectra produced by the two FPM NuSTAR detectors during the same observation, as described at the beginning of Section \ref{sec:analysis}.  We implemented a Bayesian method of combining measurements from multiple observations into a single value, that we report. While most parameters in the models are not expected to stay constant in time, the BH spin, the inclination of the inner accretion disk, and the Fe abundance are likely to not change in time. However, due to possible correlations between the assumed disk density in the model and the Fe abundance (see e.g., \citealt{2018ApJ...855....3T}), we do not report a combined Fe abundance measurement, but we do report the individual Fe abundance obtained from each observation in the tables in Section \ref{sec:figures_and_tables} for future exploration of the models. Therefore, throughout the paper, for sources that have been observed more than once, we report the spin and inclination obtained by combining the measurements from individual observations, and weight the contribution of each observation to the final measurement through the ratio of the reflected to total flux in the 3-79 keV band. For a detailed explanation of the method used to combine the results from multiple measurements and a comparison between this method and a joint, simultaneous fit of multiple observations, see Appendix \ref{sec:combining_explanation}.

\section{New Spins}\label{sec:new_measurements}
\subsection{IGR J17454-2919}
\begin{figure}[ht]
    \centering
    \includegraphics[width= 0.45\textwidth]{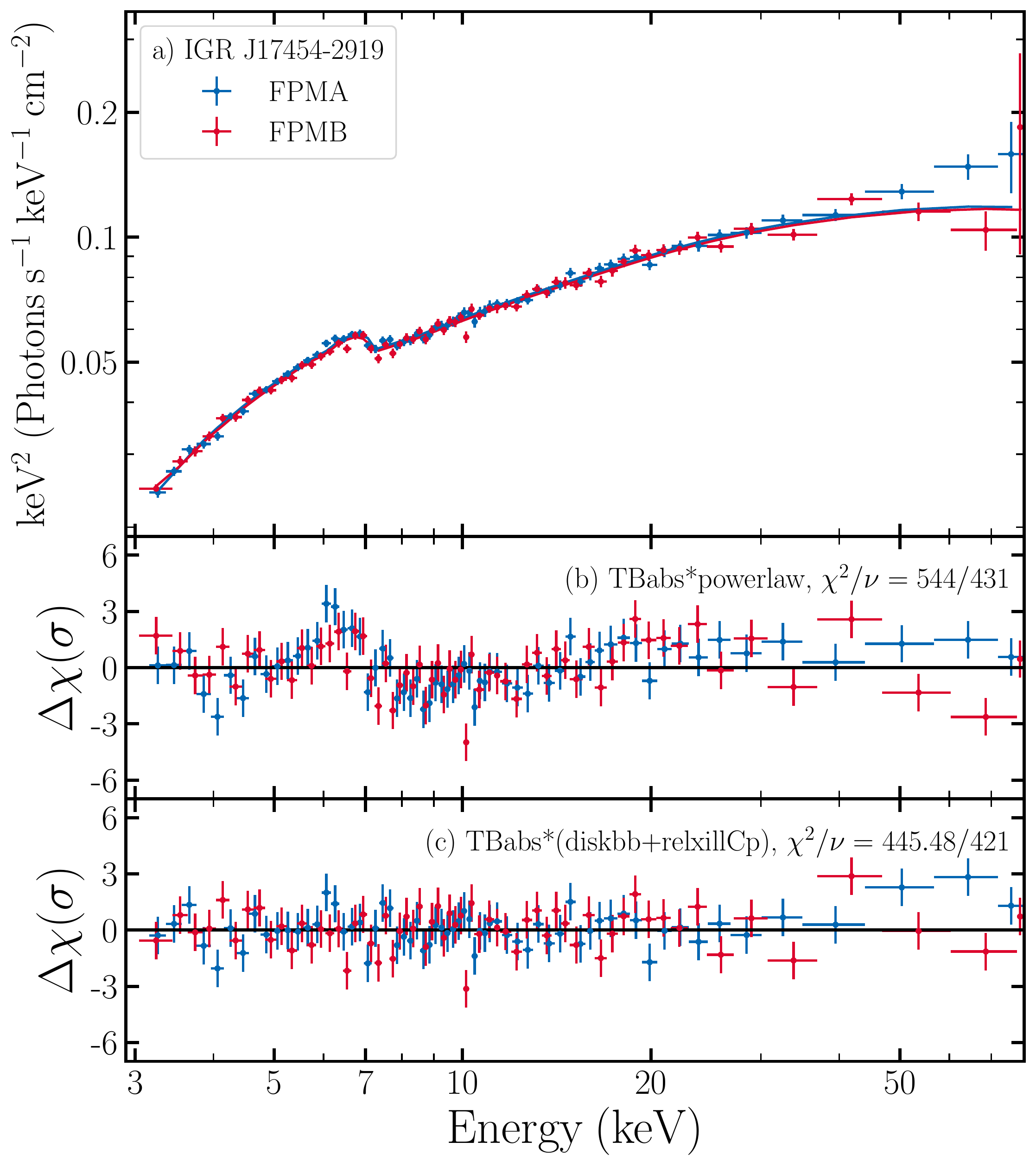}
    \caption{Top: unfolded spectrum of IGR J17454-2919, with the NuSTAR FPMA spectrum shown in blue and FPMB spectrum in red. The reported best-fit model is shown by the solid lines. The middle and bottom panels show the residuals in terms of $\sigma$ produced when fitting the spectra with \texttt{TBabs*powerlaw} (middle) and \texttt{TBabs*(diskbb+relixllCp)} (bottom), together with the statistic produced.}
    \label{fig:IGR_J17454-2919_delchi}
\end{figure}

At the time of the analysis, one NuSTAR observation of IGR J17454-2919 (ObsID 80001046002) was available in the archive. For this observation, we fit the entire 3-79~keV spectrum. Fitting with \texttt{TBabs*powerlaw} gives $\chi^2/\nu=544/433$. Reflection features and soft emission consistent with an accretion disk are present. The spectra and residuals of the fits in terms of $\sigma$ (i.e. $\frac{\rm data-model}{\rm error}$) are shown in Figure \ref{fig:IGR_J17454-2919_delchi}.

The best performing model, both in terms of $\chi^2$ and DIC, is \texttt{TBabs*(diskbb+relxillCp)}, with $\chi^2/\nu=445.48/421$. Based on the MCMC analysis performed on this model, the measured spin is $a=0.97^{+0.03}_{-0.17}$ and the inclination $\theta=60^{+8}_{-14}$ degrees. The full set of parameters determined through the MCMC analysis is presented in Table \ref{tab:table1}.

\subsection{GRS 1758-258}

\begin{figure}[ht]
    \centering
    \includegraphics[width= 0.45\textwidth]{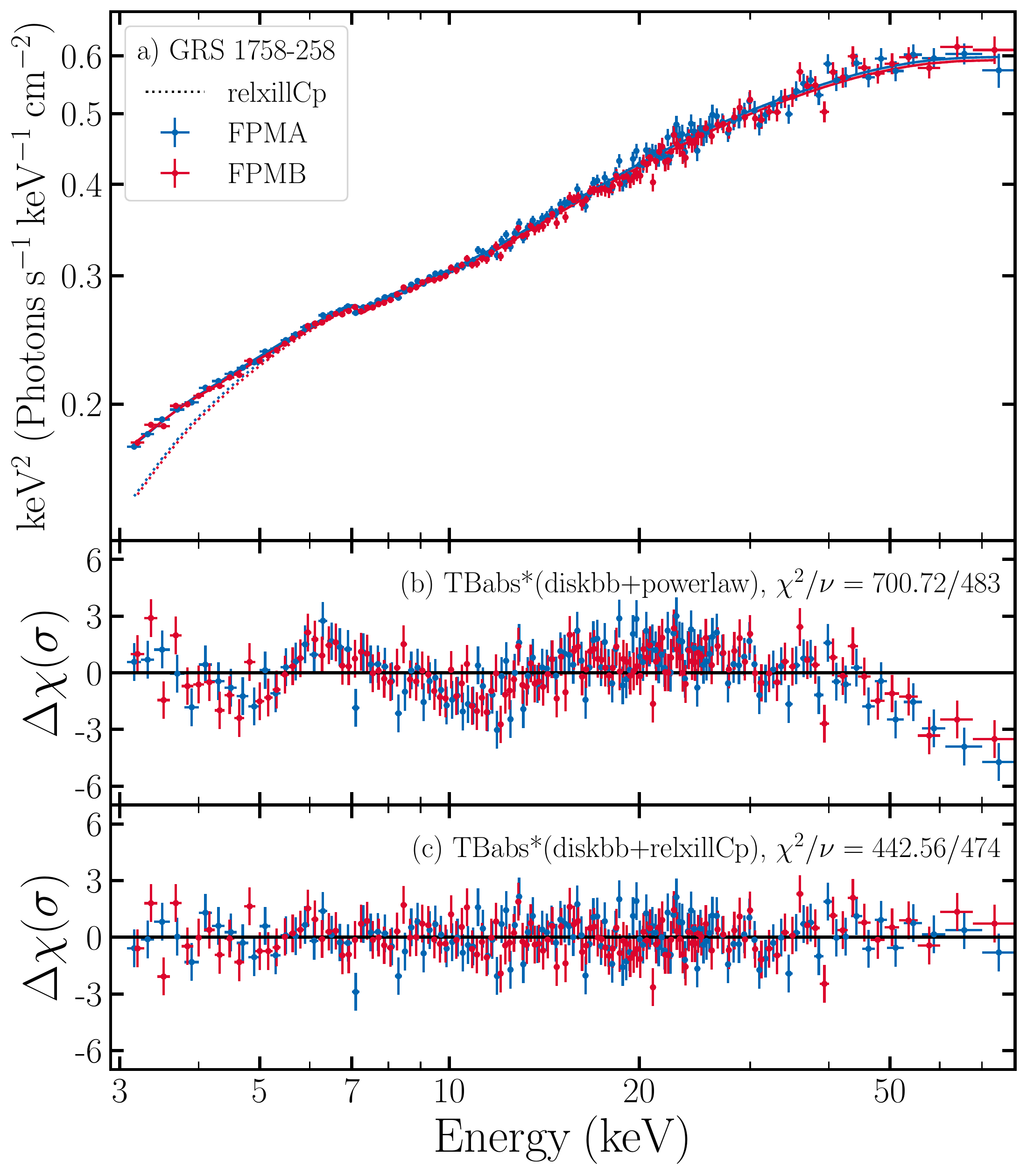}
    \caption{Top: unfolded spectrum of GRS 1758-258, with the NuSTAR FPMA spectrum shown in blue and FPMB spectrum in red. The reported best-fit model is shown by the solid lines, while the contribution of the \texttt{relxillCp} component is shown by the dotted lines. The middle and bottom panels show the residuals in terms of $\sigma$ produced when fitting the spectra with \texttt{TBabs*(diskbb+powerlaw)} (middle) and \texttt{TBabs*(diskbb+relxillCp)} (bottom), together with the statistic produced.}
    \label{fig:GRS_1758-258_delchi}
\end{figure}

Fitting the entire 3-79~keV NuSTAR spectrum of GRS 1758-258 from ObsID 30401030002 using \texttt{TBabs*(diskbb+powerlaw)} gives $\chi^2/\nu=700.72/483$, with the residuals clearly indicating reflection features (see Figure \ref{fig:GRS_1758-258_delchi}).

The three best performing reflection models in terms of $\chi^2$ and DIC are \texttt{TBabs*(diskbb+relxillCp)} ($\chi^2/\nu=442.56/474$, DIC = 467.31), \texttt{TBabs*(diskbb+relxillD)} with Log(N)=19 ($\chi^2/\nu=445.95/475$, DIC = 478.26), and \texttt{TBabs*(diskbb+relxill)} ($\chi^2/\nu=449.65/474$, DIC = 475.2). Of the three models, \texttt{TBabs*(diskbb+relxillCp)} performs best both in terms of $\chi^2$ and DIC. Therefore, we report the measurements of this model, but note that all measurements are consistent throughout the three models. Through the MCMC analysis we find $a=0.991^{+0.007}_{-0.019}$ and $\theta=66^{+8}_{-5}$ degrees. The entire set of parameters produced by the MCMC analysis is presented in Table \ref{tab:table1}. 

We note that \cite{2022arXiv220801399J} published a spin measurement for GRS 1758-258 shortly \textit{before} the submission of this paper. They measure $a>0.92$, and our measurement is within good agreement. Additionally, we find a similar ionization parameter, Fe abundance, and reflection fraction. However, the inclination measurements and emissivity profiles are different, likely owing to the fact that their analysis did not include a \texttt{diskbb} component, which they argue was not statistically required by the data. We find that for all three models presented above, the \texttt{diskbb} component is required, both in terms of reduced $\chi^2$, AIC and BIC. Therefore we continue to include the \texttt{diskbb} component in our analysis. The consistent spin measurements regardless of choice of continuum is encouraging when considering the systematic effects that act on these spin measurements. Since relativistic reflection works by disentangling the reflection features from the underlying continuum, it is encouraging that the models are able to isolate and characterize the shape of reflection even when slightly different continua are chosen. Nevertheless, due to the temporal proximity of the result of \cite{2022arXiv220801399J} to the submission of our manuscript, we chose to not exclude this source from this sample.

\subsection{MAXI J1727-203}
\begin{figure}[ht]
    \centering
    \includegraphics[width= 0.45\textwidth]{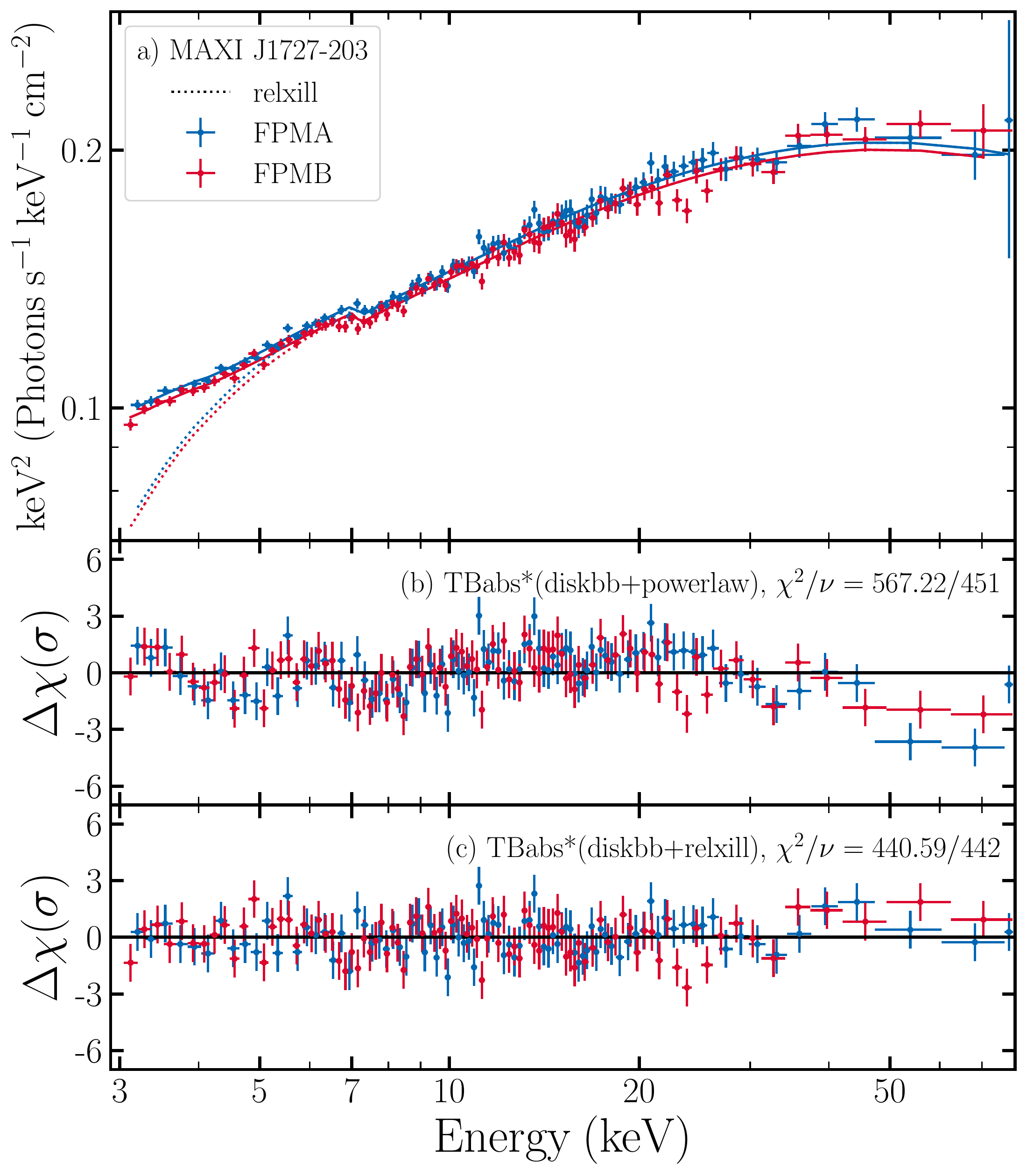}
    \caption{Top: unfolded spectrum of MAXI J1727-203, with the NuSTAR FPMA spectrum shown in blue and FPMB spectrum in red. The reported best-fit model is shown by the solid lines, while the contribution of the \texttt{relxill} component is shown by the dotted lines. The middle and bottom panels show the residuals in terms of $\sigma$ produced when fitting the spectra with \texttt{TBabs*(diskbb+powerlaw)} (middle) and \texttt{TBabs*(diskbb+relxill)} (bottom), together with the statistic produced.}
    \label{fig:MAXI_J1727-203_delchi}
\end{figure}

MAXI J1727-203 was observed by NuSTAR once, in 2018 (ObsID 90401329002). We fit the entire 3-79~keV spectra from the observation. Fitting with \texttt{TBabs*(diskbb+powerlaw)} produces $\chi^2/\nu=567.22/451$, with the residuals indicating clear reflection features. The spectra and residuals of the fits are shown in Figure \ref{fig:MAXI_J1727-203_delchi}. 

The two best performing models are \texttt{TBabs*(diskbb+relxillCp)} ($\chi^2/\nu=438.87/442$, DIC=688.31) and \texttt{TBabs*(diskbb+relxill)} ($\chi^2/\nu=440.59/442$, DIC=473.09). As it performs much better in terms of DIC, we report the values measured from the MCMC run of the \texttt{TBabs*(diskbb+relxill)} model. Through the MCMC analysis we find $a=0.986^{+0.012}_{-0.159}$ and $\theta=64^{+10}_{-7}$ degrees, and the full set of parameters is reported in Table \ref{tab:table1}.

\subsection{Swift J1753.5-0127}
\begin{figure}[ht]
    \centering
    \includegraphics[width= 0.45\textwidth]{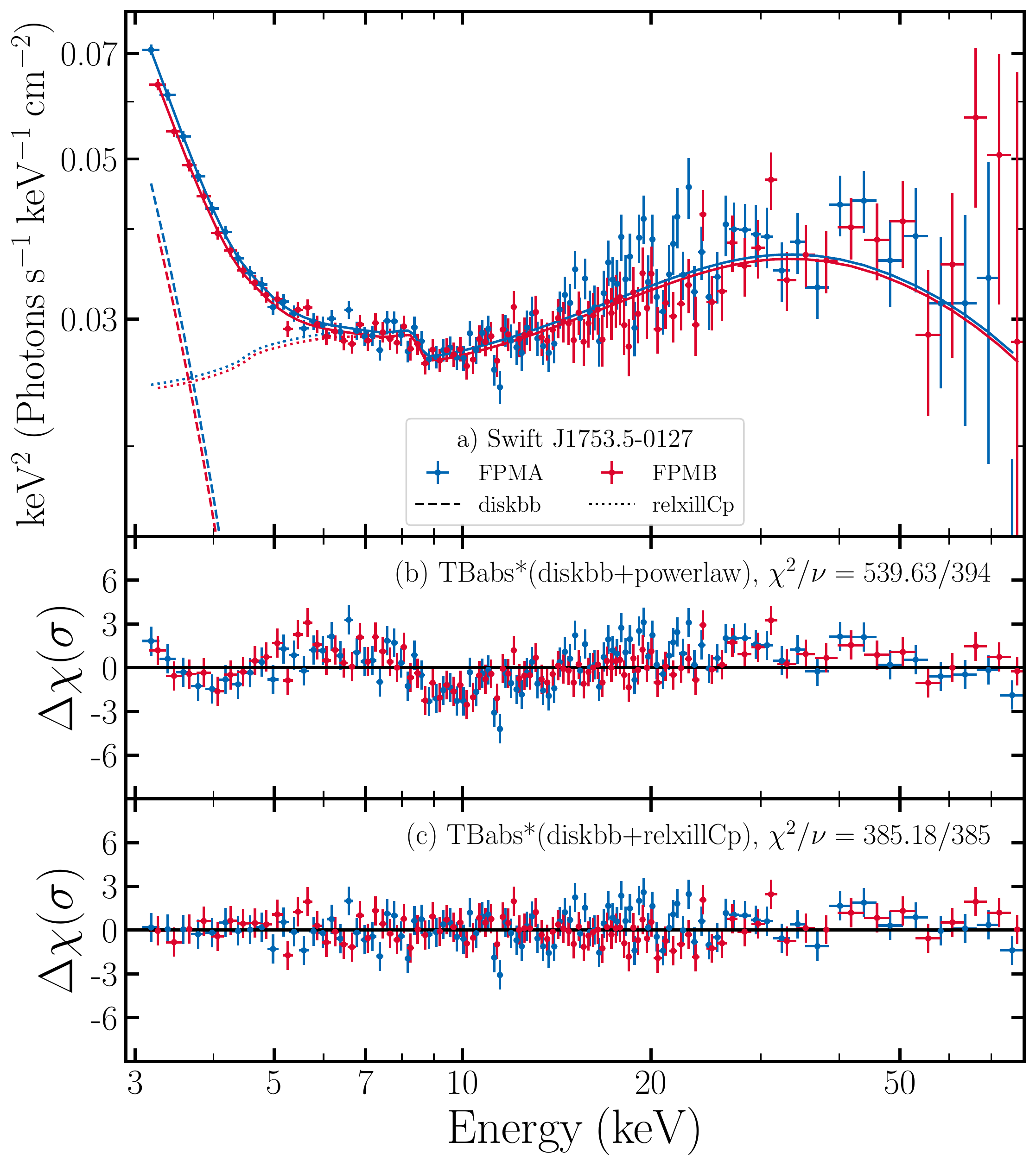}
    \caption{Top: unfolded spectrum of Swift J1753.5-0127, with the NuSTAR FPMA spectrum shown in blue and FPMB spectrum in red. The reported best-fit model is shown by the solid lines, while the contribution of the \texttt{relxillCp} component is shown by the dotted lines and the contribution of the \texttt{diskbb} component is shown through the dashed lines. The middle and bottom panels show the residuals in terms of $\sigma$ produced when fitting the spectra with \texttt{TBabs*(diskbb+powerlaw)} (middle) and \texttt{TBabs*(diskbb+relxill)} (bottom), together with the statistic produced.}
    \label{fig:Swift_J17535-0127_delchi}
\end{figure}

There are 5 NuSTAR observations of Swift J1753.5-0127. Of those, only one shows clear signs of reflection: ObsID 80001047002. Therefore, we use the spectra extracted from the observation 80001047002 in the entire 3-79 keV band. Fitting the spectra with \texttt{TBabs*(diskbb+powerlaw)} returns $\chi^2/\nu=539.63/394$, with clear indications of relativistic reflection. All reflection models with no coronal geometric assumption perform similarly, while the lamp-post model appears to converge to a different, statistically disfavored solution. The best performing reflection model, both in terms of $\chi^2$ and DIC, is \texttt{TBabs*(diskbb+relxillCp)}, with $\chi^2/\nu=385.18/385$. The posterior MCMC samples are nearly entirely concentrated at high spin values, measuring $a=0.997^{+0.001}_{-0.003}$ and $\theta=74^{+3}_{-3}$ degrees. These measurements are consistent with the MCMC run of other models without assumptions about the coronal geometry. We show the spectra and residuals of the fits in Figure \ref{fig:Swift_J17535-0127_delchi}, and the modes of the posterior distribution for each parameter in the MCMC analysis together with their 1$\sigma$ credible intervals are presented in Table \ref{tab:table1}.

The MCMC run of the lamppost model provides similar inclination constraints, but a much worse spin constraint $a=0.4^{+0.5}_{-0.8}$. This model produces a fit worse by $\Delta \chi^2=18.7$ for 2 fewer free parameters. The \texttt{relxillCp} flavor is preferred over the \texttt{relxilllp} both in terms of AIC (15.86 vs. 30.76) and BIC (79.76 vs. 86.67). An F-test returns a probability of $10^{-4}$, indicating that the non-lamppost models are indeed a statistically significant improvement. Additionally, the \texttt{relxilllp} model performs worse than the \texttt{relxillCp} model by $\Delta \rm DIC=13.57$. Lastly, visually inspecting the residuals of the fit clearly shows that the Fe K line is not fit as well by the \texttt{relxilllp} when compared to the \texttt{relxillCp} model. Therefore, we report the results of our best performing model (\texttt{TBabs*(diskbb+relxillCp)}): $a=0.997^{+0.001}_{-0.003}$ and $\theta=74^{+3}_{-3}$ degrees.

\subsection{MAXI J0637-430}\label{sec:MAXI_J0637}
\begin{figure}[ht]
    \centering
    \includegraphics[width= 0.95\textwidth]{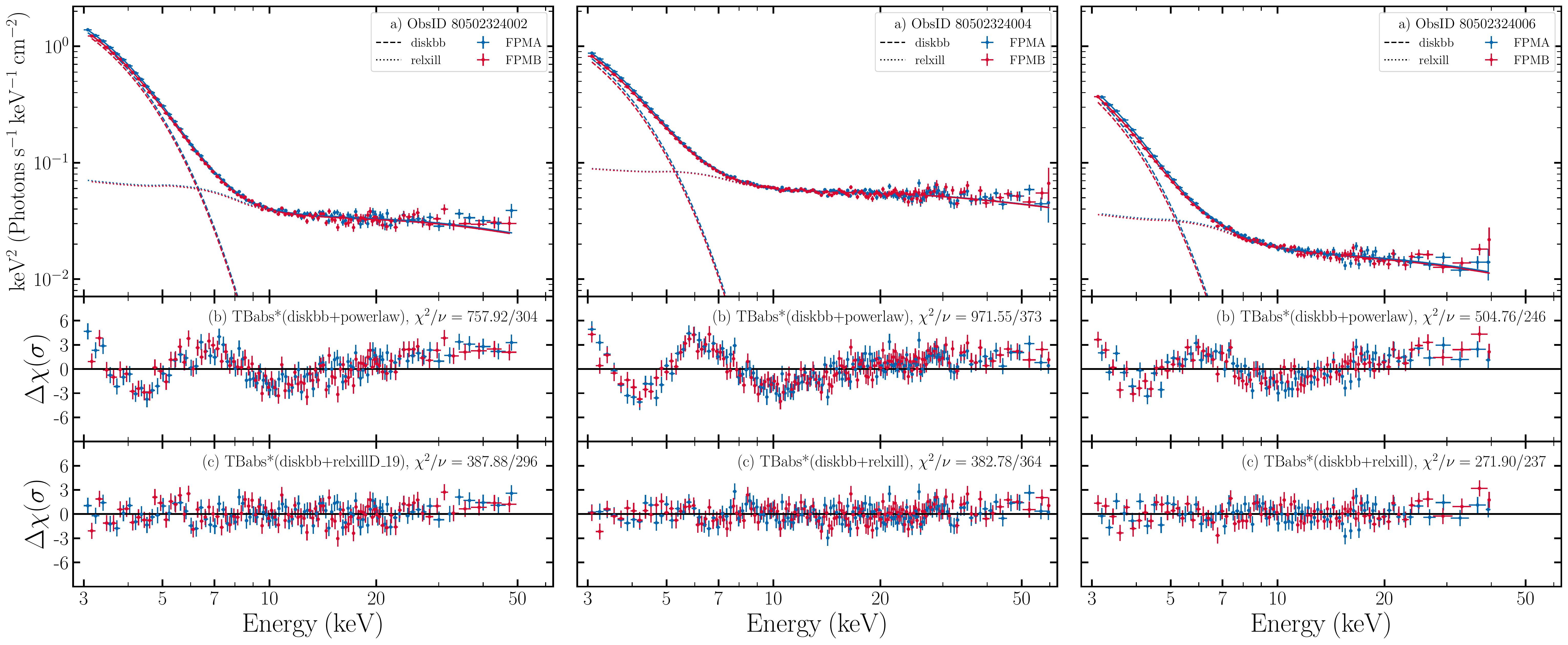}
    \caption{The three panels represent the three observations of MAXI J0637-430 analyzed. The top sub-panels show unfolded spectra, with the NuSTAR FPMA spectrum shown in blue and FPMB spectrum in red. The reported best-fit model is shown by the solid lines, while the contribution of the \texttt{relxill} component is shown by the dotted lines and the contribution of the \texttt{diskbb} component is shown through the dashed lines. The middle and bottom sub-panels show the residuals in terms of $\sigma$ produced when fitting the spectra with \texttt{TBabs*(diskbb+powerlaw)} (middle) and \texttt{TBabs*(diskbb+relxill)} (bottom), together with the statistic produced.}
    \label{fig:MAXI_J0637-430_delchi}
\end{figure}

At the time of the analysis, there were 8 NuSTAR observations of MAXI J0637-430, (ObsID 80502324002, 80502324004, 80502324006, 80502324008, 805023240010, 80502324012, 80502324014, 80502324016). The first three (80502324002, 80502324004, and 80502324006) were taken during the soft state of the late 2019 outburst of the source and are the only ones that show clear reflection features. We fit the spectra from ObsID 80502324002 in the 3--50~keV energy band, from ObsID 80502324004 in the 3--60~keV energy band, and from ObsID 80502324006 in the 3--40~keV energy band, as the spectra are background dominated at energies higher than the intervals mentioned. The unfolded spectra of the three observations are shown in the top panels of Figure \ref{fig:MAXI_J0637-430_delchi}, and the residuals of the fit to the \texttt{TBabs*(diskbb+powerlaw)} model are shown in the middle panels of Figure \ref{fig:MAXI_J0637-430_delchi}, indicating signs of relativistic reflection. 

When fitting the spectra with the models discussed in Subsection \ref{sec:models}, the \texttt{relxill} and \texttt{relxillD} with $\log(N)=19$ variants perform best in terms of $\chi^2$ for all three observations. After running the MCMC analysis on the two mentioned models for all three observations, the \texttt{relxill} variant performs best in terms of DIC for ObsID 80502324004 and 80502324006, while the \texttt{relxillD} with $\log(N)=19$ variant produces the better DIC for ObsID 80502324002. The residuals of the fits using these models are shown in the bottom panels of Figure \ref{fig:MAXI_J0637-430_delchi}, together with the fit statistic. 

\begin{figure}[ht]
    \centering
    \includegraphics[width= 0.95\textwidth]{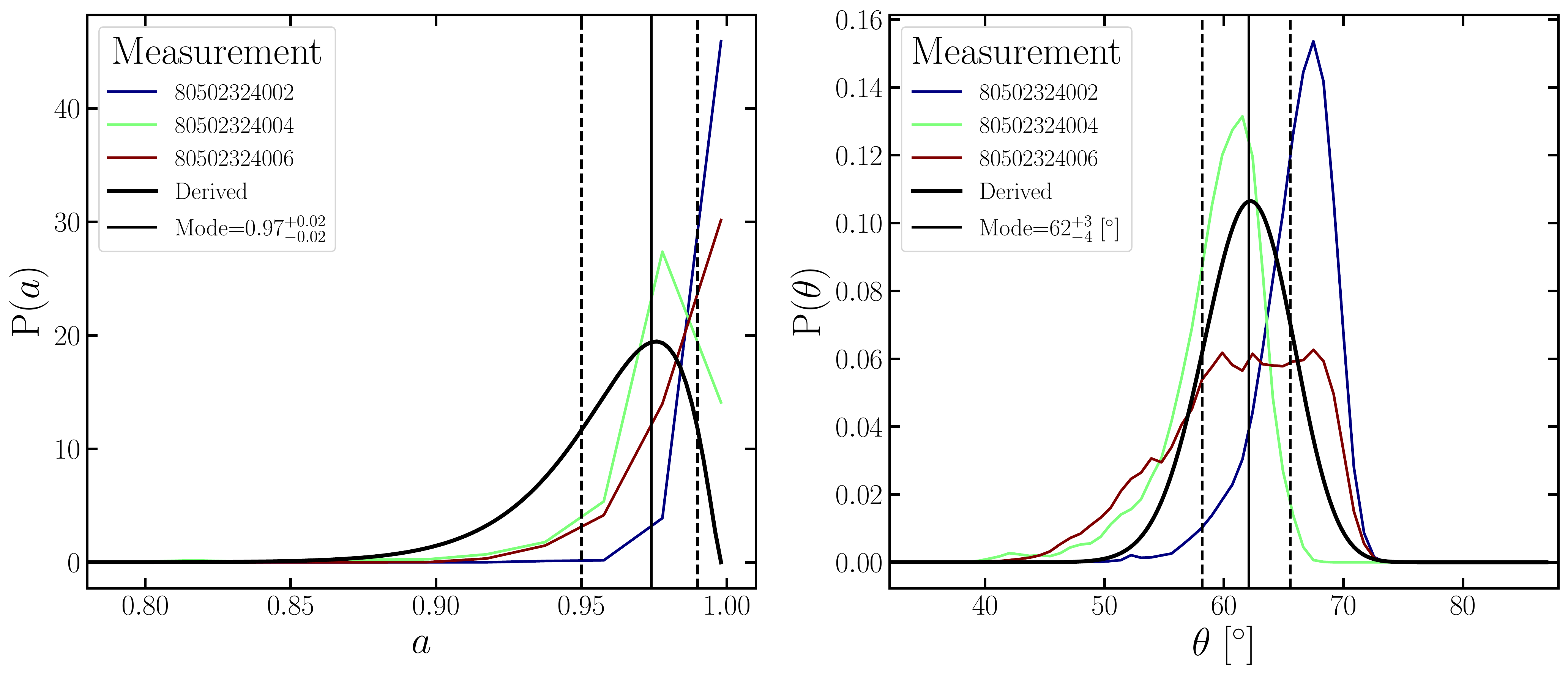}
    \caption{The left panel shows the posterior distributions resulting from the MCMC analysis of MAXI J0637-430 for spin, while the right panel shows the posterior distributions for the inclination of the inner accretion disk in the model. The different colored lines represent the three observations analyzed, and the black curves represent the combined inferred distribution, with the transparency of the lines being proportional to the ratio of reflected flux to total flux in the 3-79~keV band, which was used as weighting when combining the posterior distributions. The solid vertical black lines represent the modes of the combined distribution, and the dashed vertical black lines represent the $1\sigma$ credible intervals of the measurements. }
    \label{fig:MAXI_0637-430_combined}
\end{figure}

The modes of the posterior distributions for each parameter in the MCMC analysis, together with their $1\sigma$ credible intervals are shown in Table \ref{tab:table1}. The histograms of the posterior distributions for spin and inner disk inclination from the MCMC analysis on the three observations are shown in Figure \ref{fig:MAXI_0637-430_combined}. Running the combining algorithm described in Subsection \ref{sec:combining}, we obtain the distributions shown through the black curves in Figure \ref{fig:MAXI_0637-430_combined}, with medians and credible intervals $a=0.97\pm0.02$ and $\theta=62^{+3}_{-4}$ degrees.

\subsection{V4641 Sgr} \label{sec:v4641}
\begin{figure}[ht]
    \centering
    \includegraphics[width= 0.95\textwidth]{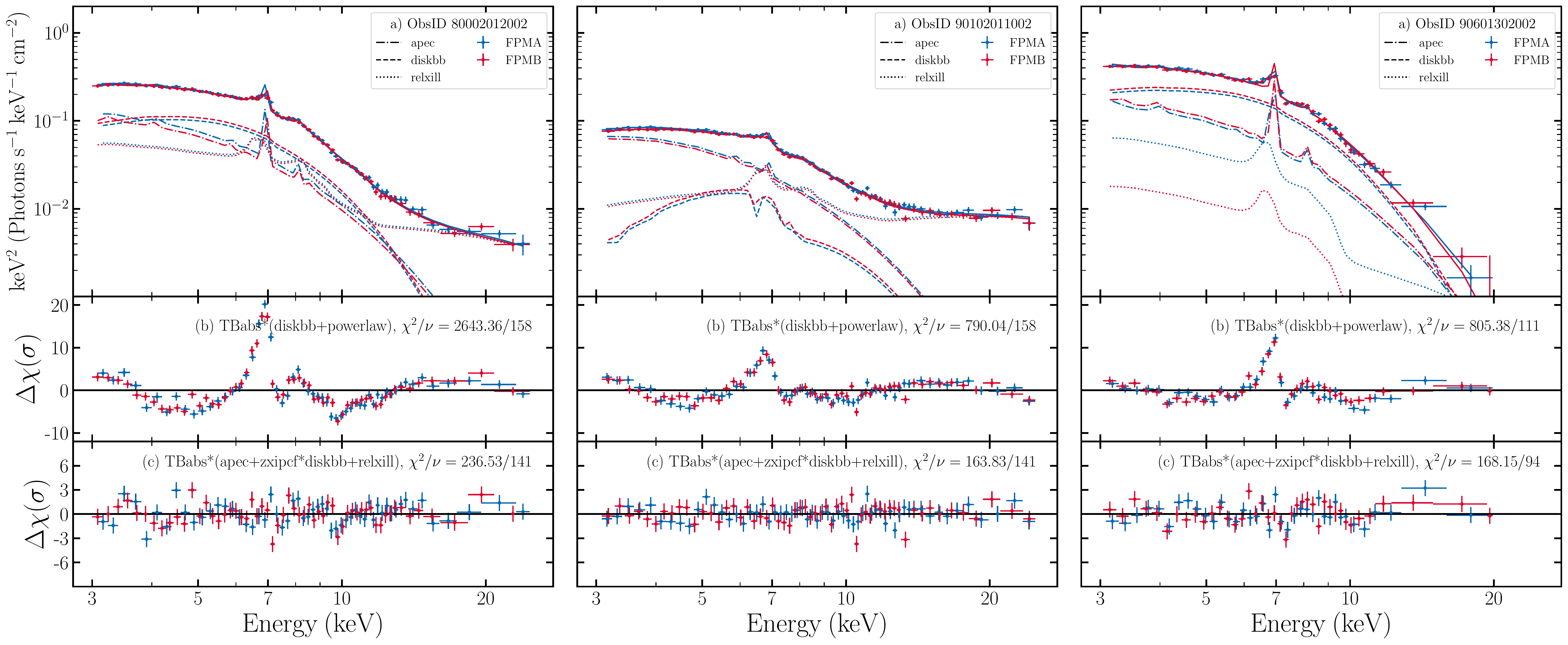}
    \caption{The three panels represent the three observations of V4641 Sgr analyzed. The top sub-panels show unfolded spectra, with the NuSTAR FPMA spectrum shown in blue and FPMB spectrum in red. The reported best-fit model is shown by the solid lines, while the contribution of the \texttt{relxill} component is shown by the dotted lines, the contribution of the \texttt{diskbb} component is shown through the dashed lines, and the contribution of the \texttt{apec} component shown in dot-dashed lines. The middle and bottom sub-panels show the residuals in terms of $\sigma$ produced when fitting the spectra with \texttt{TBabs*(diskbb+powerlaw)} (middle) and \texttt{TBabs*(apec+zxipcf*diskbb+relxill)} (bottom), together with the statistic produced.}
    \label{fig:V_4641_delchi}
\end{figure}

At the time of the analysis, four NuSTAR observations were public (ObsID 80002012002, 80002012004, 90102011002, and 90601302002). Using a mass of the black hole of $M=6.4\pm0.6\;\rm M_\odot$ and a distance to the system of $d=6.2\pm0.7\;\rm kpc$ (\citealt{2014ApJ...784....2M}), we calculate that only 3 of the 4 observations happen while the source had a luminosity within the Eddington range for which we expect the accretion disk to extend to the ISCO, with ObsID 80002012004 falling outside of that range. Therefore, we continue our analysis on the remaining 3 observations. We fit ObsID 80002012002 and 90102011002 in the 3--25~keV range, and ObsID 90601302002 in the 3--20~keV range. The top panels in Figure \ref{fig:V_4641_delchi} show the unfolded spectra from the three observations, and the middle panels show the residuals of the fit to the spectra when using \texttt{TBabs*(diskbb+powerlaw)}. The fits are poor, and the models clearly require additional components.

Simply replacing the \texttt{powerlaw} component with the six variations of the \texttt{relxill} model that we test throughout our analysis improves the quality of the fit, but not significantly. The quality of the fit is drastically improved through the addition of an \texttt{apec} (\citealt{2001ApJ...556L..91S}) component describing the emission from a collisionally-ionized, diffuse gas in the vicinity of the source, which is characterized by the plasma temperature, metal abundance, redshift, and normalization. Additionally, we allow for the presence of ionized partial covering of the \texttt{diskbb} component through a \texttt{zxipcf} (\citealt{2008MNRAS.385L.108R}) multiplicative component, which describes the effects of partially covering a fraction \emph{f} of a source by a photoionized absorber. For all three observations, the addition of the two new components improves the quality of the fit significantly. The choice for the \texttt{relxill} component does not influence the quality of the fit, so we continue using the default \texttt{relxill} variant, making the complete model \texttt{TBabs*(apec+zxipcf*diskbb+relxill)}. The residuals of the fits are shown in the bottom panels of Figure \ref{fig:V_4641_delchi}, together with the fit statistic. 

We ran the MCMC analysis starting from the best-fit solutions for each of the three observations. The modes of the posterior distributions for all the parameters in the MCMC runs, together with their $1\sigma$ credible intervals are presented in Table \ref{tab:table2}. Interestingly, in all 3 fits, the \texttt{apec} component requires a blueshift $z\sim-0.03$, far above the terminal velocity of the companion wind. This could indicate the presence of a disk wind, which has been previously hinted at by \cite{2022MNRAS.tmp.2123S}. The quality of the fit becomes significantly worse when attempting to fix the redshift of the component to zero. While the Fe abundance $A_{\rm Fe}$ in the \texttt{relxill} component is generally high, the abundance of metals in the \texttt{apec} component is low, below unity.

\begin{figure}[ht]
    \centering
    \includegraphics[width= 0.95\textwidth]{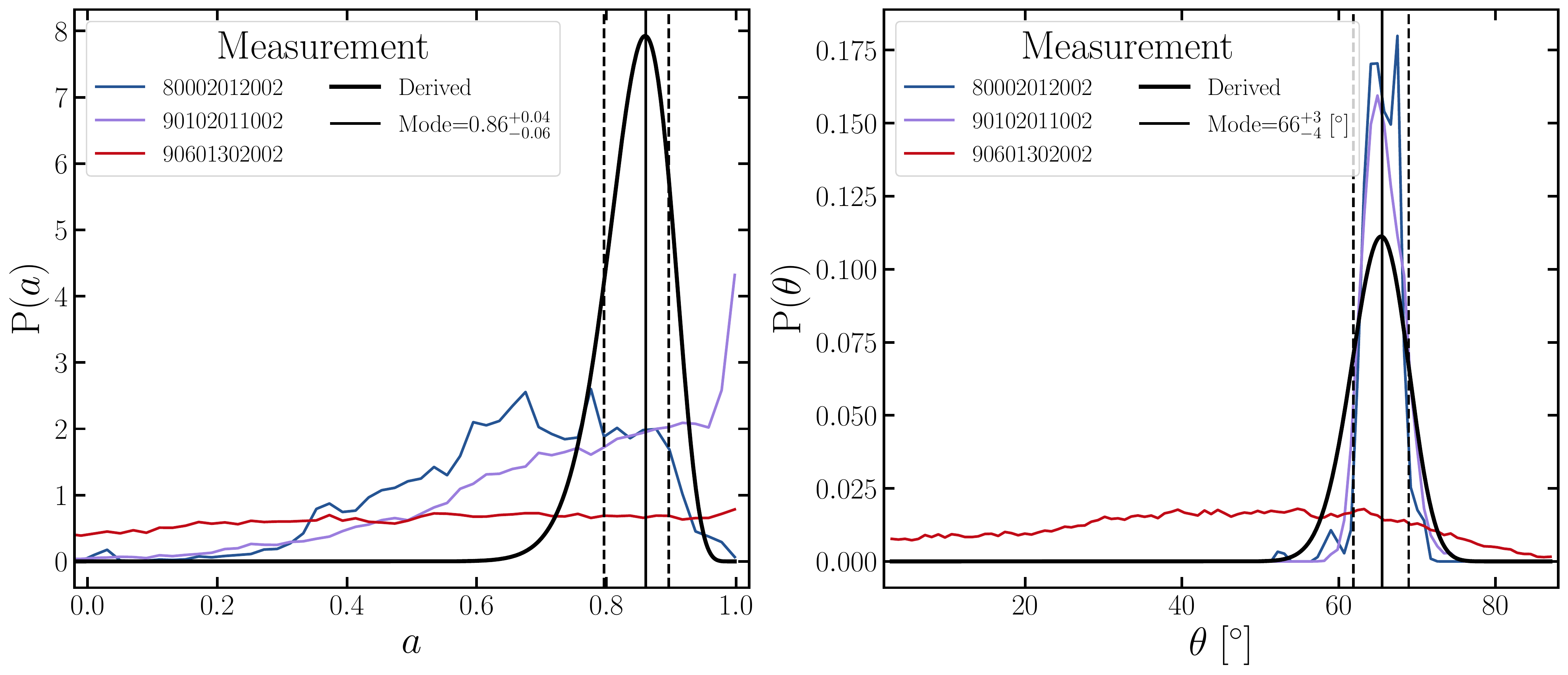}
    \caption{The left panel shows the posterior distributions resulting from the MCMC analysis of V4641 Sgr for spin, while the right panel shows the posterior distributions for the inclination of the inner accretion disk in the model. The different colored lines represent the three observations analyzed, and the black curves represent the combined inferred distribution, with the transparency of the lines being proportional to the ratio of reflected flux to total flux in the 3-79~keV band, which was used as weighting when combining the posterior distributions. The solid vertical black lines represent the modes of the combined distribution, and the dashed vertical black lines represent the $1\sigma$ credible intervals of the measurements. }
    \label{fig:V_4641_combined}
\end{figure}

Most importantly for this work, the spin of the compact object is generally high and the inclination is broadly consistent between the three observations. We combined the spin and inclination measurements from the three observations as described in subsection \ref{sec:combining}. The results are shown in Figure \ref{fig:V_4641_combined}. We measure $a=0.86^{+0.04}_{-0.06}$ and $\theta=66^{+3}_{-4}$ degrees.

Interestingly, in Figure \ref{fig:V_4641_combined}, we can see that ObsID 90601302002 does not contribute strongly to the spin or inclination measurements, with both the spin and inclination being determined by the other two observations analyzed. While for inclination, the two observations produce agreeing results, the spin measurement averages the two independent measurements. The uncertainty reported on the parameters is simply statistical, but the difference between the two posterior distributions that lead to the spin measurement of V4641 Sgr highlight the importance of understanding the magnitude of the systematic errors of spin measurements. A uniform treatment of the XB BH sample is crucial for this goal.

The inclination of the inner accretion disk in V4641 Sgr is well determined by our measurement, $\theta=66^{+3}_{-4}$ degrees, in good agreement with the orbital inclination of the system (\citealt{2001ApJ...555..489O}). Our measurement is incompatible with a low inclination, suggesting that V4641 Sgr cannot be a microblazar. However, \cite{2001ApJ...555..489O} suggested that the jet angle must be $\lesssim10^\circ$. Such a discrepancy between the inclination determined based on the jet angle and the inclination of the inner accretion disk was found by \cite{2021ApJ...920...88D} in XTE J1908+094, where it was suggested that the complexity of the local environment of the source together with an inability to correctly ``phase" different approaching/receding ejections can alter our view of inclinations determined based on the radio jet.

\subsection{4U 1543-47}
At the time of the analysis, there were 10 public NuSTAR observations of 4U 1543-47 (ObsID 80702317002, 80702317004, 80702317006, 80702317008, 90702326002, 90702326004, 90702326006, 90702326008, 90702326010 and 90702326012), all taken during the 2021 outburst, with the source flux decreasing in each new observation. However, assuming a BH mass of $M=9.4\pm2\;\rm M_\odot$ and a distance to the system of $d=7.5\pm1\;\rm kpc$ (\citealt{2004ApJ...610..378P}), all 10 observations occur while the source is at an Eddington fraction larger than 0.3. Still, we analyze the last 4 observations taken (ObsID 90702326006, 90702326008, 90702326010 and 90702326012), as they happen while the source was at the lowest Eddington fraction of the 10 observations, and also had the highest hardness of the 10 available observations. Within the allowed range of BH mass and distance to the system, the Eddington fraction during the last four observations varies between $\sim 0.32$ for the most favorable parameter combination for the last observation (ObsID 90702326012) and $\sim 0.8$ for the worst parameter combination for the first of the four observations analyzed (ObsID 90702326006). It is important to note that the first of the ten observations (ObsID 80702317002) gives an estimate between $2-5$ for the Eddington fraction during the observation, suggesting that perhaps our mass and distance estimates are not accurate. As the source was outside of the Eddington ratio range for which we expect that the accretion disk extends to the ISCO, we allowed the inner disk radius $r_{\rm in}$ to vary during our spectral fitting. For all four observations we fit the spectra over the entire NuSTAR pass band of 3--79~keV.

\begin{figure}[ht]
    \centering
    \includegraphics[width= 0.95\textwidth]{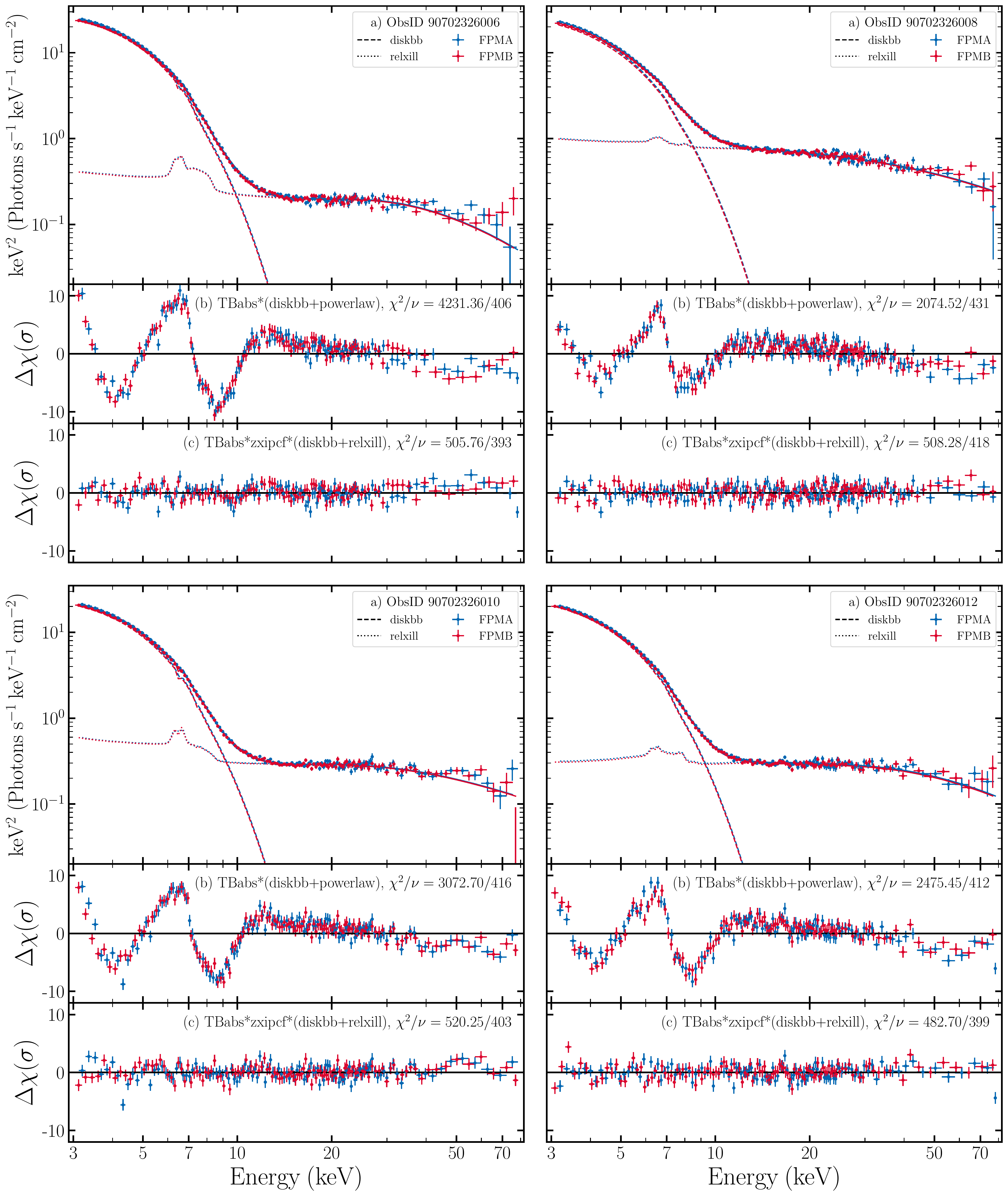}
    \caption{The four panels represent the four observations of 4U 1543-47 analyzed. The top sub-panels show unfolded spectra, with the NuSTAR FPMA spectrum shown in blue and FPMB spectrum in red. The reported best-fit model is shown by the solid lines, while the contribution of the \texttt{relxill} component is shown by the dotted lines and the contribution of the \texttt{diskbb} component is shown through the dashed lines. The middle and bottom sub-panels show the residuals in terms of $\sigma$ produced when fitting the spectra with \texttt{TBabs*(diskbb+powerlaw)} (middle) and \texttt{TBabs*zxipcf*(diskbb+relxill)} (bottom), together with the statistic produced.}
    \label{fig:4U_1543_delchi}
\end{figure}

Figure \ref{fig:4U_1543_delchi} shows the unfolded spectra of the four observations in the top sub-panels of each panel. The middle sub-panels show the residuals and statistics produced when fitting the spectra with the simple \texttt{TBabs*(diskbb+powerlaw)} model, clearly indicating the presence of relativistic reflection. Similarly to the case of V4641 Sgr presented in subsection \ref{sec:v4641}, simply replacing the \texttt{powerlaw} component with any of the six variations of the \texttt{relxill} family that we discussed in subsection \ref{sec:models} does not lead to a good fit. 

We started fitting the observation with the hardest spectrum (ObsID 90702326008). The initial fit using \texttt{TBabs*(diskbb+powerlaw)} returns $\chi^2/\nu=2074.52/431$, while replacing the \texttt{powerlaw} component with \texttt{relxill} returns $\chi^2/\nu=623.42/422$ showing additional broad residuals. To fit additional features, models are often expanded to include an additional \texttt{xillver} component that has the role to account for distant, unblurred reflection (see e.g., \citealt{2018ApJ...860L..28M}). In this case, the addition of a \texttt{xillver} component does not improve the quality of the fit, with the solution simply converging to the same parameter combination as before by fitting the reflection fraction of the \texttt{xillver} component to $R=0$. 

Since the residuals indicate the presence of \textit{broad} spectral features, we replaced the \texttt{xillver} component with a second \texttt{relxill} component in the model, with the role of mimicking a torn accretion disk. To do that, we linked the black hole spin between the two components, the Fe abundance, the power law indices $\Gamma$, the high-energy cutoff of the underlying power law continuum, and the normalization of the two components. The reflection fractions of the two components were allowed to vary, with the ``inner" \texttt{relxill} component taking positive values for $R$, while the ``outer" component taking only negative reflection fractions. Positive values of reflection fraction in \texttt{relxill} components ensure that the model includes the direct coronal emission, while negative reflection fraction values force the components to only include the reflected emission. This way, the underlying continuum is included by the ``inner" \texttt{relxill} component. We allowed the inner disk radius of the ``inner" \texttt{relxill} component to vary, the outer disk radius of the ``outer" \texttt{relxill} component to $r_{\rm out}=1000\;\rm r_g$, and linked the outer radius of the first \texttt{relxill} component to the inner radius of the second \texttt{relxill} component. The disk inclination in the two \texttt{relxill} components was allowed to vary. This model improves the quality of the fit, producing $\chi^2/\nu=540.68/416$. The residuals of the fit clearly indicate a narrow absorption feature around 7.5~keV, and the measured inclination of the two \texttt{relxill} components is different, but not well constrained: the inner component measures $\theta\sim3^\circ$ while the outer component measures $\theta\sim85^{\circ}$. When accounting for the absorption feature using a \texttt{gaussian} additive component, and also linking the inclination in the two \texttt{relxill} components, the fit improves much further, producing $\chi^2/\nu=501.13/416$. With the inclination of the two reflection components linked, the main difference between them is the ionization, with the inner component taking large values $\log(\xi)\sim4.6$, while the outer one measures $\log(\xi)\sim0$. 

An alternative treatment that can be used to test for variable ionization throughout the accretion disk is the \texttt{relxilllpion} model. We replaced the two reflection components in our model with this new model, making the model \texttt{TBabs*(diskbb+relxilllpion)}. This also produces an improved fit when compared to the the simple \texttt{relxill} or \texttt{relxilllp} variants, returning $\chi^2/\nu=523.85/424$. Lastly, similarly to the case of V4641 Sgr, we tested the addition of ionized partial covering (through \texttt{zxipcf}) and re-emission (through \texttt{apec}). While the \texttt{apec} component did not improve the quality of the fits in this case, the addition of the \texttt{zxipcf} component to the default model improves the quality of the fit significantly ($\chi^2/\nu=508.48/418$) for the model \texttt{TBabs*zxipcf*(diskbb+relxill)}. The absorber requires moderate ionization ($\log\xi\sim2$) and covering fraction $f\sim0.4$, consistent with a disk wind. The inner disk radius was allowed to vary, however the fit converges to a value of $r_{\rm in}$ consistent with the size of the ISCO. 

\begin{figure}[ht]
    \centering
    \includegraphics[width= 0.95\textwidth]{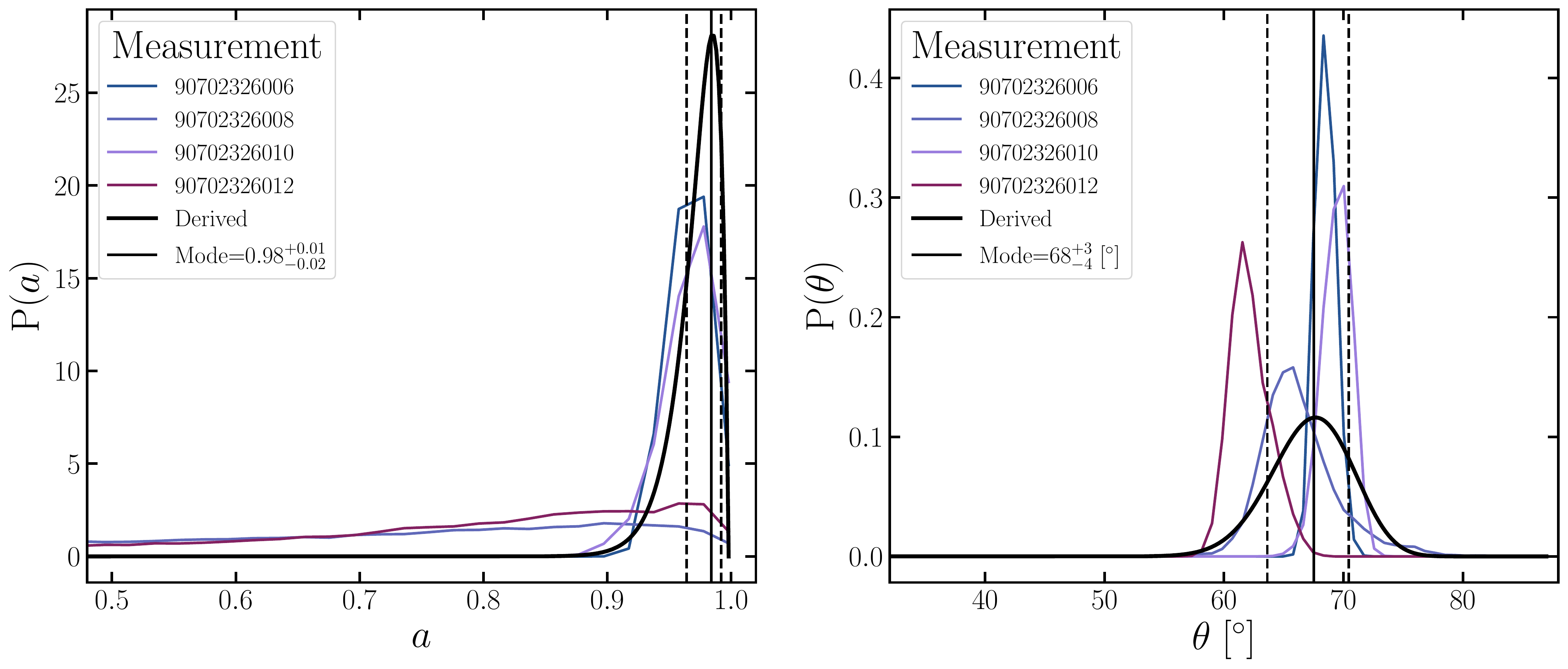}
    \caption{The left panel shows the posterior distributions resulting from the MCMC of 4U 1543-47 analysis for spin, while the right panel shows the posterior distributions for the inclination of the inner accretion disk in the model, with the transparency of the lines being proportional to the ratio of reflected flux to total flux in the 3-79~keV band, which was used as weighting when combining the posterior distributions. The different colored lines represent the four observations analyzed, and the black curves represent the combined inferred distribution. The solid vertical black lines represent the modes of the combined distribution, and the dashed vertical black lines represent the $1\sigma$ credible intervals of the measurements. }
    \label{fig:4U_1543_combined}
\end{figure}

We tested the three mentioned models on all four observations. In all four cases, the last model (including \texttt{zxipcf}) was preferred in terms of AIC and BIC, likely due to it producing a better statistic than the variable ionization model and having a smaller number of free parameters than the model with two reflection components. We mention that modifying the \texttt{relxill} variant in these models does not strongly affect the quality of the fit or the measured spin. Therefore we continue our analysis using the model \texttt{TBabs*zxipcf*(diskbb+relxill)}. The residuals produced by this model are shown in the bottom panels of Figure \ref{fig:4U_1543_delchi}. We ran the MCMC analysis on these four observations with the given model. The modes and $1\sigma$ credible intervals for all parameters in the model are presented in Table \ref{tab:table2} for each observation. The 1D posterior distributions of the spin and inclination parameters are shown in Figure \ref{fig:4U_1543_combined}. By combining the posterior distributions as explained in Section \ref{sec:combining}, we measure $a=0.98^{+0.01}_{-0.02}$ and $\theta=68^{+3}_{-4}$ degrees. Interestingly, this spin measurement disagrees with previous measurements, namely $a=0.67^{+0.15}_{-0.08}$, measured through relativistic reflection on RXTE data (\citealt{2020MNRAS.493.4409D}) and $a=0.80\pm0.05$, measured through continuum fitting (\citealt{2006ApJ...636L.113S}).

\subsection{4U 1957+11}
\begin{figure}[ht]
    \centering
    \includegraphics[width= 0.75\textwidth]{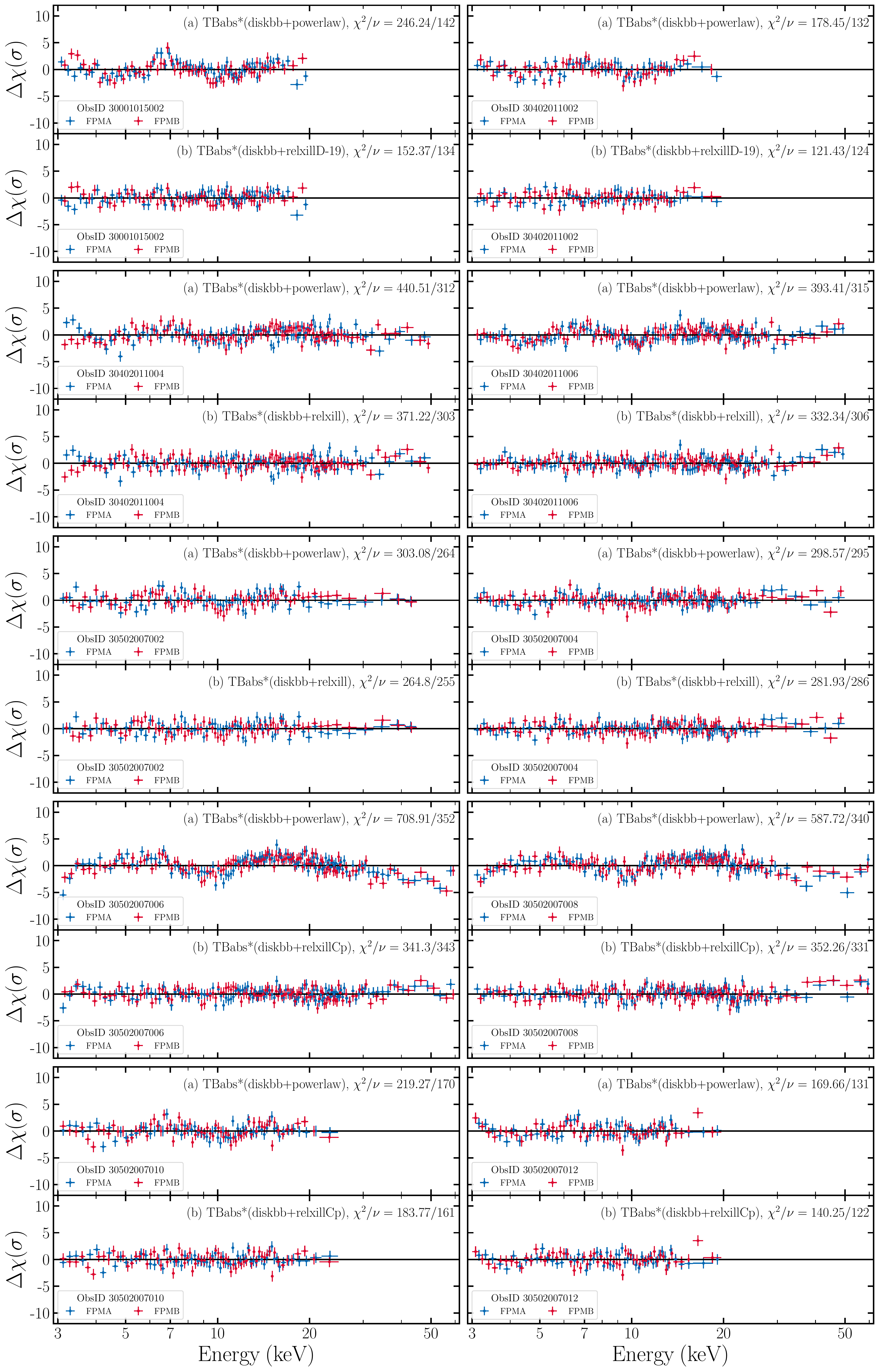}
    \caption{The 10 panels represent the 10 NuSTAR observations of 4U 1957+11 analyzed. The top sub-panels show the residuals in terms of $\sigma$ produced when fitting the spectra with \texttt{TBabs*(diskbb+powerlaw)}, and the bottom sub-panels show the residuals when fitting with the best performing reflection model. The blue and red points represent the FMPA and FMPB spectra from each observation, respectively. The sub-panel titles show the model used to fit and the statistic produced, and the sub-panel legends mention the ObsID number of the observation shown.}
    \label{fig:4U_1957_delchi}
\end{figure}

Using NuSTAR data, \cite{2021RAA....21..214S} estimated the black hole mass and distance to 4U 1957+11 to be $M=5\pm1\;\rm M_\odot$ and $d\sim7\;\rm kpc$ and measured a spin of $a\sim0.85$. Using Suzaku data, \cite{2012ApJ...744..107N} estimated a BH spin $a>0.90$ for $M=3\;\rm M_\odot$ and $d\sim10\;\rm kpc$, and \cite{2014ApJ...794...85M} measured a spin $a>0.98$. All measurements were performed using the continuum fitting method. 4U 1957+11 is a particularly interesting source, due to its relatively low mass estimates and large distance to the system, which makes this source consistently soft while at a relatively low Eddington fraction. Regardless of the choice of black hole mass and distance to the system of the two combinations, all 10 existing NuSTAR observations are taken while the source was within the Eddington ratio limits for which we expect the accretion disk to extend to the ISCO. We fit all existing observations in the following energy bands: 30001015002 (3--20~keV), 30402011002 (3--20~keV), 30402011004 (3--50~keV), 30402011006 (3--50~keV), 30502007002 (3--45~keV), 30502007004 (3--50~keV), 30502007006 (3--60~keV), 30502007008 (3--60~keV), 30502007010 (3--25~keV), 30502007012 (3--20~keV). The residuals when fitting the 10 spectra with \texttt{TBabs*(diskbb+powerlaw)} are shown in the top panels in Figure \ref{fig:4U_1957_delchi}, along with the statistic. The spectra of 4U 1957+11 are generally dominated by disk emission and the reflection features are not always immediately obvious, but fitting with models that account for relativistic reflection always improves the quality of the fit.

\begin{figure}[ht]
    \centering
    \includegraphics[width= 0.95\textwidth]{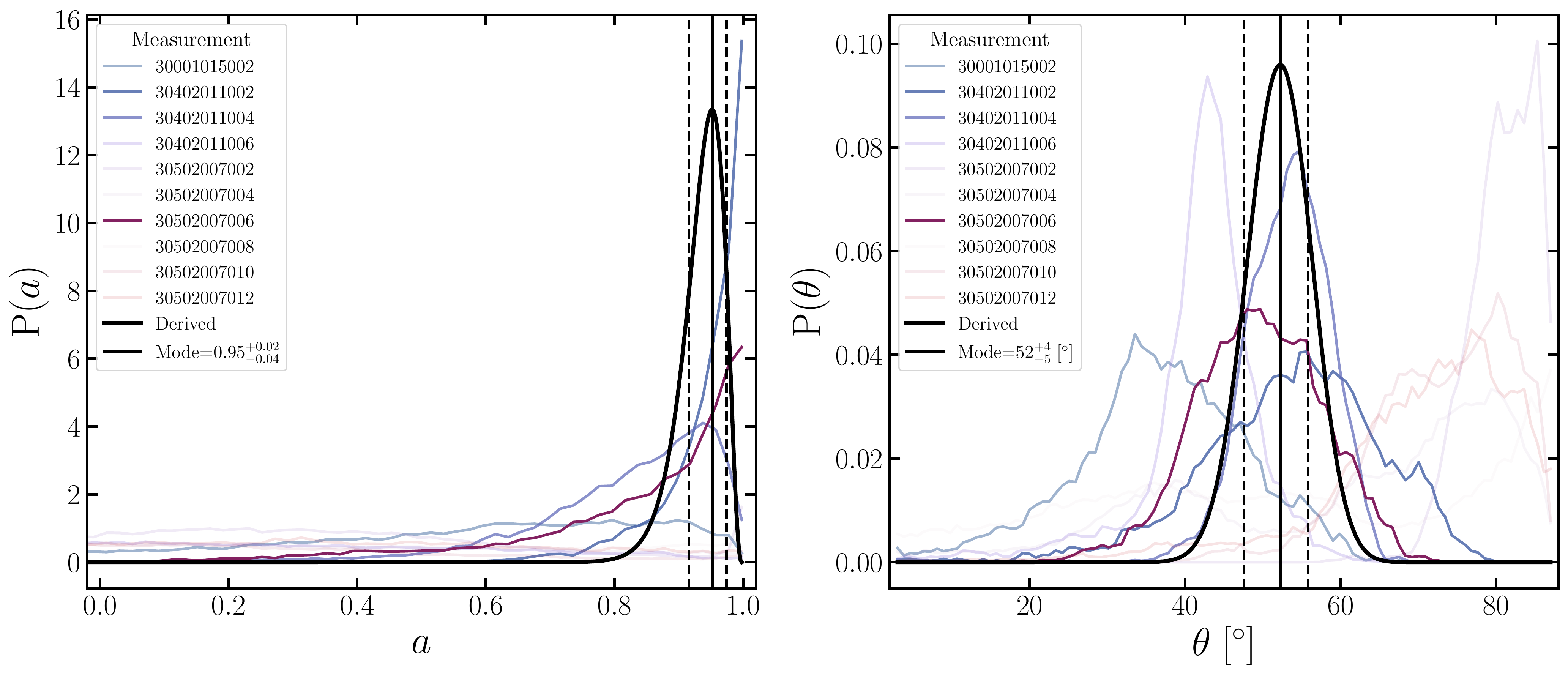}
    \caption{The left panel shows the posterior distributions resulting from the MCMC of 4U 1957+11 analysis for spin, while the right panel shows the posterior distributions for the inclination of the inner accretion disk in the model, with the transparency of the lines being proportional to the ratio of reflected flux to total flux in the 3-79~keV band, which was used as weighting when combining the posterior distributions. The different colored lines represent the different observations analyzed, and the black curves represent the combined inferred distribution. The solid vertical black lines represent the modes of the combined distribution, and the dashed vertical black lines represent the $1\sigma$ credible intervals of the measurements. }
    \label{fig:4U_1957_combined}
\end{figure}

Replacing the \texttt{powerlaw} component in our model with the six variants of \texttt{relxill} discussed in Subsection \ref{sec:models} produces better fits. We ran the MCMC analysis on the two best performing models that account for reflection and selected the one that produced the best DIC. The residuals of the best performing models in terms of DIC are shown in the bottom panels of Figure \ref{fig:4U_1957_delchi}, and the modes and $\pm1\sigma$ credible intervals of the posterior distributions for each parameter are presented in Table \ref{tab:table3}. The 1D posterior distributions for the spin and inclination are shown in Figure \ref{fig:4U_1957_combined}, with the transparency of the lines corresponding to each observation being proportional to the weights used when combining the measurements. We measure a spin of $a=0.95^{+0.02}_{-0.04}$ and an inclination of $\theta=52^{+4}_{-5}$ degrees. For this source in particular, it is extremely important to weight the posterior distributions from each individual measurement when combining them into a single measurement, and the combined spin and inclination values are dominated by the distributions inferred from the observations where reflection was strongest. This spin measurement broadly agrees with the measurements made using continuum fitting, and it could be used to place better constraints on the mass of the black hole and the distance to the system.

\subsection{H 1743-322}
There were ten NuSTAR archival observations of H 1743-322 at the time of the analysis. However, the spectra from ObsID 80002040004 are well fit by \texttt{TBabs*(diskbb+powerlaw)} and the source is not detected in ObsID 80002040006. Using a black hole mass of $M=11.21^{+1.65}_{-1.96}\;\rm M_{\odot}$ (\citealt{2017ApJ...834...88M}) and a distance to the system of $d=8.5\pm0.8\;\rm kpc$ (\citealt{2012ApJ...745L...7S}), all remaining eight observations (80001044002, 80001044004, 80001044006, 80002040002, 80202012002, 80202012004, 80202012006, 90401335002) fall within the $10^{-3}\lesssim L_{disk}/L_{Edd}\lesssim 0.3$ Eddington fraction range. Therefore, we continue analyzing the eight NuSTAR observations of H 1743-322, in the entire $3-79\;\rm keV$ NuSTAR band.

\begin{figure}[ht]
    \centering
    \includegraphics[width= 0.75\textwidth]{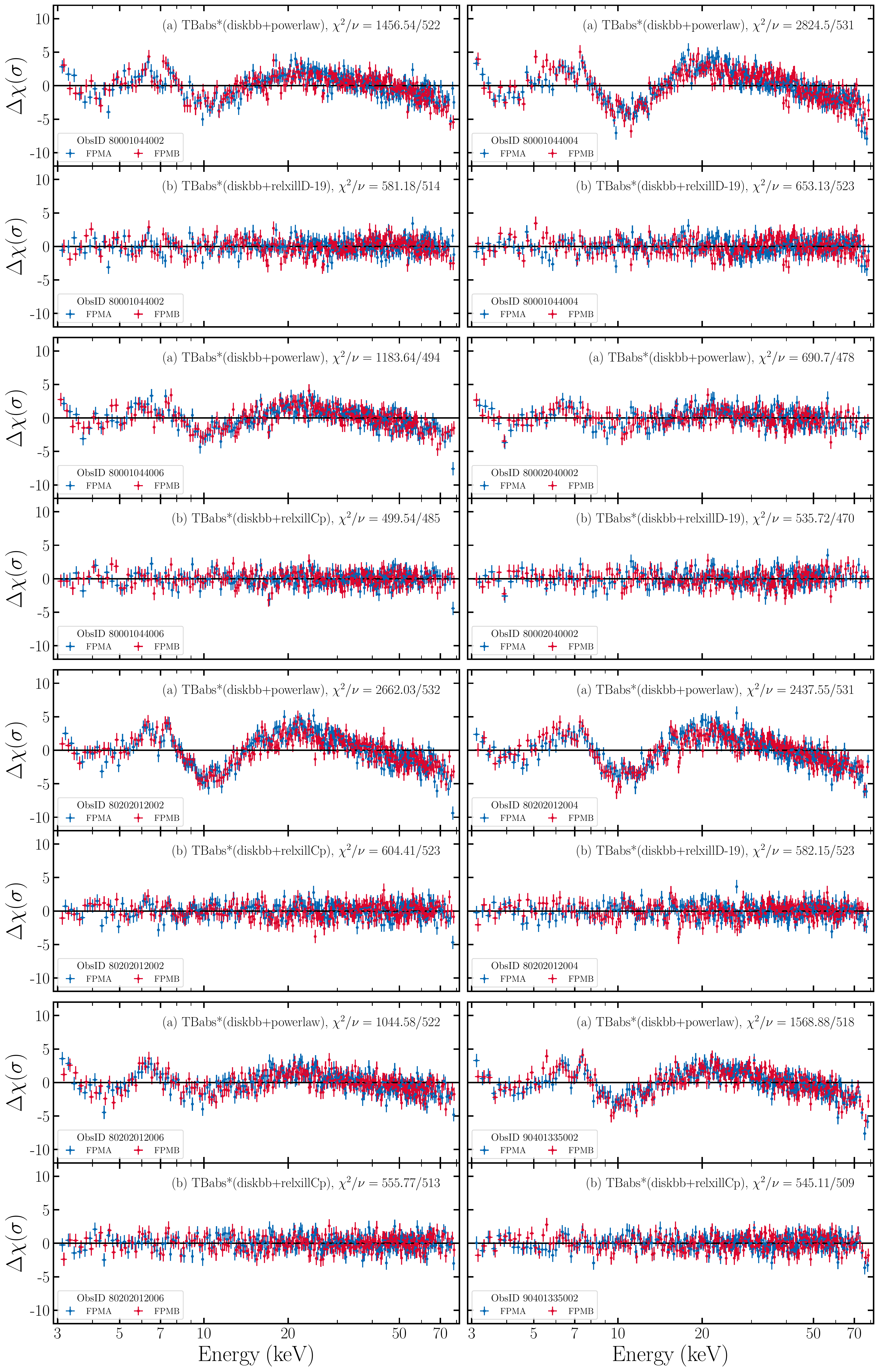}
    \caption{The 8 panels represent the 8 NuSTAR observations of H 1743-322 analyzed. The top sub-panels show the residuals in terms of $\sigma$ produced when fitting the spectra with \texttt{TBabs*(diskbb+powerlaw)}, and the bottom sub-panels show the residuals when fitting with the best performing reflection model. The blue and red points represent the FMPA and FMPB spectra from each observation, respectively. The sub-panel titles show the model used to fit and the statistic produced, and the sub-panel legends mention the ObsID number of the observation shown.}
    \label{fig:H_1743_delchi}
\end{figure}

\begin{figure}[ht]
    \centering
    \includegraphics[width= 0.95\textwidth]{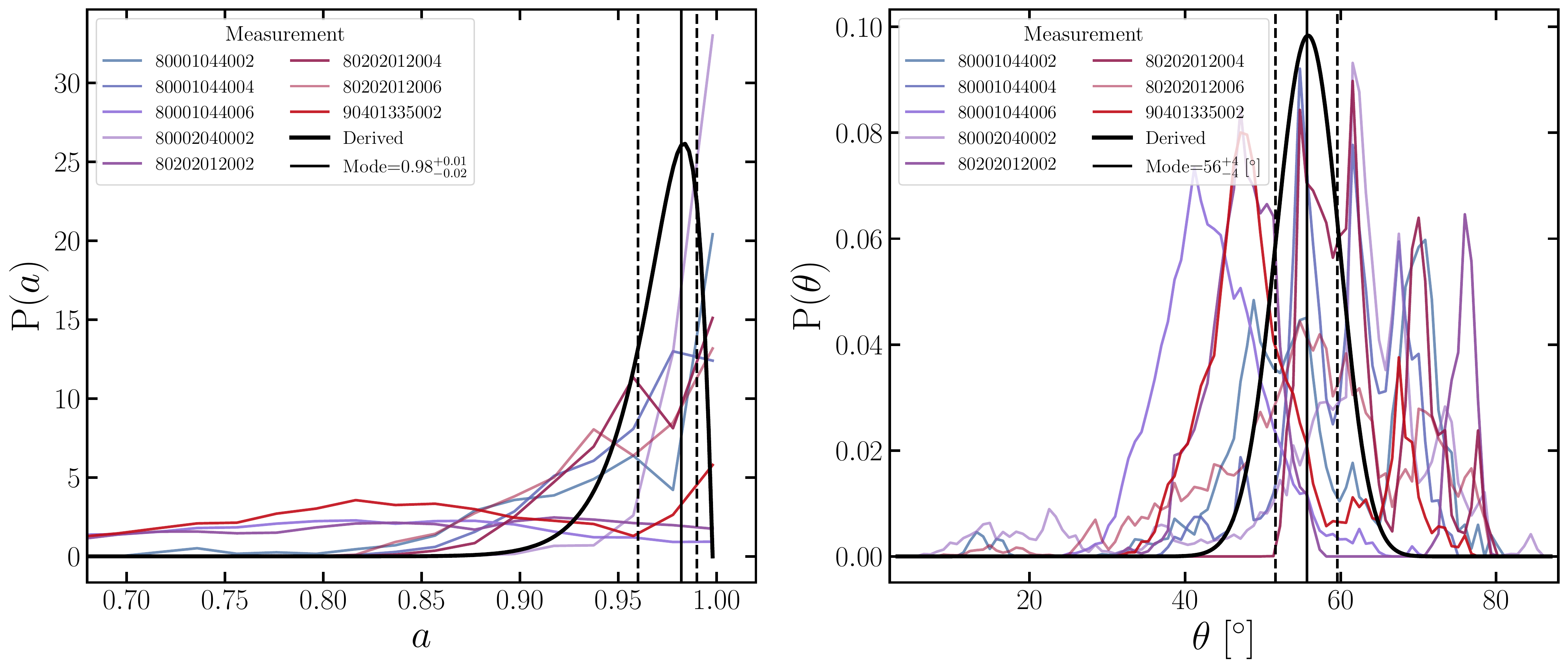}
    \caption{The left panel shows the posterior distributions resulting from the MCMC of H 1743-322 analysis for spin, while the right panel shows the posterior distributions for the inclination of the inner accretion disk in the model, with the transparency of the lines being proportional to the ratio of reflected flux to total flux in the 3-79~keV band, which was used as weighting when combining the posterior distributions. The different colored lines represent the different observations analyzed, and the black curves represent the combined inferred distribution. The solid vertical black lines represent the modes of the combined distribution, and the dashed vertical black lines represent the $1\sigma$ credible intervals of the measurements. }
    \label{fig:H_1743_combined}
\end{figure}

Using the \texttt{TBabs*(diskbb+powerlaw)} model to fit the spectra from the eight NuSTAR observations indicates the presence of relativistic reflection. The residuals of the fits are shown in the top panels of Figure \ref{fig:H_1743_delchi}, together with the fit statistic. We accounted for reflection through the six models discussed in Subsection \ref{sec:models}, ran the MCMC analysis on the best performing two models for each observation, and selected the best performing model of the two (in terms of DIC) for each observation. The residuals of the fits using the best reflection model for each observation are shown in the bottom panels in Figure \ref{fig:H_1743_delchi}, together with the $\chi^2$ statistic produced by the models. The modes and $1\sigma$ credible intervals of the posterior distributions for each parameter produced by the reflection models returning the best DIC for each observation are presented in Table \ref{tab:table4}, and the posterior distributions for the black hole spin and inclination of the inner accretion disk are shown in Figure \ref{fig:H_1743_combined}. Combining the spin and inclination measurements produced by each observation as described in Subsection \ref{sec:combining} returns a spin of $a=0.98^{+0.01}_{-0.02}$ and an inclination of $\theta=56\pm4$ degrees.

Both the spin and inclination measurements of our analysis disagree with those determined by \citealt{2012ApJ...745L...7S} through continuum fitting ($a=0.2\pm0.3$) and from radio and X-ray jets ($\theta=75\pm3$). While it is expected that the jet axis coincides with the black hole spin axis and while that often is the case in observations (see e.g., Cyg X-1 - \citealt{2022arXiv220609972K}, or MAXI J1820+070 in Subsection \ref{sec:MAXIJ1820}), disparities between the inclination inferred through jet morphology and through relativistic reflection have previously been found (see e.g., XTE J1908+094 - \citealt{2021ApJ...920...88D}). Interestingly, \citealt{2018PAN....81..279T} estimated that the black hole in H 1743-322 has a spin of $a=0.4\pm0.2$ from quasiperiodic oscillations (QPO). However, also from QPOs, \citealt{2010ApJ...708.1442M} find that the black hole in H 1743-322 must have a spin $a>0.68$. It is important to note that there is no commonly accepted model to predict BH masses and spins based on QPOs, and models often disagree (\citealt{2017arXiv171111398S}). Furthermore, as QPO frequencies often shift, it is likely that they do not properly trace the BH ISCO.

\subsection{MAXI J1820+070}\label{sec:MAXIJ1820}
NuSTAR observed MAXI J1820+070 27 times in 2018 and 2019. Using a black hole mass of $M=8.48\pm0.79\;\rm M_{\odot}$ (\citealt{2020ApJ...893L..37T}) and a distance of $d=2.96\pm0.33\;\rm kpc$ (\citealt{2020MNRAS.493L..81A}), we find that 19 of the 27 observations happened while the source was at an Eddington fraction between $10^{-3}\lesssim L_{disk}/L_{Edd}\lesssim 0.3$: ObsID 90401309002, 90401309004, 90401309006, 90401309008, 90401309010, 90401309012, 90401309013, 90401309014, 90401309016, 90401309018, 90401309019, 90401309021, 90401309023, 90401309025, 90401309026, 90401309027, 90401309033, 90401309035, and 90401324002. We ran our analysis on these 19 observations, fitting in the entire $3-79\;\rm keV$ NuSTAR band pass in each case. The residuals produced by fitting the spectra from each observation with the model \texttt{TBabs*(diskbb+powerlaw)}, and the fit statistic are shown in the top panels in Figures \ref{fig:MAXI_J1820_delchi_1} and \ref{fig:MAXI_J1820_delchi_2}. 

The residuals generally indicate the presence of relativistic reflection, so we fit the spectra from the 19 observations with the six models discussed in Subsection \ref{sec:models}. Due to the large number of observations, we only ran the MCMC analysis on the best performing models for each observation in terms of $\chi^2_r=\chi^2/\nu$, where $\nu$ represents the number of degrees of freedom in the model. The bottom panels in Figure \ref{fig:MAXI_J1820_delchi_1} and continued in Figure \ref{fig:MAXI_J1820_delchi_2} show the residuals of the best performing reflection models. We ran the MCMC analysis on the fits to the spectra from the 19 observations. Table \ref{tab:table5} shows the modes of the posterior distributions for each parameter in the MCMC analysis for the 19 observations, along with the best performing \texttt{relxill} variant, that we used in our analysis. The posterior distributions for spin and inclination are shown in Figure \ref{fig:MAXI_J1820_combined}. When combining the posterior distributions while weighting them by the ratio of reflected to total flux in the $3-79$~keV band, we obtain a spin of $a=0.988^{+0.006}_{-0.028}$ and an inclination of $\theta=64^{+3}_{-4}$ degrees. 

Our inclination measurement is in good agreement with that determined by \cite{2020MNRAS.493L..81A} from the radio jet ($\theta=63\pm3^\circ$). However, our spin measurement disagrees with the values found using continuum fitting: $a=0.14\pm0.09$ (\citealt{2021ApJ...916..108Z}) and $a=0.2^{+0.2}_{-0.3}$ (\citealt{2021MNRAS.504.2168G}). However, \cite{2021MNRAS.508.3104B} used a Relativistic Precession Model for QPO frequencies on NICER data to measure a spin of $a=0.799^{+0.016}_{-0.015}$, and \cite{2022MNRAS.514.6102P} analyzed multiple Swift/XRT (\citealt{2005SSRv..120..165B}), NICER (\citealt{2012SPIE.8443E..13G}), NuSTAR, and AstroSat (\citealt{2006AdSpR..38.2989A}) observations and found that when fitting the reflection spectra, the spin was always high, so they fixed it at the maximum allowed value throughout their analysis. While \cite{2021MNRAS.504.2168G} argue that given their low spin measurement, the strong jet seen in MAXI J1820+070 must be powered by the accretion disk, our measurement allows for the jet to be powered by the high back hole spin through the Blandford–Znajek mechanism (\citealt{1977MNRAS.179..433B}). This would be expected, especially given the good agreement between the inclination of the black hole spin axis measured from relativistic reflection in this work and the inclination of the observed radio jet. 

\section{Spin Distribution}\label{sec:distribution}
The new era of gravitational wave (GW) signals from binary black hole (BBH) mergers holds tremendous promise.  Currently, two challenges keep the full potential of black hole mergers from being realized: (1) a degeneracy between the mass ratio in the binary and a combination of the spins of the black holes, and (2) a second degeneracy between the two spins themselves, which makes it difficult to measure individual spins (\citealt{2016PhRvD..93h4042P}). It was estimated that the degeneracies can be broken for signal to noise ratios (SNR) of 100, but only a few such observations are likely to occur every year (\citealt{2016PhRvD..93h4042P}).  Moreover, even in this situation, the lowest mass black hole in the system can only be classified as having "high" or "low" spin, as opposed to an actual measurement.  Given current event rates, the most pragmatic path forward is to obtain informative priors, based on a robust derivation of the spins of stellar-mass black holes in X-ray binary systems: the posterior distributions for GW spin measurements are correlated to the assumed prior distribution (\citealt{2017PhRvL.119y1103V, 2019PhRvX...9c1040A, 2020arXiv201014527A}), making it crucial to have educated predictions for the prior distributions.  

We compiled all XB BH spin measurements in literature obtained through continuum fitting and relativistic reflection in Table \ref{tab:all_spins} and plotted them on the same scale in Figure \ref{fig:all_spins}, with the orange points showing measurements made through continuum fitting, yellow points representing relativistic reflection measurements made on data from instruments other than NuSTAR, blue points represent reflection measurements obtained using NuSTAR data, and the green points showing the 10 measurements of this paper. 

A few interesting points can be made regarding the existing sample of BH spins in XB. Firstly, it is worth nothing that there are no precise negative BH spin measurements, only upper limit measurements suggesting $a\leq0$. Of the seven measurements consistent with negative spins, five are made using continuum fitting and two are made using relativistic reflection, but not using NuSTAR data. Also, only four of the seven measurements that allow negative spins exclusively require negative spins. A few of those constraints are particularly interesting. For GX 339-4, the continuum fitting method finds $a<0.9$ (\citealt{2010MNRAS.406.2206K}), while relativistic reflection on NuSTAR data finds $a=0.95_{-0.08}^{+0.02}$ (\citealt{2016ApJ...821L...6P}). While the continuum fitting measurement nominally allows low and negative spins, simultaneously considering the two measurements made using the two methods indicates that the BH spin in GX339-4 is indeed high. Another particularly interesting source is IGR J17091-3624, for which continuum fitting measures $a<0.2$ (\citealt{2012ApJ...757L..12R}) while relativistic reflection on NuSTAR data finds $a=0.07\pm0.20$ (\citealt{2018MNRAS.478.4837W}). Again, while considering only the continuum fitting measurement negative spins are allowed, treating the two measurements together indicates the BH is likely slowly rotating, with spin consistent with zero.

Secondly, there appears to be a trend of decreasing measurement uncertainty with increasing BH spin. This trend applies both to spins obtained through the reflection method and through continuum fitting. As the size of the ISCO is set by the BH spin, a truncated disk around a rapidly spinning BH would produce similar spectral features as a disk that extends all the way to the ISCO around a slowly rotating BH (e.g., a maximally spinning BH with a disk truncated at $6\;r_g$ would produce similar features as a non-spinning BH with a disk that extends to the ISCO). However, since XB spin measurements assume inner disk radii consistent with the ISCO and since possible disk truncation would have the effect of biasing spin measurements to lower values, the high measured spins cannot be systematically mistaken and are likely accurate. The increased precision of high measured spins could be owed to higher both direct and reflected emissivity with increased proximity to the BH, making the spectral features easier to observe and characterize, leading to smaller uncertainties on spin measurements. Works such as \cite{2016MNRAS.458.1927B} and \cite{2018A&A...614A..44K} test the accuracy of spin measurements in AGN, finding that for values above 0.8, the spin is better constrained both in terms of accuracy and precision.

Thirdly, it seems like the overall distribution of spin measurements is concentrated around high values. Using the same method that we used to combine multiple measurements into a single distribution, we combined the multiple independent spin measurements in Table \ref{tab:all_spins} and Figure \ref{fig:all_spins} in an attempt to understand the entire \textit{observed} BH spin distribution in XB. It is crucial to acknowledge that this distribution is likely influenced by observational and selection biases, which can be better understood through a uniform treatment of the entire spin sample. Additionally, it is important to note that the measurements in literature vary widely in terms of the energy range and resolution of the instrument used and of theoretical assumptions and numerical resolution of the models used, which can likely lead to biases in the measurements. 

We plot the combined distributions in Figure \ref{fig:inferred_distributions}. The orange curve shows the distribution of XB BH spin measurements obtained through continuum fitting. The green curve shows the distribution inferred based on the 10 measurements presented in this paper, which is a subset of the blue curve, showing the distribution inferred based on all measurements made using relativistic reflection on NuSTAR data. The blue curve is a subset of the yellow curve, which shows the distribution of all XB spin measurements obtained through relativistic reflection. The black curve shows the combined distribution of all XB spin measurements, through both relativistic reflection (yellow) and continuum fitting (orange). For reference, we compare this with the spin distribution inferred by \cite{2021arXiv211103634T} based on all the GW signals from BBH mergers presented in GWTC-3, shown in purple. 

The spin distributions in BBH and in XB are clearly in disagreement. \cite{2022ApJ...929L..26F} compared the distribution of spins in the previous version of the GWTC (GWTC-2, \citealt{2021PhRvX..11b1053A}) to the distribution of BH spins in high mass X-ray binaries (HMXB), inferred from 3 sources (LMC X-1, Cyg X-1, and M33 X-7). Despite the very limited sample size and despite only using the values measured through continuum fitting, \cite{2022ApJ...929L..26F} conclude that the HMXB and BBH spin distributions disagree at the $\geq 99.9\%$ level. However, they conclude that the spin distribution of BHs in low Mass X-ray binaries (LMXB) agrees with that of HMXB. There is value in attempting to compare the BBH sample to HMXB only, since HMXB are candidates to evolving to produce BBH systems. However, it is worth mentioning that studies have shown that neither Cyg X-1 (\citealt{2011ApJ...742L...2B}) nor M33 X-7 (\citealt{2022arXiv220807773R}) are likely to evolve to produce a BBH system. Additionally, works such as  \cite{2022arXiv220714290G} find that at most 20\% of BBH originate from HMXB systems and at most 11\% of HMXB evolve into BBH systems. Here, we highlight the comparison of all BH spins in XB obtained through continuum fitting and relativistic reflection to the spins in BBH. We performed a two-sample Kolmogorov-Smirnov test (K-S test - \citealt{smirnov1939estimation}) in order to quantify the probability that the two observed distributions of spins in BBH and XB are drawn from the same unknown underlying distribution. Using the distributions shown in Figure \ref{fig:inferred_distributions}, we generated $10^5$ samples of 50 ``measurements" randomly drawn from the XB population and 200 ``measurements" drawn from the BBH population, simulating the existing sample sizes. Then, we performed the two-sample K-S test and analyzed the distribution of probabilities produced. When comparing the BBH distribution with all spin measurements in XB, we find that the highest probability across the $10^5$ tests is $p\sim2\times10^{-5}$, while when comparing it with the distribution of spins obtained using relativistic reflection, the highest probability was $p\sim5\times10^{-13}$. The median probability across the $10^5$ samples in the two cases agrees, $p\sim9\times10^{-16}$. These results clearly state that the distribution of spins in BBH is different from the observed distribution of BH spins in XB. Since the formation channels that lead to the different spin distributions are likely different, it is important to consider the need to explain both distributions when attempting to provide a unified formation mechanism for all black holes.

\section{Discussion}\label{sec:discussion}
In this work, we analyzed the archival NuSTAR observations of a sample of 24 X-ray binary sources containing BHs that do not have previous spin constraints using relativistic reflection on NuSTAR data. We measured 10 new BH spins. In 14 cases, we were unable to obtain a BH spin measurement. The 10 new measurements represent an increase in the sample size of BH spin measurements obtained through the relativistic reflection method using NuSTAR data of nearly 40\%, and an increase of nearly 25\% in the sample size of all relativistic reflection measurements.
Four of the ten sources wherein we measure the spin using reflection have previous spin constraints from continuum fitting. Interestingly, in three of the four sources, our relativistic reflection measurements disagree with the spins obtained through continuum fitting (MAXI J1820+070, H 1743-322, and 4U 1543-47), while for 4U 1957+11 our measurement is consistent within the uncertainties with that obtained through continuum fitting. 

With the exception of V4641 Sgr, the new spin measurements have very high values, $a>0.95$, with all values being larger than 0.8 within $1\sigma$. This is consistent with the spin distribution in X-ray binaries (\citealt{2022ApJ...929L..26F}). It is unlikely that the relativistic reflection method is biased to measure high black hole spins, as low spins produce spectral shapes that are less blurred, and therefore easier to distinguish from the underlying continuum. Additionally, low and moderate spins are still detected by works such as \cite{2018MNRAS.478.4837W, 2021ApJ...920...88D, 2021MNRAS.507.4779J} or \cite{2022MNRAS.511.3125J}. However, one cannot exclude the possibility that there is an \textit{observational} bias, making highly spinning BHs easier to detect. \citealt{2016MNRAS.458.2012V} explain that due to selection effects of flux-limited observations, the BH spin distribution measured in Active Galactic Nuclei (AGN) is biased toward high values, but they also argue that this effect should not be directly applicable to stellar mass galactic BHs. \cite{2021PASA...38...56H} explain that high mass X-ray binaries (HMXB) should only become observable if the companion star has a Roche lobe filling factor above 0.8-0.9. A BH rotating in retrograde direction or slowly in the prograde direction will have a large ISCO radius; when paired with a high Roche lobe filling factor of the companion, this might not allow enough room for the formation of an accretion disk around the BH, with matter plunging directly towards the compact object. If an accretion disk cannot be formed, the X-ray binary will not be observed, biasing our observations toward high spin values. 

At the same time, one cannot exclude the possibility that the intrinsic spin distribution of BHs in X-ray binaries strongly favors high spins. Due to the short lifetimes of HMXBs (\citealt{1976IAUS...73...35V, 1995ApJS..100..217I}), we know that the observed high spins (see e.g., \citealt{2015ApJ...808....9P, 2021MNRAS.507.4779J}) must be natal. Works such as \cite{2015ApJ...800...17F} argue that the high spins in galactic low mass X-ray binaries (LMXB) can be explained through long episodes of highly efficient accretion. However, it is important to acknowledge that a BH must accrete 80$\%$ of its initial mass to increase its spin from $a=0$ to $a=0.9$ and 122$\%$ of its initial mass to increase its spin from $a=0$ to $a=0.99$ (\citealt{1970Natur.226...64B, 2008ApJ...682..474B}), so a limited reservoir of matter might prevent significant spin changes even in LMXB. Additionally, works such as Draghis et al. (2022) find that even the BHs in intermediate or low mass X-ray binaries can be born with high spins, not requiring significant accretion.

We measured ten high BH spins, with nine of them approaching the maximal possible value. However, only three of the ten sources show resolved radio jets or "cannonball" ejections: MAXI J1820+070 (\citealt{2020ApJ...891L..29H}), H 1743-322 (\citealt{2009ApJ...702.1648H}), and GRS 1758-258 (\citealt{2002A&A...386..571M}). This suggests that a high BH spin is not sufficient for production of resolvable radio jets, and that the observability of the ejecta is likely more strongly influenced by system properties such as the mass accretion rate in the system, the orbital period of the binary, the surrounding interstellar medium (ISM), and even the distance to the system.

The reported uncertainties of our measurements are purely statistical, and the systematic uncertainties are likely to be smaller, yet not negligible. As the BH spin and inclination of the inner disk influence spectral features similarly, we show all the individual spin and inclination measurements for all observations treated in this work in Figure \ref{fig:All_spin_vs_incl}, aiming to highlight any possible correlation between the two parameters in the models. While there appears to be a positive trend, that is largely owed to the measurements of MAXI J1820+070. 

In Figure \ref{fig:MAXI_J1820_spin_vs_incl}, we show the spin and inclination measurements obtained from the 19 observations of MAXI J1820+070, with the transparency of the points being proportional to the ratio of reflected to total flux in the observation, which was used to weight the measurements when computing the combined measurement. The combined measurements for spin and inclination are shown through solid lines, while their uncertainties are presented through the shaded regions. Previously, any of those measurements could have been reported in literature, but by looking at all the observations we obtain a value that is not biased by observation-specific peculiarities and more clearly reflects the true value of the parameters. MAXI J1820+070 was one of the brightest X-ray binary systems ever observed, and with it being extensively observed in a comprehensive manner across the electromagnetic spectrum, the existing data set makes this source an anchor in our understanding of electromagnetic emission from stellar mass BHs.

Figure \ref{fig:All_spin_vs_incl} also shows that few spin measurements take low or negative values, most originating from observations of 4U 1957+11. In Figure \ref{fig:4U_1957_spin_vs_incl}, we show the inclination vs spin parameter space for the 10 observations of 4U 1957+11, with the transparency of the points being proportional to the ratio of reflected to total flux. Similar to Figure \ref{fig:MAXI_J1820_spin_vs_incl}, the solid lines and shaded regions represent the values and uncertainties obtained while combining the individual measurements. While there are some low and negative spin measurements, those come from observations in which reflection is not as strong, with the combined measurements being dominated by the observations in which reflection is more strongly present. Figures \ref{fig:MAXI_J1820_spin_vs_incl} and \ref{fig:4U_1957_spin_vs_incl} highlight the importance of analyzing all available observations with the aim of reducing any possible systematic uncertainties of our measurements.

Having such a large data set produced through consistent models and assumptions and with uniform systematic uncertainties also allows us to look into the systematic effects of parameters in the models on the spin measurements. In Figures \ref{fig:AFe_vs_a} and \ref{fig:R_vs_a}, we show all the individual spin measurements presented in this paper, in relation with the Fe abundance ($A_{\rm Fe}$) and reflection fraction ($R$) predicted by the model, respectively. While it appears that the models prefer high Fe abundance and low reflection fraction, there is no apparent direct correlation between these values and the spin measurement. Additionally, we can begin to look into the general behavior of the models by highlighting correlations between other parameters. For example, in Figure \ref{fig:AFe_vs_R} we show the Fe abundance in all measurements in comparison with the reflection fraction. In this case, no direct correlation is present. However, such parameter correlations can also be treated at the level of each individual fit through the posterior distributions obtained through the MCMC runs. While inspecting the corner plots showing trends in the posterior distributions of parameters, no obvious correlations affecting the spin were found. Still, on occasion, degeneracies between parameters were present. Such examples include the correlation between the outer emissivity index $q_2$ and the breaking radius $R_{\rm br}$ parametrizing the coronal emissivity, which often lead to unphysically low values of $q_2$ and high values of $R_{\rm br}$ due to the inability of the data to constrain the parameters. Due to the large number of parameters, the scale of a complete treatment of the parameter space is substantial and beyond the scope of this work and will be treated in a future paper, in which we also plan to expand the sample size of spin measurements with existing measurements reanalyzed using the same assumptions of this paper. 

%concluding paragraph about BBH vs XB. See Zevin & Bavera 2022 
While based on the entire GWTC-3 sample, \cite{2021arXiv211103634T} concluded that most BHs in BBH systems are slowly rotating. However, a few rapidly rotating BHs were identified in this sample, such as GW190412 (either the spin of the more massive component is $a_1=0.44^{+0.16}_{-0.22}$ - \citealt{2020PhRvD.102d3015A}, or the spin of the less massive component is $a_2=0.88^{+0.11}_{-0.24}$ - \citealt{2020ApJ...895L..28M}), GW190517\_055101 ($a_1=0.86^{+0.13}_{-0.35}$ - \citealt{2021PhRvX..11b1053A}), GW190403 051519 ($a_1=0.92^{+0.07}_{-0.22}$ - \citealt{2021arXiv210801045T}), or GW191109\_010717 ($a_1=0.83^{+0.15}_{-0.58}$ - \citealt{2021arXiv211103606T}). The formation of BBH systems is a topic of debate, and the current preferred view is that the entire observed distribution cannot be explained through a single formation channel (see e.g., \citealt{2021ApJ...910..152Z} for a detailed view on BBH formation channels). \cite{2022ApJ...935L..26F} find that based on the existing sample of GW measurements, at most 26\% of the underlying BBH population originates from "hierarchical mergers" (repeated mergers of smaller BHs). While works such as \cite{2021ApJ...921L...2O} argue that the observed distribution of BBH is consistent with formation from isolated binaries, \cite{2022arXiv220714290G} find that at most 20\% of BBH originate from HMXB systems and at most 11\% of HMXB systems evolve into BBH, suggesting that the spin distributions in BBH and XB are not only observed to be different (\citealt{2022ApJ...929L..26F}), but intrinsically different. In the future, expanding and comparing the two spin samples will allow placing better constraints on BH formation and evolution mechanisms.

\begin{acknowledgments}
AZ is supported by NASA under award number 80GSFC21M0002. JAT acknowledges partial support from NASA through NuSTAR Guest Observer grants 80NSSC20K0644 and 80NSSC22K0059.\\

\textit{Software:} 
\texttt{Astropy} (\citealt{astropy:2013, astropy:2018}), 
\texttt{emcee} (\citealt{2013PASP..125..306F}), 
\texttt{numpy} (\citealt{harris2020array}), 
\texttt{matplotlib} (\citealt{Hunter:2007}), 
\texttt{scipy} (\citealt{2020SciPy-NMeth}), 
\texttt{pandas} (\citealt{jeff_reback_2022_6702671, mckinney-proc-scipy-2010}), 
\texttt{corner} (\citealt{corner}), 
\texttt{iPython} (\citealt{PER-GRA:2007}), 
\texttt{Xspec} (\citealt{1996ASPC..101...17A}), 
\texttt{relxill} (\citealt{2014MNRAS.444L.100D, 2014ApJ...782...76G}).
\end{acknowledgments}

%\newpage

\bibliography{paper}{}
\bibliographystyle{aasjournal}

\appendix
\section{Choice of source and background regions}\label{sec:regions}
The choice of background region adopted in this work is somewhat uncommon when handling NuSTAR data. The common practice in the community is to adopt a circular source region of a certain radius ($\sim100''$) and a circular background region of the same size, placed in an area of the detector that is not contaminated by the source flux or any stray light that is being detected (for a discussion on stray light in NuSTAR data, see \citealt{2017arXiv171102719M, 2021ApJ...909...30G}). This process requires the user to manually identify the position of the source and background region in relation to the image coordinates on the detector. 

In order to simplify, automate, and increase the reproducibility of the spectral extraction process, we adopted circular regions of radius equal to 120'' for extraction of the source spectrum, centered at the position of the source in the header of events file of the observation, for each of the detectors. This choice for the position of the source region was made in order to avoid the possibility that the coordinates available in literature for some sources are uncertain and therefore bias the spectral extraction process. For the background region, we used annular regions centered at the position of the source, with an inner radius of 200'' and an outer radius of 300''. Firstly, this reduces the arbitrariness of the choice of position of the background region. Secondly, this method samples a much larger detector area when constructing the background spectrum. However, for bright sources, the 200'' inner radius of the annulus might be too small, leading to source photons being included in the background. This is potentially important for sources that have a low flux at high energies, as an increase in flux in the background spectrum would lead to the source spectra becoming background dominated at lower energies.

We tested this effect by extracting source and background spectra from the observation 80502324004 of MAXI J0637-430,  using five combinations of source and background regions. The spectra obtained are shown in Figure \ref{fig:region_comp}, together with the residuals of the fit to the best performing reflection model for this observation (see Subsection \ref{sec:MAXI_J0637}). The error bars on the background points were omitted for visual clarity. Additionally, we binned the spectra using three different methods: the ``optimal" binning scheme (\citealt{2016A&A...587A.151K}), or by requiring minimum a signal to noise ratio of 5 and of 10 in each spectral bin. The five combinations of source and background regions are as follows: Case 1 - 120'' radius circular source region and annular background region with 200'' inner radius and 300'' outer radius, centered at the position of the source; Case 2 - 120'' radius circular source region and 120'' radius circular background region, placed as far as possible on the detector; Case 3 - 90'' radius circular source region and 90'' radius circular background region, placed as far as possible on the detector; Case 4 - 150'' radius circular source region and 150'' radius circular background region, placed as far as possible on the detector; Case 5 - 120'' radius circular region and annular background region with 300'' inner radius and 600'' outer radius, centered at the position of the source.

The source spectrum is practically unchanged regardless of the size of the source region, with no significant differences even at high energies. The background spectra show changes between the five different cases, mainly at low energies. The background spectra extracted using the region in Case 1 (used throughout the analysis in this paper) produces a background rate at 3~keV an order of magnitude larger than e.g., Case 2 and 3 (i.e. a 120'' or  90''  circular background region placed far on the detector from the source region). However, when comparing to Cases 4 and 5, the background rate is only $\sim5$ times higher in Case 1. Given that the sources treated in this paper have high fluxes at low energies, the difference in background spectra should not influence the quality of the data and the obtained results. Given the goal of automatizing of the spectral extraction process, Case 5 would produce the ideal combination of source and background spectra. However, in Case 5, with the significantly increased size of the annular extraction region, the risk of the region overlapping with regions on the detector dominated by stray light during observations is increased, likely influencing the background spectra more than the difference between Case 1 and 5 would.

At high energies, the background spectra are consistent regardless of choice of extraction region. However, different binning schemes can lead to the appearance of the source spectra becoming background dominated at lower energies, the clearest being Case 3 when requiring all bins to have a minimum signal to noise ratio of 10. For the same case, different binning techniques clearly indicate that the background spectra dominate over the source spectra above $\sim60$ keV. Interestingly, the binning techniques requiring a minimum signal to noise ratio often have difficulties in treating the background spectra, creating apparent streaks. However, the error bars on the fluxes in the background bins are larger than the variation trend seen in Cases 1-4. Nevertheless, it appears that optimal binning better handles background spectra.

With the different cases of source and background region and different binning schemes producing visually no difference at high energies and changes of little importance at low energies, it is important to quantify the difference produced when fitting the data with a model. We fit all spectra shown in Figure \ref{fig:region_comp} with the best performing \texttt{relxill} model for this particular observation in our analysis. The best-fit values for the temperature of the accretion disk, BH spin, inner accretion disk inclination, power-law index, and disk ionization are shown in Figure \ref{fig:region_par_comp}. The different colors of the points represent the five cases presented in Figure \ref{fig:region_comp}, and the different marker represent the binning scheme used, with circles, squares, and diamonds representing the optimal binning scheme, the signal to noise ratio of 5 and of 10, respectively. The filled markers indicate the results of fitting the spectra in the 3-60~keV band, while the empty markers indicate the results of fitting the spectra in the entire 3-79~keV NuSTAR band. The left panels show only the value of the parameters for which the best-fit solution was achieved, while the right panels also show the uncertainty of the parameters, computed as the square root of the diagonal element in the covariance matrix of the fit. This uncertainty is used to produce the proposal distribution for the walkers in the MCMC analysis. 

It is important to mention that these measurements are made only by fitting one of the two NuSTAR spectra obtained during one observation (FPMA in this case), and fitting both spectra jointly is likely to reduce the variation between the parameters and to decrease the size of the uncertainties shown in the right panels. Truncating the spectra at 60~keV as opposed to 79~keV has no obvious impact on the best-fit parameters, with the only noticeable difference being that fitting the spectra up to 79~keV produces slightly smaller uncertainties on some parameters. The difference between the best-fit parameters found when fitting spectra extracted using the five different cases for the region size, shape, and location is comparable with the difference produced when using different binning techniques. However, the variability in the best-fit parameters is negligible when compared to the size of the errors. As these fit parameters are only used for the proposal distribution in the MCMC analysis, the results will not be significantly influenced by this variability induced by the different choices of source and background region and of binning scheme used. 

Therefore, throughout the analysis, we continue using Case 1 for extraction of source and background spectra, as it increases the ability to automate our spectra extraction pipeline when compared to Cases 2-4, and it reduces the risk of producing biased background spectra due to stray light when compared to Case 5, at the cost of producing increased background spectra at low energies, which appear to have no significant impact on the result. We continue using the optimal binning scheme, as it leads to similar results with the other binning techniques, while producing spectra with fewer spectral bins compared to the other techniques, which helps reduce the time required to fit the spectra. We mention that the optimal binning scheme still produces bins with enough counts that using $\chi^2$ statistics is still appropriate. 

\section{Algorithm to combine measurements}\label{sec:combining_explanation}
The script used to combine measurements takes the posterior distributions of the individual parameters obtained through the MCMC runs and maps them to the $[0,1]$ interval. Then, it then generates a beta distribution with the probability density function described by:
\begin{equation}
    PDF(x)=\frac{x^{\alpha-1}(1-x)^{\beta-1}}{B(\alpha,\beta)}
\end{equation}
where $B(\alpha,\beta)=\frac{\Gamma(\alpha)\Gamma(\beta)}{\Gamma(\alpha+\beta)}$ and $\Gamma$ is the Gamma function, $\Gamma(a)=\int_{0}^{\infty}t^{a-1}e^{-a}dt$. This distribution has two parameters, $\alpha$ and $\beta$, that we determine through a Bayesian analysis in order to describe the combined measurement. We assume uniform priors for $\alpha$ and $\beta$, but we require that $\alpha>1$, $\beta>1$, and $(\alpha+\beta)<171.5$. The first two requirements come from the definition of the beta distribution, and the third is dictated by numerical limitations of the used programming language: $\Gamma(171.5)\sim9.5\times10^{307}$. This number is so large that when computing the value of $B(\alpha,\beta)$, any combination of $\alpha,\beta$ for which $\alpha+\beta\geq171.5$ returns $B(\alpha,\beta)=0$ due to division by a very large number. This then leads to division by zero when computing the value of the PDF at a given location.

The total likelihood of a parameter combination is evaluated by summing the (logarithm of the) likelihood of the parameter combination describing each individual measurement. As the posterior distributions for the parameter are normalized in the same way as the Beta distribution, we quantify how well the Beta distribution together with a parameter combination describes the spin measurement by taking the ratio of the are under the minimum of the two curves to the area under the maximum of the two curves, evaluated over the $[0,1]$ interval. When computing the total likelihood by summing the (logarithm of the) individual likelihoods for each measurement, we weight each measurement in the sum by the ratio of the reflected flux to the total flux in the 3-79~keV band during the observation. This way, we weight observations with stronger reflection more than those with weaker reflection, as they are more likely to allow us to separate the reflection component from the underlying continuum and better constrain the parameters.

As an example of how our likelihood function behaves, Figure \ref{fig:combine_likelihood} illustrates our attempt to fit a Gaussian function defined on the $[0,1]$ interval, with a mean of $\mu=0.8$ and a standard deviation $\sigma=0.1$. The left panels show the Gaussian function (in black) and three attempted models shown in green: a beta distribution (top panel), Delta distribution (middle panel), and a uniform distribution (bottom panel). All curves are normalized in the same way. On the right, we show the minimum of the two curves in blue and the maximum of the two in red. In order to maximize likelihood, our function attempts to maximize the minimum of the two curves and to minimize the maximum of the two. The method becomes more intuitive upon inspection of the top right panel in Figure \ref{fig:combine_likelihood}: our code aims to increase the blue curve (minimum of the two) and to decrease the red one (maximum of the two). This way, the minimum and maximum are most similar, translating to a better fit of our beta distribution to the initial assumed distribution (in this example) and to the data (in real usage).

To test the ability of our script to combine measurements and recover the underlying distribution, we adopted a Gaussian distribution of some given mean and standard deviation and generated a number of Gaussian measurements with position probabilities given by the initial assumed probability density function. We then used our script to combine all ``measurements" in order to infer the underlying distribution, with the main goal of recovering its mode and credible interval. 

Four examples are shown in the four panels of Figure \ref{fig:tests_combining}. The black curves represent the assumed underlying Gaussian distributions, and the dashed green curves show the Beta distributions representing the best fit of our algorithm to the distribution underlying all ``measurements". The top left panel simulates an attempt to combine ten measurements around a mean of $\mu=0.7$, with a standard deviation $\sigma=0.1$. Both the mode and the credible interval of our inferred distribution matches the true values very closely. We note that due to numerical artifacts caused by the limited number of computation iterations, the truth mode and credible intervals of the assumed distribution (black curve) differ very slightly from the values assumed when generating the distributions. This example is well applicable to combining multiple inclination measurements. While the example shows distributions defined on the $[0,1]$ interval, one could combine inclination measurements through a change of variable.

The top right panel shows an attempt to combine five measurements generated around a mean of $\mu=0.97$ and standard deviation $\sigma=0.05$. This example illustrates the usability of combining spin measurements clustered around the maximal value. In this case, the inferred distribution (green dashed line) again closely recovers the assumed underlying distribution, with the mode and credible intervals of the assumed and derived distributions being in good agreement. In this case, the derived distribution appears to be slightly shifted toward lower spin values, likely due to the requirement of the Beta distribution to return to a value of zero at the edge of the interval.

The bottom panels of Figure \ref{fig:tests_combining} show an attempt to combine three measurements drew from a Gaussian of mean $\mu=0.5$ and standard deviation $\sigma=0.05$. The left panel shows the combined distribution when all three measurements are weighted equally, well recovering the assumed distribution. The right panel weights the second measurement (shown through the solid green line) eight times more than the other two measurements. The inferred distribution much better recovers the second ``measurement", which was more heavily weighted, while also slightly accounting for the other two measurements. 

Often, when multiple observations of the same source are available, an option is to fit the spectra jointly, linking parameters that are expected to stay constant between observations. In this case, the BH spin and the inclination of the inner accretion disk could be linked between observations. The constraints on the parameters would simultaneously account for all the data that is being fit, providing more credible measurements. However, fitting multiple NuSTAR spectra with complicated models quickly becomes a task too computationally intensive, which requires unfeasible times to run. In order to test the robustness of our result combining method against the possibility of fitting the spectra jointly with parameters linked, we fit two of the three observations of MAXI J0637-430 used in the analysis presented in Subsection \ref{sec:MAXI_J0637} jointly. We fit the spectra of the two observations in the 3-50~keV band, linking the BH spin, the inclination of the inner disk, and the column density along the line of sight, and allowed all other parameters to vary freely, as described in Subsection \ref{sec:models}.

Fitting the spectra jointly using \texttt{TBabs*(diskbb+powerlaw)} returns $\chi^2/\nu=1651.15/634$, with clear signs of reflection. Similarly to the individual fits, the best performing models in terms of $\chi^2$  that accounts for refection are \texttt{TBabs*(diskbb+relxill)}, returning $\chi^2/\nu=707.67/619$ and \texttt{TBabs*(diskbb+relxillD)} with Log(N)=19, returning $\chi^2/\nu=710.5/622$. The latter performs significantly better in terms of DIC, with $\Delta$DIC=200 better than the former. Therefore, we report the measurements based on the \texttt{TBabs*(diskbb+relxillD)} with Log(N)=19 MCMC run. We find $a=0.980^{+0.005}_{-0.009}$ and $\theta=59^{+3}_{-2}$ degrees. In comparison, when fitting the spectra individually and combining the results using our algorithm, we find $a=0.97\pm0.02$ and $\theta=62^{+3}_{-4}$ degrees. The results formally agree well within their uncertainties, suggesting that our method produces results that are robust and precise. However, it is worth noting that the combined result also takes advantage of the third available observation. The quoted statistical uncertainties on both the spin and inclination measurements are larger when combining the measurements when compared to the measurements based on a joint fit. This is linked to the fact that the errors quoted are simply statistical, and the combined measurement begins to include systematical errors, which are likely to contribute significantly.

\section{Sources wherein no measurement was possible}\label{sec:no_measurement}
\subsection{XTE J1859+226}
NuSTAR observed XTE J1859+226 once (ObsID 90701305002) for 41 ks. This observation was described in \cite{2021ATel14512....1D}. The spectra are background dominated above $\sim20\;\rm keV$. Fitting with the power law model returns $\chi^2/\nu=187/208$ when the data is binned to have a minimum signal/noise ratio SNR=5, and $\chi^2/\nu=304/340$ when using optimal binning. No reflection features are apparent and a spin measurement is not possible.

\subsection{4U 1755-338}

4U 1755-338 was observed once by NuSTAR (ObsID 90601313002), and the observation was first described by \cite{2020ATel13665....1D}. During the observation, the source was in a ``high/soft" state, with the emission being dominated by the accretion disk. The source is detected up to an energy $\sim25\;\rm keV$. Fitting the spectra using \texttt{diskbb+powerlaw} produces a good fit, returning $\chi^2/\nu=416/374$. No reflection features are apparent.

\subsection{M31 ULX-1 (CXOM31 J004253.1+411422)}
The position of M31 ULX-1 was covered by 11 NuSTAR observations: ObsID 50026001002, 50026001004, 50111001002, 50110001002, 50110001004, 50101001002, 50110001006, 50201001002, 50302001002, 50302001004, 50302001006. The source is not detected in any of the 11 observations.

\subsection{M31 ULX-2 (XMMU J004243.6+412519)}
There are 4 NuSTAR observations of the region containing M31 ULX-2: ObsIDs 50302001002, 50302001004, 50302001006, and 30365002002. The source is not detected in any of the four observations.

\subsection{XTE J1748-288}
NuSTAR observation 40012019001 was pointed at $\sim8'$ from XTE J1748-288, placing the known location of the source close to the edge of the detectors. However, the source was quiescent during the NuSTAR observation and not detected.

\subsection{SS 433}
SS 433 is an X-ray binary system that is best known for launching extended relativistic jets (v = 0.26c, \citealt{1989ApJ...347..448M}) that are loaded with baryons (see, e.g., \citealt{2013ApJ...775...75M}).  The X-ray spectrum of the source is dominated by emission from the variable and precessing jets, complicating efforts to detect emission from the inner disk.  Moreover, the jets may be launched because SS 433 harbors a compact object that is accreting far above its Eddington limit (e.g. \citealt{2015NatPh..11..551F}).  Especially in this case, the central engine may be enshrouded within an optically-thick spherical flow, or at least partially obscured by a vertically extended disk that further serves to obscure signatures from the innermost accretion flow.  Therefore, despite emerging evidence that SS 433 may harbor a black hole rather than a neutron star (e.g., \citealt{2020A&A...640A..96P}), SS 433 fails several of our criteria for examination, and it is excluded from our analysis.

% \subsection{IGR J17379-3747}  %this is actually a neutron star
% There are 2 NuSTAR observations of IGR J17379-3747 (ObsID 90401312002 and 90601333002). However, the source is only detected during ObsID 90401312002. Fitting the spectra using \texttt{powerlaw} returns $\chi^2/\nu=384/351$, with no reflection features apparent.
   
\subsection{XTE J0421+560}
XTE J0421+560 is the X-ray counterpart of CI Camelopardalis, which is believed to be a B[e] star. The nature of the compact object in the system is not constrained, and \cite{2004ApJ...601.1088I} argued that the emission from the system during quiescence does not allow placing constraints on the nature of the compact object, but that the ASCA observations taken during the 1998 outburst of the source broadly point toward the compact object being a white dwarf. Constraints on the nature of a compact object can be placed through spin measurements: a spin higher than $a=0.7$ (\citealt{2011ApJ...728...12L}) indicates that it must be a black hole (see e.g., Draghis et al. 2022 in prep for an example analysis). Therefore, we chose to analyze this source despite a lack of clear evidence of it harboring a black hole.

The spectra from ObsID 90201040002 of XTE J0421+560 do not require a \texttt{diskbb} component. We fit the spectra in the 3-20~keV band. Fitting with \texttt{powerlaw} gives $\chi^2/\nu=103.2/95$. The two best performing models are \texttt{TBabs*relxillD} with Log(N)=19 with $\chi^2/\nu=92.37/87$ and \texttt{TBabs*relxill} with $\chi^2/\nu=93.41/87$.  

Running MCMC on both models, the chains produce similar DIC, 109.07 and 109.42 respectively. In both cases, the spin is low, $a<0.26$. However, running \texttt{steppar} between -0.998 and 0.998 in spin shows differences of $\Delta \chi^2=0.1$ across the entire parameter space. Even though reflection is evident in this observation, the spin cannot be constrained given the quality of the data.

\subsection{IC 10 X-1}
There is one NuSTAR observation of IC 10 X-1 (ObsID 30001014002). Using the continuum fitting method, \cite{2016ApJ...817..154S} measured the spin of the BH in IC 10 X-1 to be $a=0.85^{+0.04}_{-0.07}$ (90\% confidence).

Fitting the 3-20~keV spectrum with \texttt{TBabs*(diskbb+powerlaw)} produces a good fit, with $\chi^2/\nu=111.91/96$. Replacing the power-law component with models from the \texttt{relxill} family significantly improves the fit, with the best performing model \texttt{TBabs*(diskbb+relxilCp)} returning $\chi^2/\nu=81.10/86$. We find that the fits are unable to constrain the BH spin. 68\% of the posterior samples in the MCMC run take spin values $a\leq0.17$. This measurement could be biased towards low values by constraining the accretion disk to extend all the way to the ISCO. Still, assuming the mass and distance estimates from \cite{2016ApJ...817..154S} ($M\sim15\;\rm M_{\odot}$ and $d\sim800~\rm kpc$ and given the 3-20~keV flux in the NuSTAR observation, this would result in an Eddington fraction of $\sim3\%$, well within the expected theoretical range for which the inner disk is expected to extend to the ISCO. Additionally, the environment around the black hole is likely to be significantly more complicated than the simplistic assumption of this model. As the signal to noise of the observation is low, we do not pursue more complicated models. 

\subsection{M33 X-7}
There are 2 NuSTAR observations that cover the position of the source (ObsID 50310001002 and 50310001004). M33 X-7 is not detected during the two observations, taken in March and July 2017. No spin measurement was possible.

\subsection{M33 X-6}
In the same two NuSTAR observations as for M33 X-7 (ObsID 50310001002 and 50310001004), the source M33 X-6 is clearly detected. For ObsID 50310001002 we fit the spectra in the 3-15~keV band, while for ObsID 50310001004 we fit in the 3-10~keV band. For the former, fitting the 3-15~keV spectra with \texttt{TBabs*powerlaw} returns $\chi^2/\nu=95.91/80$, while the best performing reflection model \texttt{TBabs*relxilllp} returns $\chi^2/\nu=90.75/73$. Similarly, for ObsID 50310001004, fitting with \texttt{TBabs*powerlaw} produces a good fit, returning $\chi^2/\nu=54.72/50$, while the best reflection model \texttt{TBabs*relxill} returns $\chi^2/\nu=50.48/41$, with the improvement in statistic not being justified by the increase in the number of free parameters. In both cases, the addition of a reflection component is not justified statistically. 

\subsection{M33 X-8}
There are 2 NuSTAR observations of M33 X-8 (ObsID 50310002001 and 50310002003) in which the source is clearly detected. Fitting the 50310002001 spectra in the 3-15~keV band using \texttt{TBabs*(diskbb+powerlaw)} returns $\chi^2/\nu=118.97/89$, while the best performing reflection model \texttt{TBabs*(diskbb+relxilCp)} returns $\chi^2/\nu=116.20/80$. For the 50310002003 observation, also fitting in the 3-15~keV band, the \texttt{TBabs*(diskbb+powerlaw)} model returns a good fit, with $\chi^2/\nu=60.72/86$, while the best performing reflection model \texttt{TBabs*(diskbb+relxilD)} with Log(n)=19 fixed returns $\chi^2/\nu=55.89/78$. The statistical improvement owed to the addition of reflection is not justified by the increase in the number of free parameters, suggesting that reflection is not statistically significant. 

% Still, it is important to note that this source is in a hard state and that the best-fit normalization of the \texttt{diskbb} component is low, with the power law component dominating even at low energies. If we fit the spectra with \texttt{TBabs*powerlaw}, then $\chi^2/\nu=69.89/89$, while the best reflection model \texttt{TBabs*relxillCp} returns $\chi^2/\nu=60.04/80$. The p value from an F test indicates a probability of 0.18 that the improvement can be attributed to chance. While formally this value is above the classical limit for significance of 0.05 and would suggest that reflection is not significant in this case, we recommend further observations of this source during future outbursts.

\subsection{IGR J17451-3022}
There is one NuSTAR observation in the direction of this source (ObsID 30601014003), but the offset between the position of IGR J17451-3022 and the pointing of NuSTAR was $\sim 9.3'$, meaning that the system was near the edge of the field of view of NuSTAR. This observation cannot be used for a spin measurement. 

\subsection{Swift J1357.2-0933}
There are 3 NuSTAR observations of Swift J1357.2-0933. For ObsID 90201057002, we fit the 3-79~keV spectrum; for 90301005002, we fit the 3-50~keV spectrum; and for 90501325002, we fit the 3-20~keV spectrum. We fit all three observations jointly. The quality of the fits does not improve through the addition of a \texttt{diskbb} component nor by including absorption through \texttt{TBabs}, as expected given that the estimated column density along the line of sight is relatively small, $\sim3\times10^{20}\;\rm cm^{-2}$.

Fitting the six spectra with \texttt{powerlaw} gives a cumulative $\chi^2/\nu=928.89/808$. Visually inspecting the residuals does not reveal clear signs of reflection. When fitting the spectra with a power-law with a high-energy cutoff (\texttt{cutoffpl}), the quality of the fit improves significantly, to $\chi^2/\nu=874.08/805$. We tested whether reflection is present by fitting the spectra first with \texttt{xillver} (producing $\chi^2/\nu=858.98/797$), and with the models in the \texttt{relxill} family. Of those, \texttt{relxillCp} produced the best statistic, returning $\chi^2/\nu=861.18/787$. Note that the value of $\chi^2$ is worse than in the case of unblurred reflection (\texttt{xillver}), suggesting that any reflection does not originate from the innermost regions of the accretion disk. 

We computed the AIC and BIC for the four mentioned models. Of those, \texttt{cutoffpl} and \texttt{xillver} produce the lowest values, suggesting stronger statistical significance. Comparing the two models, \texttt{cutoffpl} performs better than \texttt{xillver} in terms of both AIC and BIC producing AIC=79.17 and BIC=135.64 versus AIC=80.93 and BIC=175.05 for \texttt{xillver}. The lower values of both AIC and BIC produced by fitting the data with \texttt{cutoffpl} when compared to \texttt{xillver} suggests that reflection is not statistically significant in this source, during the three observations. %Still, the quality of the fit does improve by adding adding reflection, therefore we suggest monitoring of this source during future outbursts.

\subsection{Cyg X-3} \label{sec:cyg_x3}
There are four archival NuSTAR observations of Cyg X-3: ObsIDs 10102002002, 90202051002, 90202051004, and 90502319002. Using the estimates for the black hole mass of $2.4^{+2.1}_{-1.1}\;\rm M_\odot$ (\citealt{2013MNRAS.429L.104Z}) and for distance to the system of $7.4\pm1.1\;\rm kpc$ (\citealt{2016ApJ...830L..36M}), we conclude that all four observations happened while the source was at an Eddington fraction between the limits for which we expect a the accretion disk to extend all the way to the ISCO. Therefore, in principle, all four observations could be used in our analysis. 

However, Cyg X-3 is a small black hole in a binary with a Wolf-Rayet (WR) star, that is feeding the accretion onto the compact object through stellar winds. Therefore, in addition to the high column density along the line of sight due to the location of Cyg X-3 in the plane of the galaxy, we expect a significant amount of absorption and re-emission within the intra-binary medium. Due to the high amount of obscuration within the system, we chose to only analyze the observation obtained while the source was in the hardest state of the ones available. Of the four available NuSTAR observations, ObsID 10102002002 has both the lowest soft flux and highest hard flux, suggesting both that the soft emisison from the stellar winds is reduced and that the flux from the corona is enhanced (or less absorbed). This is the same observation treated by \cite{2020A&A...639A..13K}.

Fitting the spectra obtained from ObsID 10102002002 with the same series of models that we apply to our entire sample returns unacceptable fits. No model is able to properly fit the enhanced emission in the Fe K region. When using the \texttt{TBabs*(diskbb+powerlaw)} model, $\chi^2/\nu=69088.50/456$, and when attempting to account for relativistic reflection by replacing the \texttt{powerlaw} component with \texttt{relxill} returns $\chi^2/\nu=11347.22/447$. Clearly, the simplistic treatment does not suffice for this source.

A significantly better fit is obtained when introducing two \texttt{apec} components and two \texttt{zxipcf} absorption components to the model, and replacing the \texttt{powerlaw} component with a power law with a high energy cutoff component (\texttt{cutoffpl}). The first pair of \texttt{apec} and \texttt{zxipcf} was added to describe the effect of the stellar wind from the WR companion, while the second pair was added in order to describe the effects of an accretion disk wind. This model returns $\chi^2/\nu=1101.06/442$. While nominally a poor fit, this is a very significant improvement from previous models. 

However, while this model includes the contribution from an accretion disk through the \texttt{diskbb} component and from a compact corona through the \texttt{cutoffpl} component, it does not include reflection features. Therefore, we added a \texttt{xillver} component to represent unblurred reflection to test the statistical significance of the presence of reflection. In \texttt{Xspec} parlance, the complete model is \texttt{TBabs*zxipcf*zxipcf*(diskbb+cutoffpl+xillver+apec+apec)} and returns $\chi^2/\nu=620.09/436$. In this model, we set the reflection fraction of the \texttt{xillver} component to negative values in order to only model the contribution of reflection. For both \texttt{apec} components, the redshift was fixed to zero, and the metal abundance was linked between components, taking a moderate value $\sim 1.5$. For the \texttt{zxipcf} components, the one describing the stellar wind had a low ionization ($\log\xi\sim0.9$), higher covering fraction ($f\sim0.9$), and lower column density ($N_H\sim5\times10^{23}\;\rm cm^{-2}$), while the component describing the accretion disk wind had a higher ionization ($\log\xi\sim4$), lower covering fraction ($f\sim0.7$), higher column density ($N_H\sim1.8\times10^{24}\;\rm cm^{-2}$), and allowed to have a negative redshift (blueshift), taking a value of $z\sim-0.043$, which if real would correspond to a wind being launched at $\sim4\%$ of the speed of light. The photon index $\Gamma\sim2.4$ and the high energy cutoff $E_{\rm cut}\sim10\;\rm keV$ were linked between the \texttt{xillver} and the \texttt{cutoffpl}. The predicted ionization of the \texttt{xillver} component is moderately low ($\log\xi\sim1.3$), indicative of distant reflection, while the Fe abundance is subsolar $A_{Fe}\sim0.62$. The inclination $\theta$ and reflection fraction $R$ are unconstrained. The values reported here come from the direct \texttt{Xspec} fit, and no MCMC analysis was run on the parameter space, which is why no uncertainties are reported for the parameters, and we report these values only for completeness.

When substituting the \texttt{xillver} component with the relativistically blurred \texttt{relxill} version, the quality of the fit decreases to $\chi^2/\nu=641.13/431$. While it is possible that the fit converged to a local minimum, the quality of the fit appeared insensitive to the relativistic reflection parameters. Using a model with both \texttt{relxill} and \texttt{xillver} produces $\chi^2/\nu=619.78/427$, with the improvement in $\chi^2$ not being justified by the increase in the number of free parameters. Since the quality of the fit decreases through the addition of relativistically blurred reflection when compared to unblurred reflection, we conclude that while reflection is present, it likely happens at relatively large distances from the compact object, making a spin measurement from the current NuSTAR data unachievable. In the future, pairing NuSTAR or HEX-P (\citealt{2018SPIE10699E..6MM}) data with XRISM (\citealt{2018SPIE10699E..22T}) and eventually ATHENA (\citealt{2018SPIE10699E..1GB}) will allow distinguishing between the emission from stellar winds, accretion disk, disk winds, and the relativistic reflection and enable spin measurements, and Cygnus X-3 is a prime and exciting candidate for this work.

%\newpage
\section{Figures and Tables}\label{sec:figures_and_tables}
This appendix contains all the figures and tables in the paper.

\begin{figure}[ht]
    \centering
    \includegraphics[width= 0.75\textwidth]{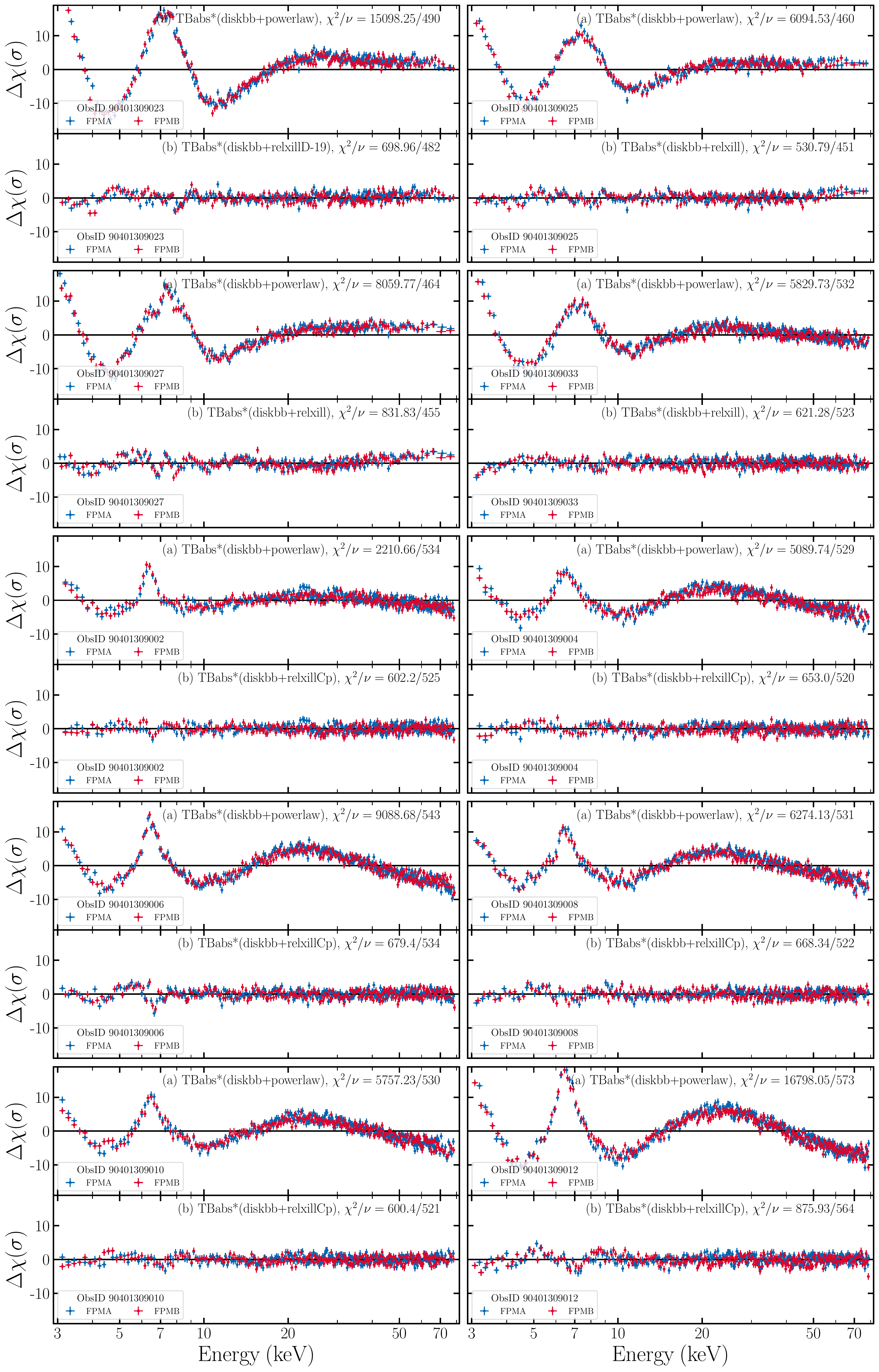}
    \caption{The 10 panels represent the first 10 NuSTAR observations of MAXI J1820+070 analyzed. The top sub-panels show the residuals in terms of $\sigma$ produced when fitting the spectra with \texttt{TBabs*(diskbb+powerlaw)}, and the bottom sub-panels show the residuals when fitting with the best performing reflection model. The blue and red points represent the FMPA and FMPB spectra from each observation, respectively. The sub-panel titles show the model used to fit and the statistic produced, and the sub-panel legends mention the ObsID number of the observation shown.}
    \label{fig:MAXI_J1820_delchi_1}
\end{figure}

\begin{figure}[ht]
    \centering
    \includegraphics[width= 0.75\textwidth]{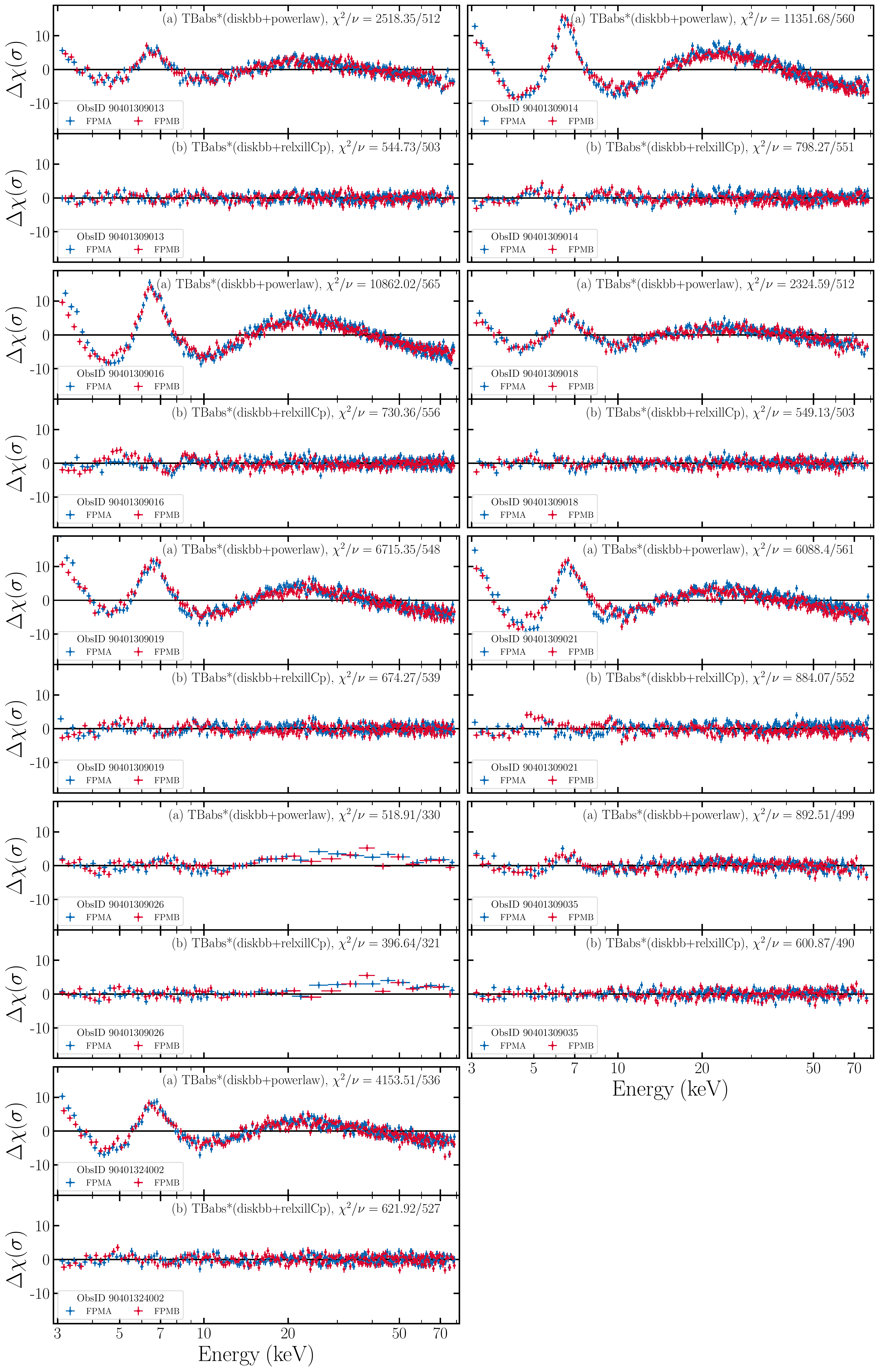}
    \caption{Continuation of Figure \ref{fig:MAXI_J1820_delchi_1} for the remaining 9 observations of MAXI J1820+070 analyzed.}
    \label{fig:MAXI_J1820_delchi_2}
\end{figure}

\begin{figure}[ht]
    \centering
    \includegraphics[width= 0.85\textwidth]{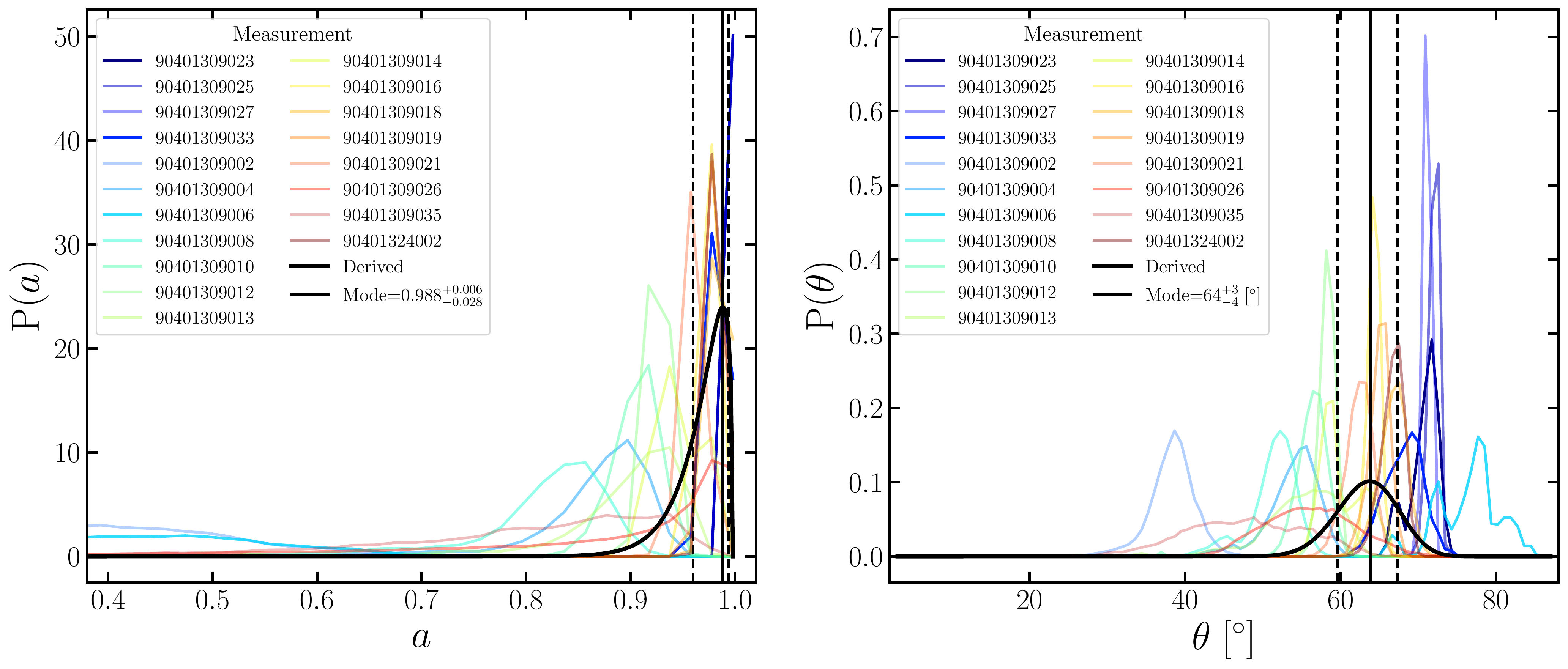}
    \caption{The left panel shows the posterior distributions resulting from the MCMC of MAXI J1820+070 analysis for spin, while the right panel shows the posterior distributions for the inclination of the inner accretion disk in the model, with the transparency of the lines being proportional to the ratio of reflected flux to total flux in the 3-79~keV band, which was used as weighting when combining the posterior distributions. The different colored lines represent the different observations analyzed, and the black curves represent the combined inferred distribution. The solid vertical black lines represent the modes of the combined distribution, and the dashed vertical black lines represent the $1\sigma$ credible intervals of the measurements. }
    \label{fig:MAXI_J1820_combined}
\end{figure}

\begin{figure}[ht]
    \centering
    \includegraphics[width= 0.85\textwidth]{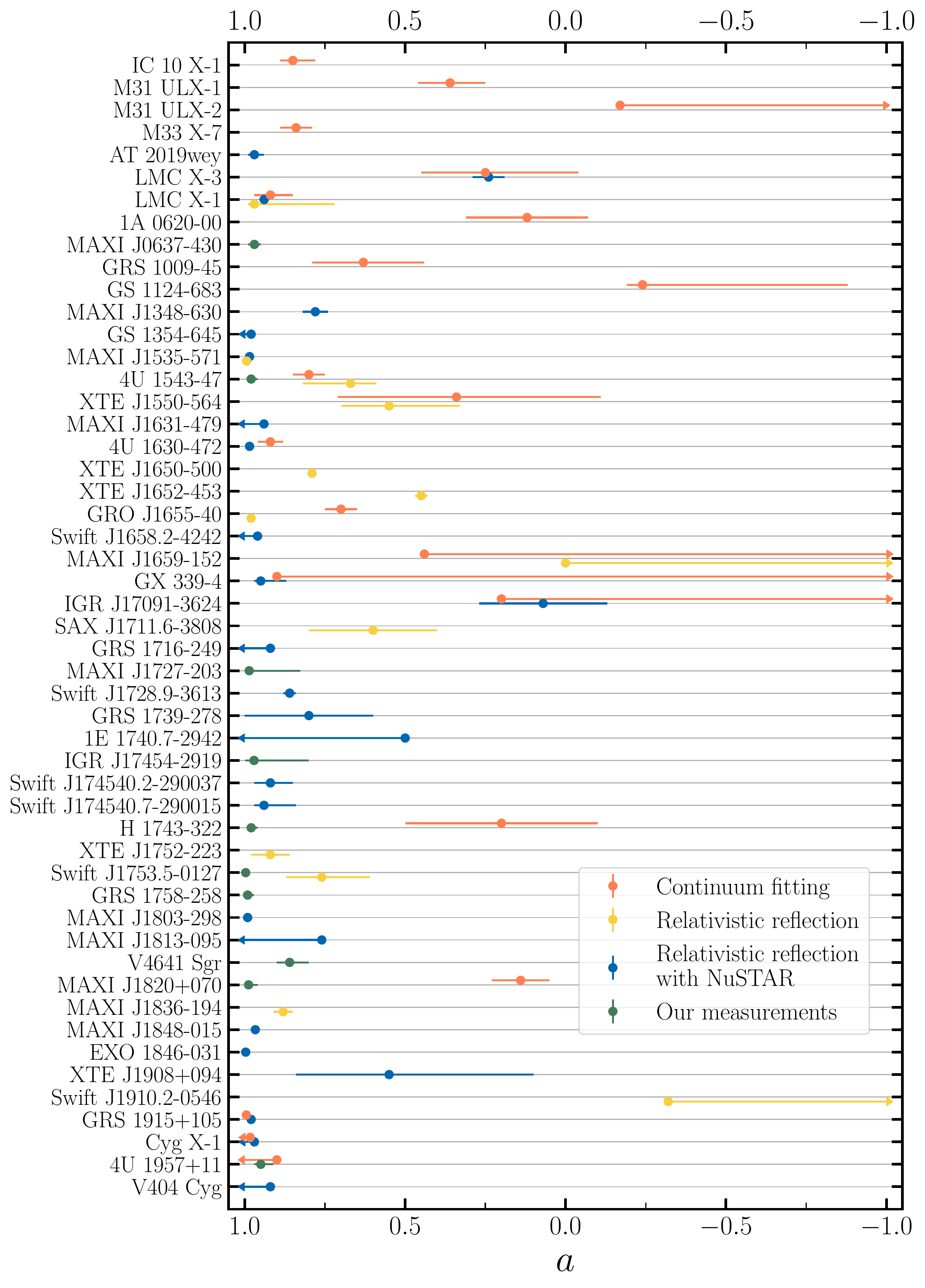}
    \caption{All current measurements of BH spins in X-ray binaries, obtained through continuum fitting (orange) or relativistic reflection. Spin measurements obtained through relativistic reflection using measurements from NuSTAR are indicated in blue, while reflection measurements using data from other instruments are shown in yellow. The new measurements of this paper are shown in green. The sources are ordered in increasing right ascension, and the values and references for all measurements are presented in Table \ref{tab:all_spins}. The arrows indicate upper or lower limits.}
    \label{fig:all_spins}
\end{figure}

\begin{figure}[ht]
    \centering
    \includegraphics[width= 0.85\textwidth]{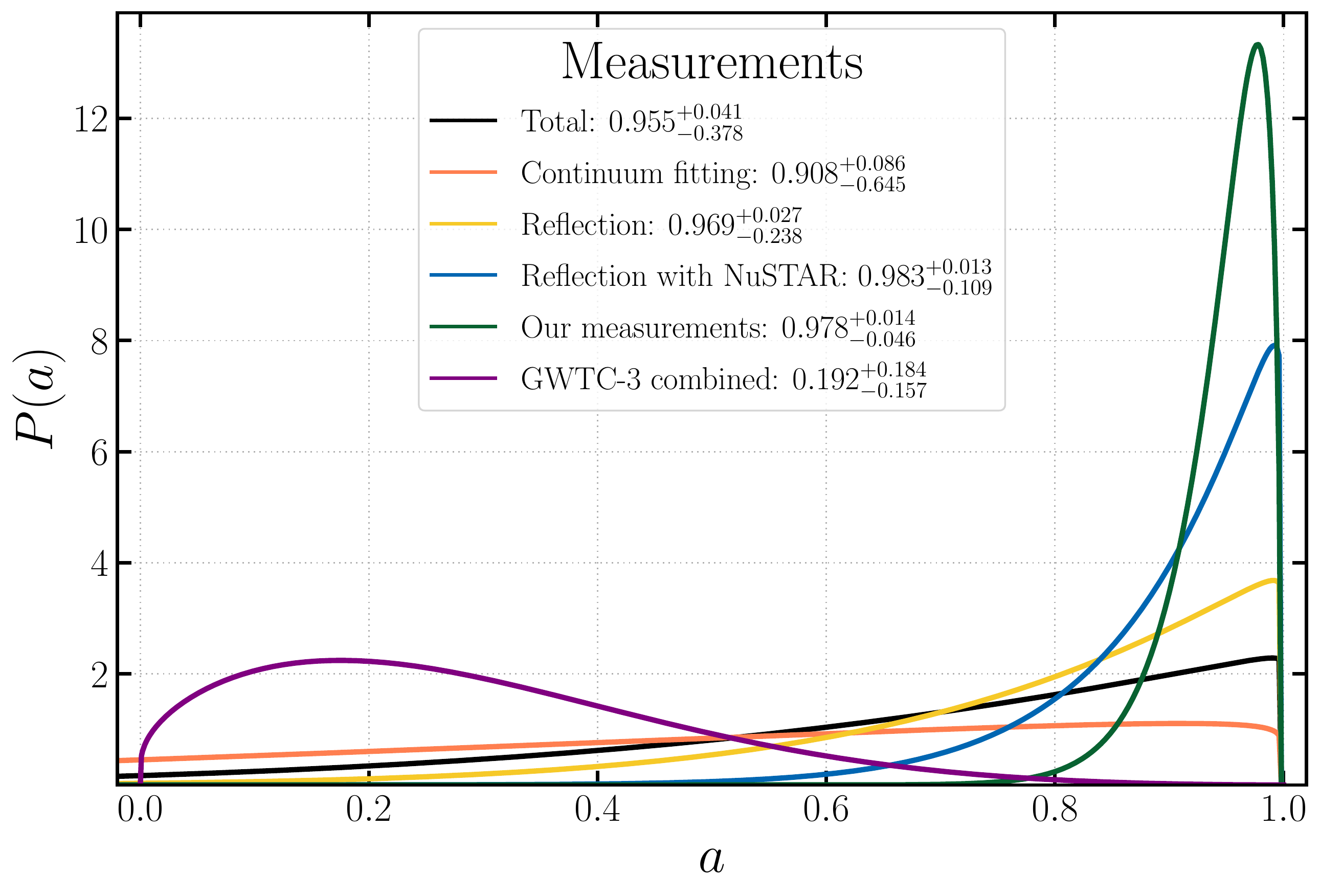}
    \caption{The spin distribution in X-ray binaries inferred based on all the measurements presented in Figure \ref{fig:all_spins} and in Table \ref{tab:all_spins}. The black curve shows the combined distribution of all spins obtained through continuum fitting and relativistic reflection. The orange curve shows the distribution of continuum fitting measurements, and the yellow curve shows the distribution of all measurements obtained through the relativistic reflection method. The blue curve shows the subset of all reflection measurements (yellow curve) that were obtained using NuSTAR data. The green curve represents the distribution inferred using only the ten measurements in this paper and is a subset of the blue curve. For comparison, we indicate the spin distribution measured in BBH mergers, presented in GWTC-3 in purple. The values quoted in the figure legend represent the modes and $1\sigma$ credible intervals of the distributions.}
    \label{fig:inferred_distributions}
\end{figure}

\begin{figure}[ht]
    \centering
    \includegraphics[width= 0.75\textwidth]{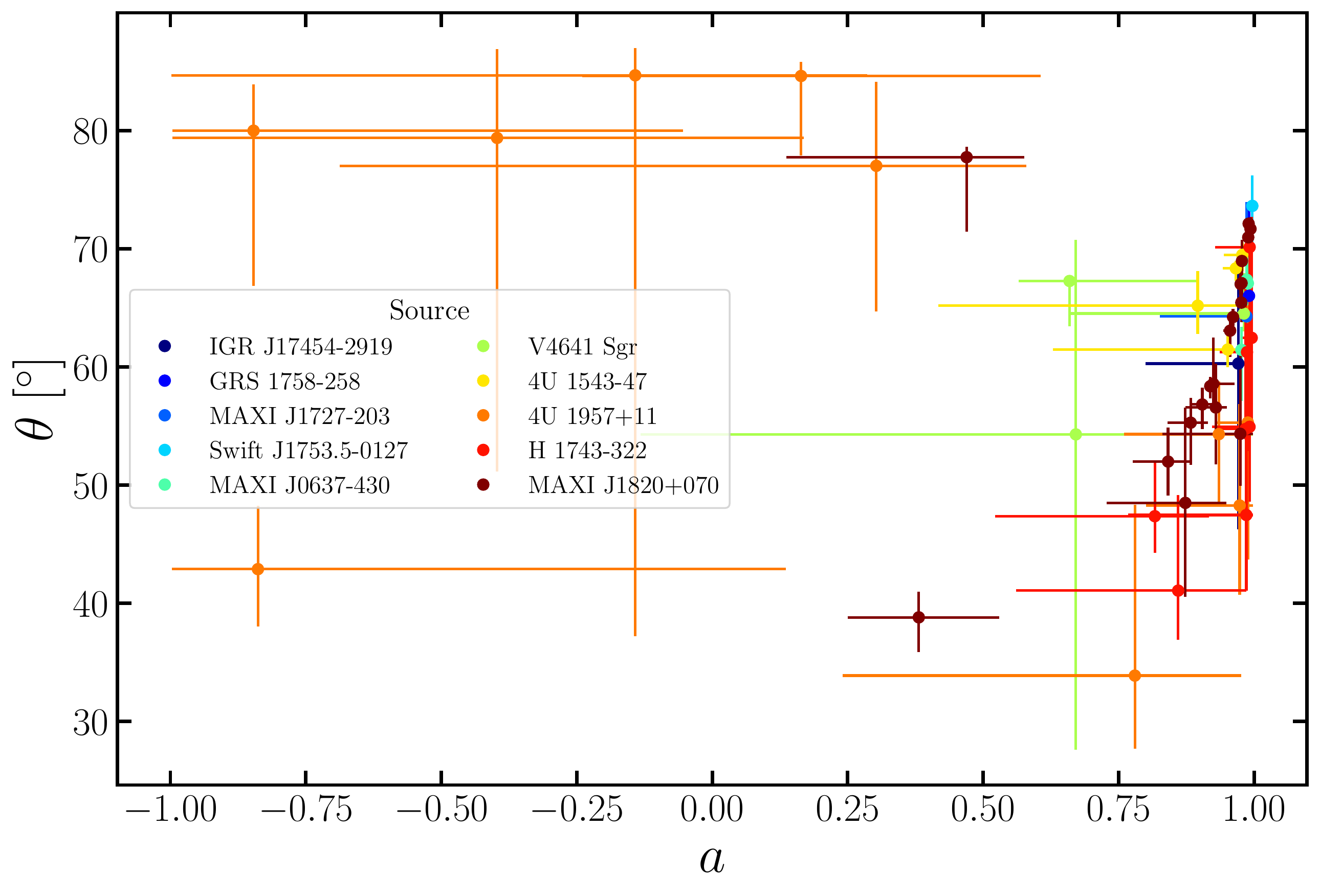}
    \caption{Inclination of the inner accretion disk $\theta$ vs. spin $a$ measurements resulting from all observations analyzed in this paper. The different colors represent measurements of the spin and inclination of different sources.}
    \label{fig:All_spin_vs_incl} 
\end{figure}

\begin{figure}[ht]
    \centering
    \includegraphics[width= 0.75\textwidth]{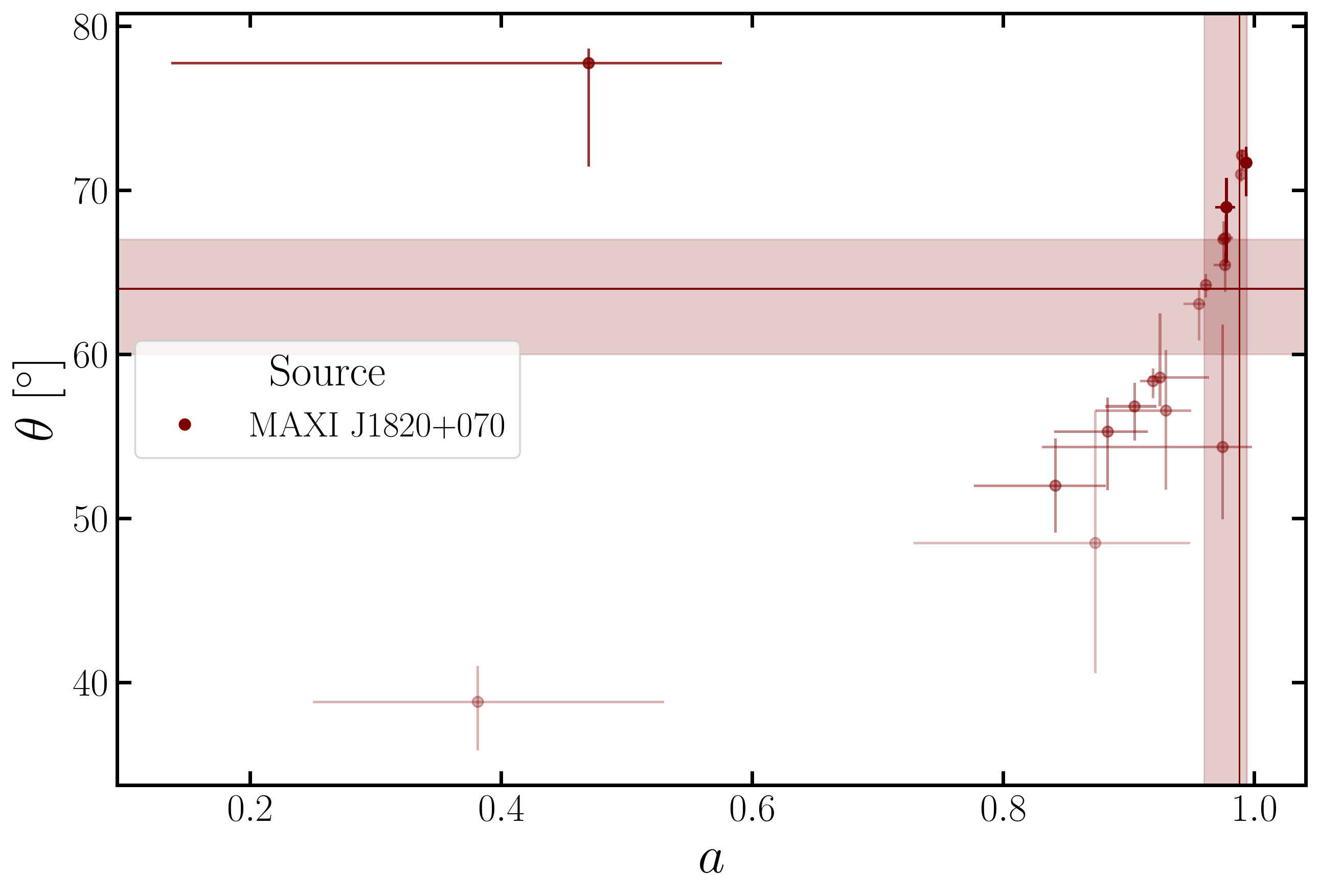}
    \caption{Inclination of the inner accretion disk $\theta$ vs. spin $a$ measurements resulting from all observations of MAXI J1820+070. The transparency of the points is proportional to the ratio of reflected flux to total flux in the 3-79~keV band, which was used as weighting when combining the posterior distributions. The solid horizontal and vertical lines represent the modes of the combined distributions for the two parameters, with the shaded regions indicating the $1\sigma$ credible intervals of the combined measurements.}
    \label{fig:MAXI_J1820_spin_vs_incl}
\end{figure}

\begin{figure}[ht]
    \centering
    \includegraphics[width= 0.75\textwidth]{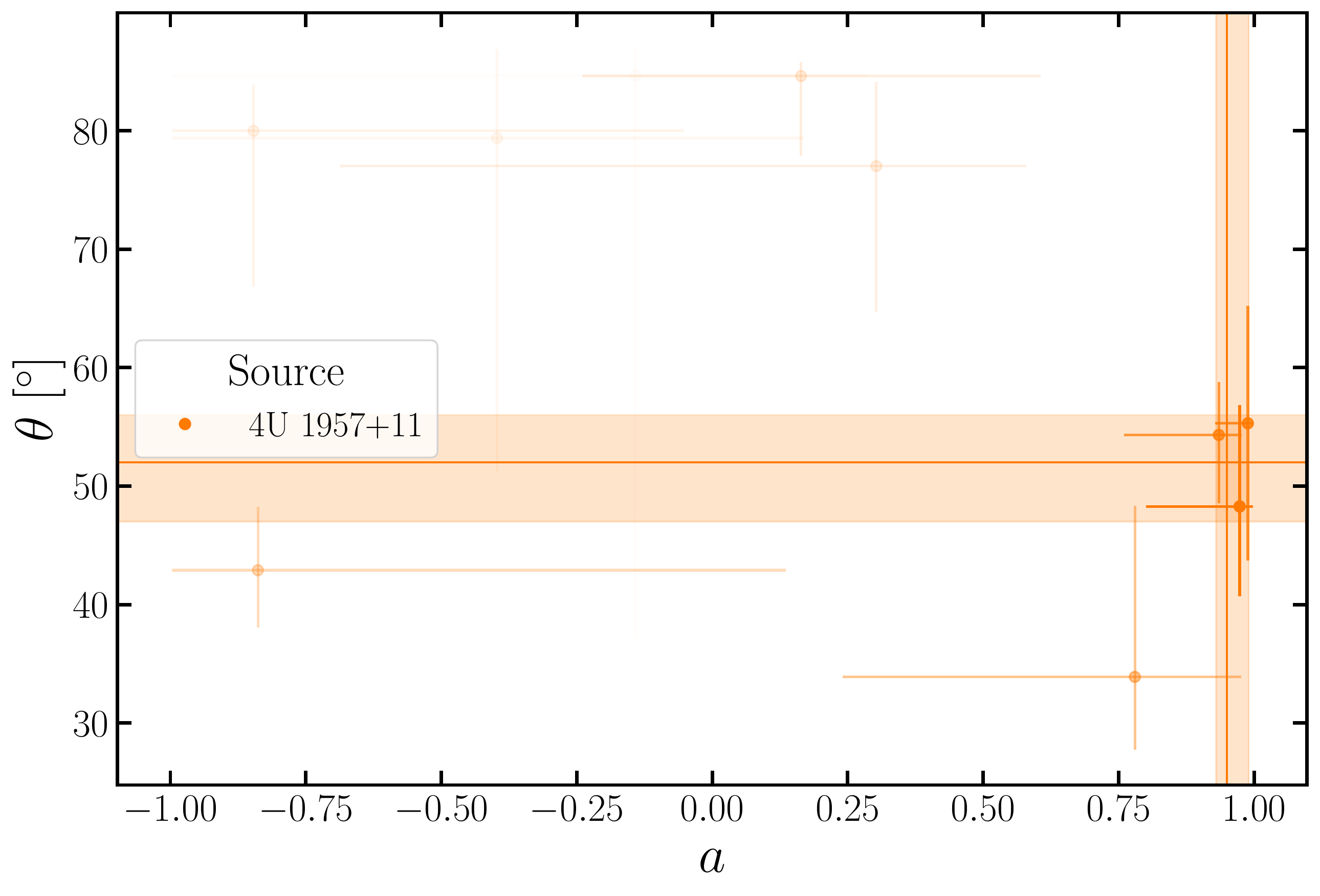}
    \caption{Inclination of the inner accretion disk $\theta$ vs. spin $a$ measurements resulting from all observations of 4U 1957+11. The transparency of the points is proportional to the ratio of reflected flux to total flux in the 3-79~keV band, which was used as weighting when combining the posterior distributions. The solid horizontal and vertical lines represent the modes of the combined distributions for the two parameters, with the shaded regions indicating the $1\sigma$ credible intervals of the combined measurements.}
    \label{fig:4U_1957_spin_vs_incl}
\end{figure}

\begin{figure}[ht]
    \centering
    \includegraphics[width= 0.75\textwidth]{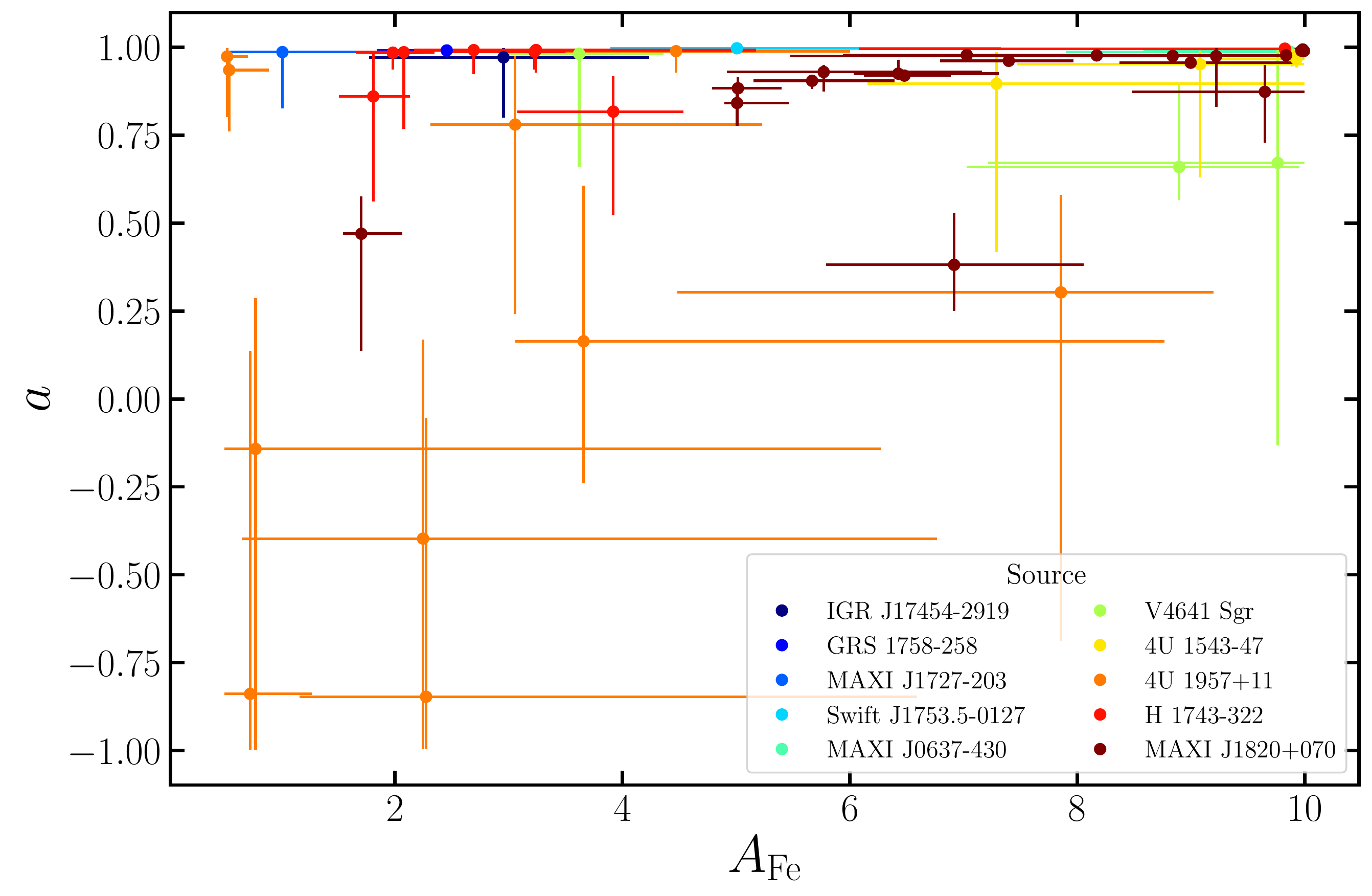}
    \caption{Spin $a$ vs iron abundance $A_{\rm Fe}$ measurements resulting from all observations analyzed in this paper. The different colors represent measurements of the spin and Fe abundance for different sources.}
    \label{fig:AFe_vs_a}
\end{figure}

\begin{figure}[ht]
    \centering
    \includegraphics[width= 0.75\textwidth]{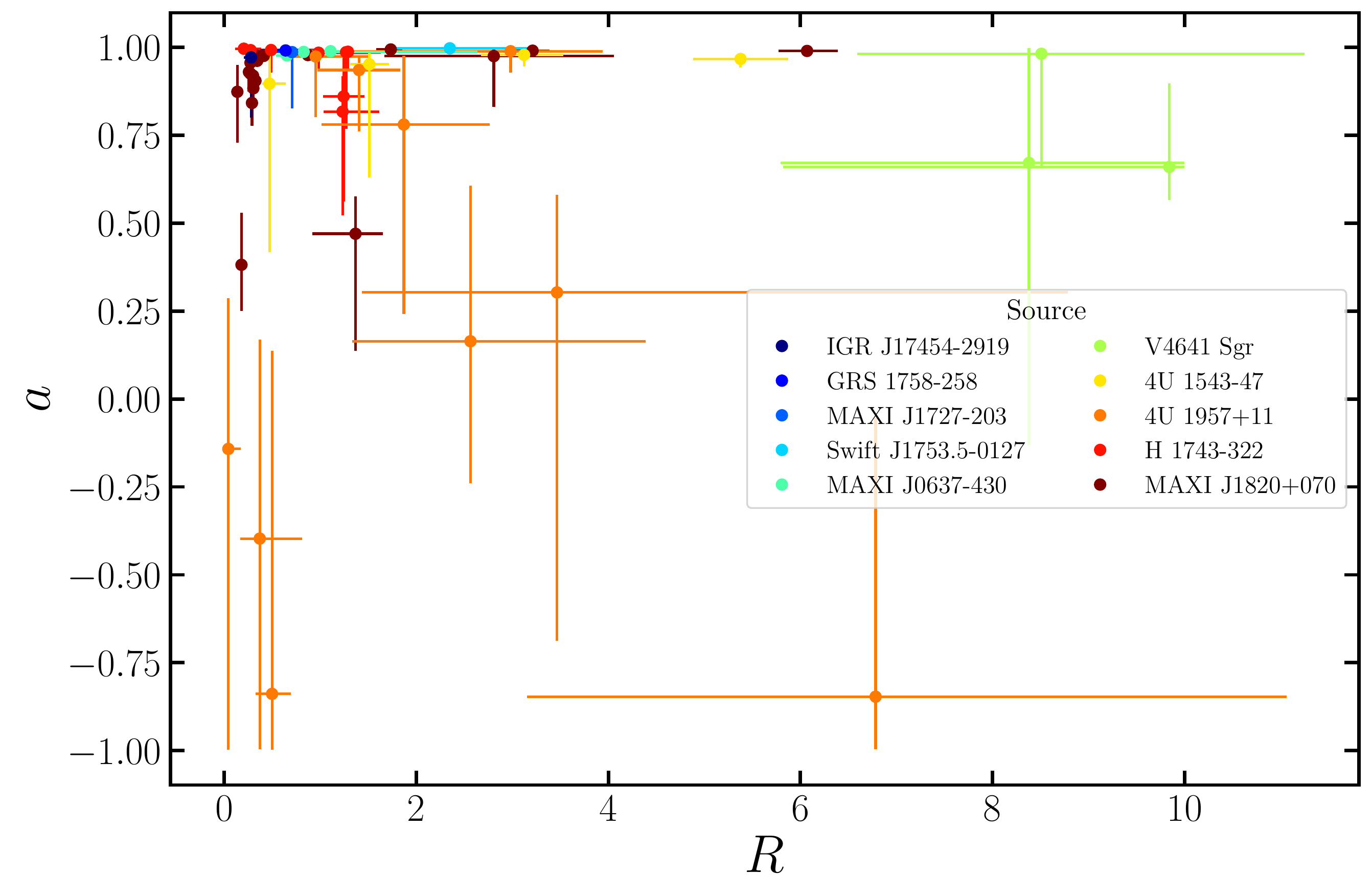}
    \caption{Spin $a$ vs reflection fraction $R$ measurements resulting from all observations analyzed in this paper. The different colors represent measurements of the spin and reflection fraction for different sources.}
    \label{fig:R_vs_a}
\end{figure}

\begin{figure}[ht]
    \centering
    \includegraphics[width= 0.75\textwidth]{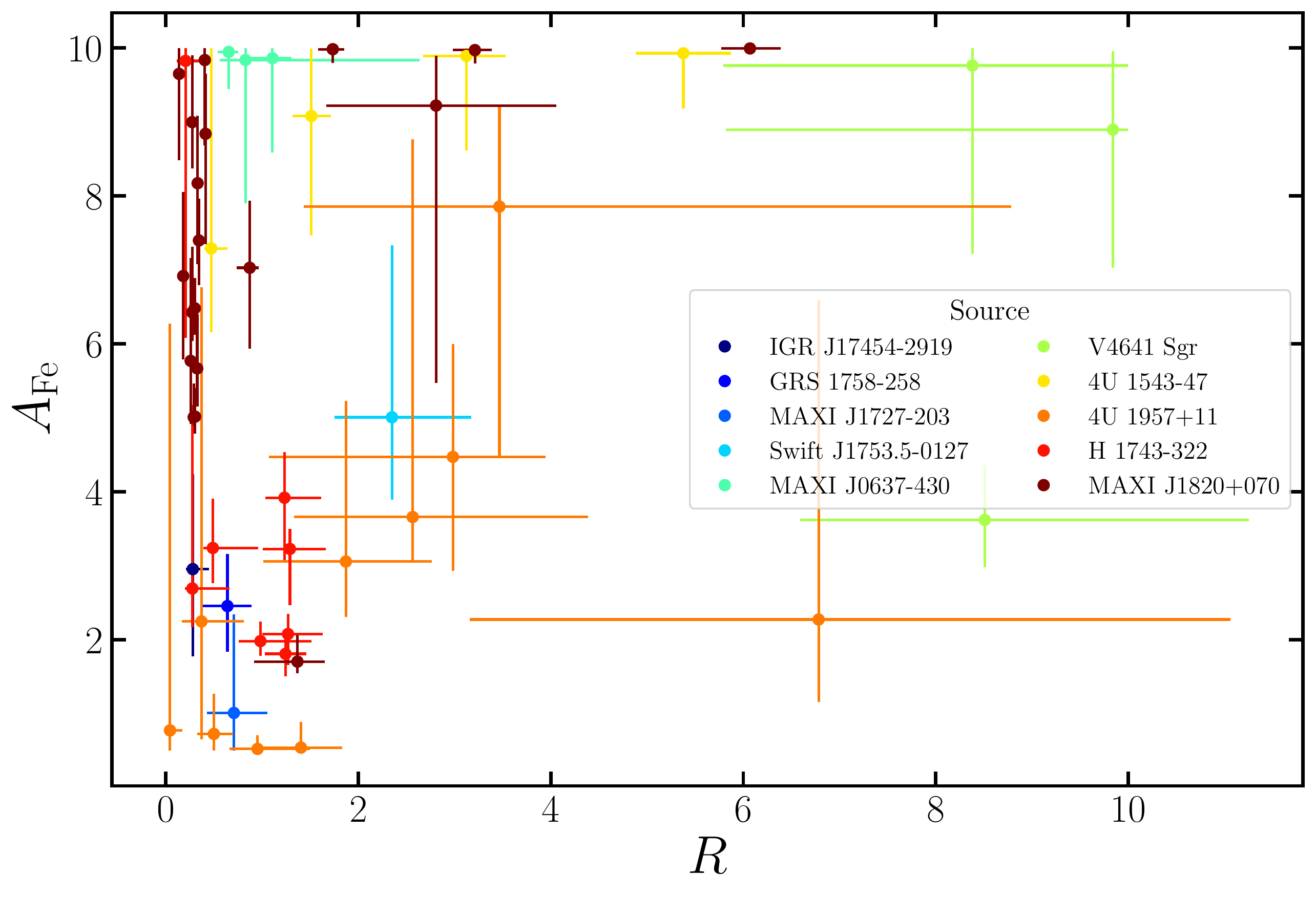}
    \caption{Iron abundance $A_{\rm Fe}$ vs reflection fraction $R$ measurements resulting from all observations analyzed in this paper. The different colors represent measurements of the reflection fraction and Fe abundance for different sources.}
    \label{fig:AFe_vs_R}
\end{figure}

\begin{figure}[ht]
    \centering
    \includegraphics[width= 0.65\textwidth]{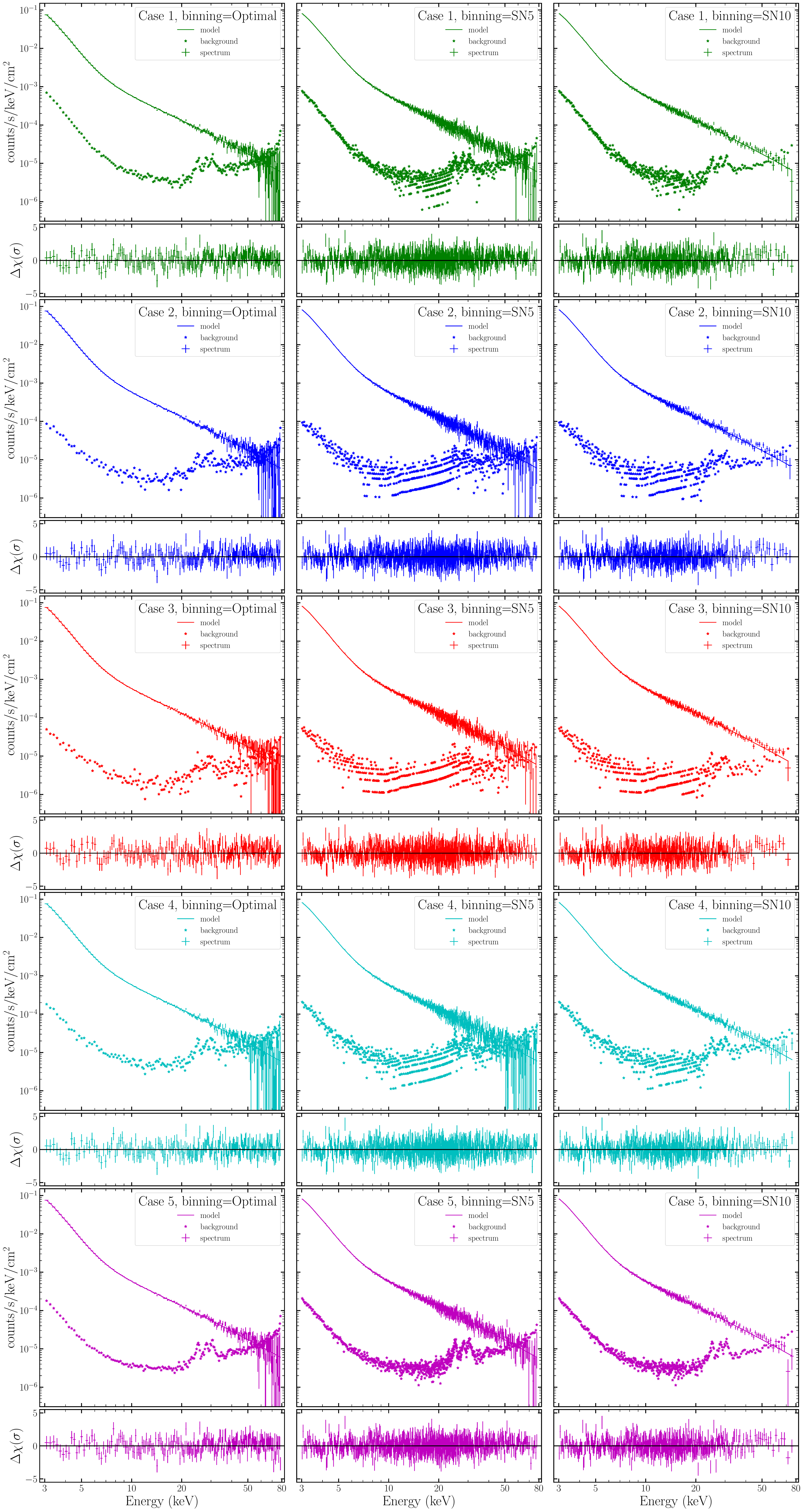}
    \caption{Source and background FPMA spectra from observation 80502324004 of MAXI J0637-430, together with the residuals produced when fitting with with the best performing reflection model, obtained using five different cases of source and background spectra extraction region (indicated by different colors, in different rows), and three different binning techniques (shown in the three different columns). See the text in Appendix \ref{sec:regions} for additional explanations.}
    \label{fig:region_comp}
\end{figure}

\begin{figure}[ht]
    \centering
    \includegraphics[width= 0.75\textwidth]{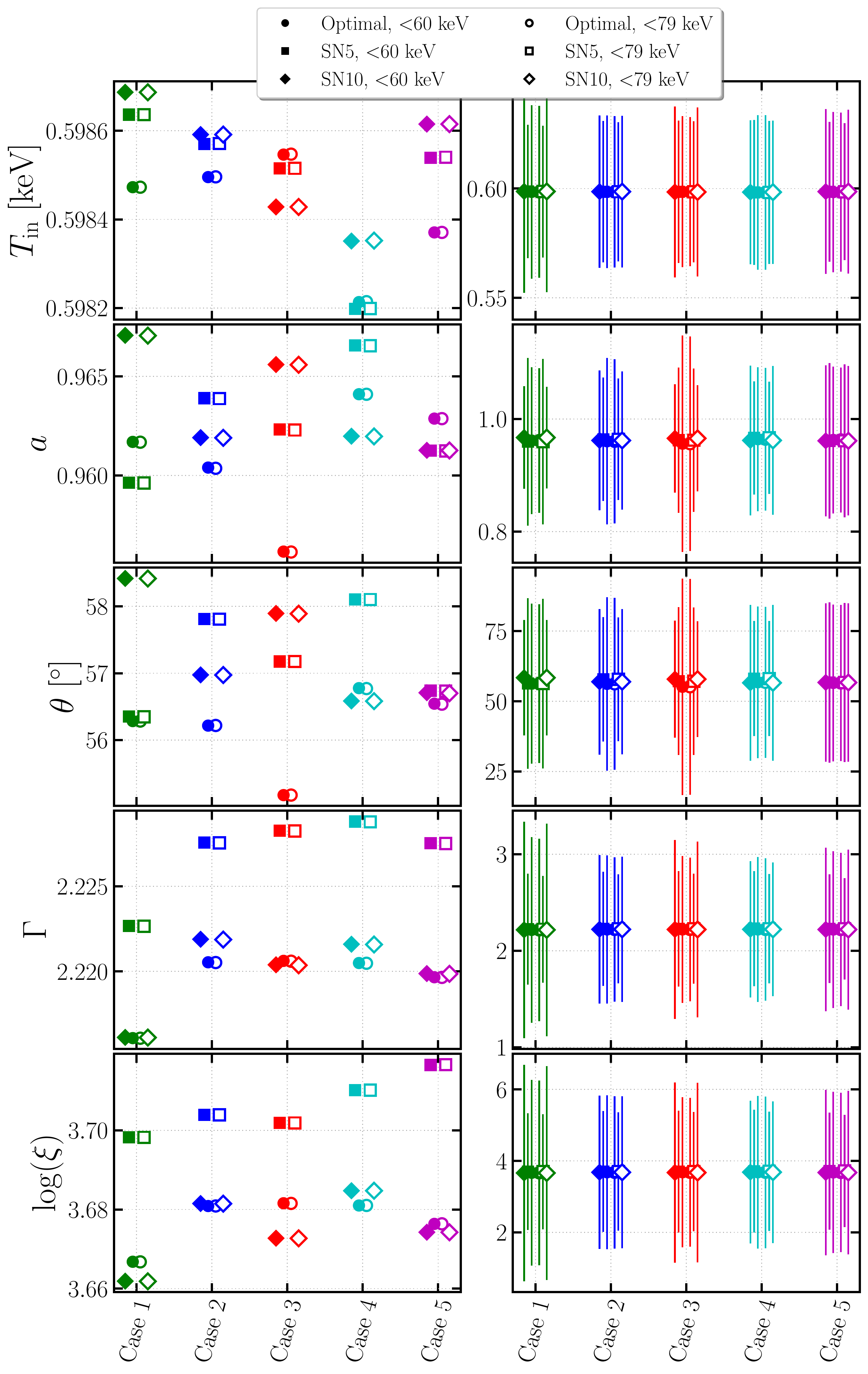}
    \caption{Comparison of parameter values that produce the best fit to the spectra obtained in the five source and background region extraction cases highlighted in Figure \ref{fig:region_comp} and discussed in Appendix \ref{sec:regions}, for different spectral binning schemes and high energy truncation of the fitted spectrum. The left panels show only the best-fit values, while the right panels also show the uncertainty on the measurements obtained directly from the covariance matrix of the fit.  See the text in Appendix \ref{sec:regions} for additional explanations.}
    \label{fig:region_par_comp}
\end{figure}

\begin{figure}[ht]
    \centering
    \includegraphics[width= 0.85\textwidth]{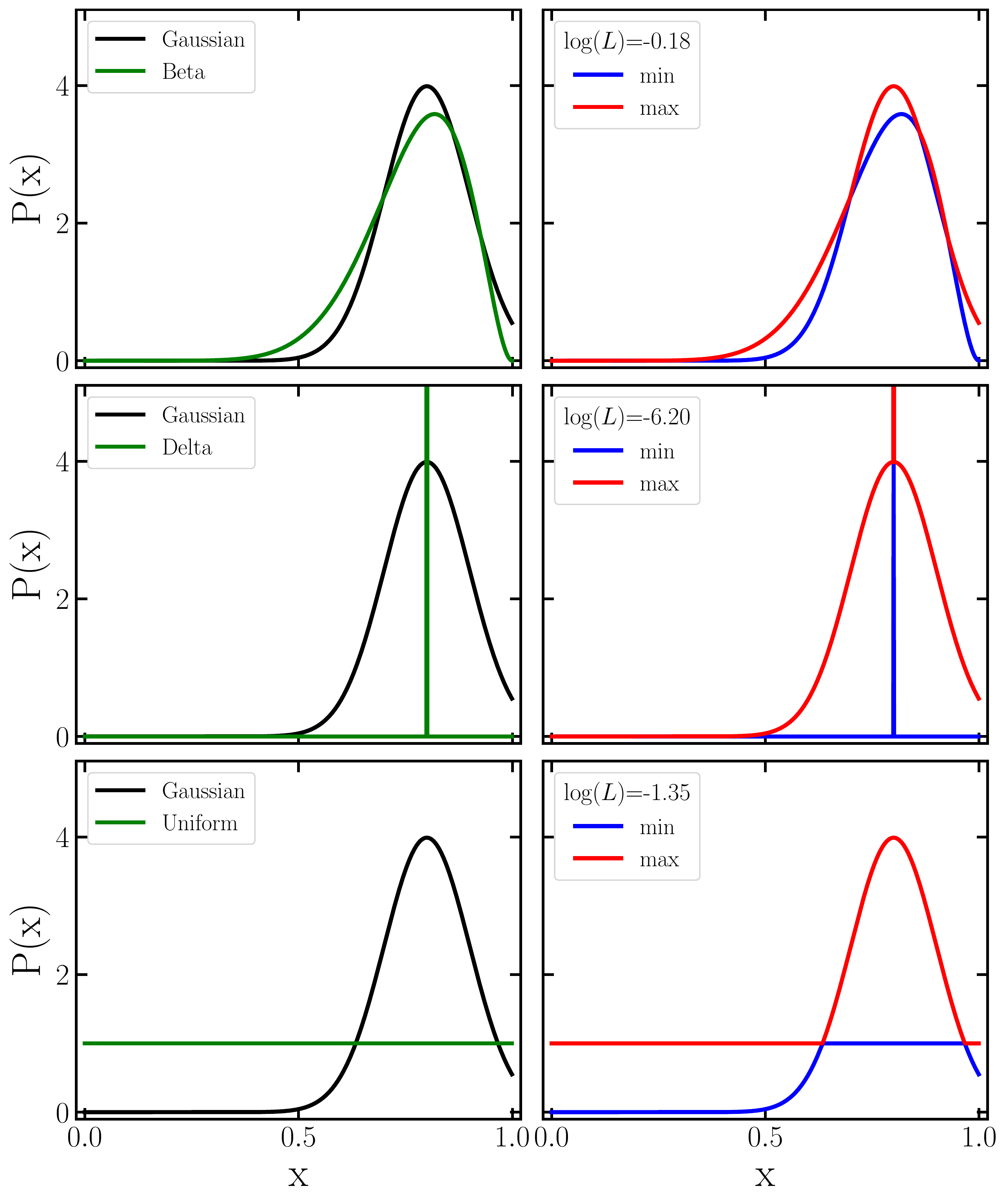}
    \caption{Illustration of the behavior of the likelihood function used when combining measurements. The left panels show a Gaussian distribution with a mean of 0.8 and standard deviation of 0.1 (in black), for which we compute the likelihood of overlap using three distribution, shown in green: Beta (top), Delta (middle) and uniform (bottom). The right panels show the minimum overlap of the two distributions presented in the right panels in blue and the maximum of the two in red. The legends in the right panels contain the logarithm of the computed likelihood, which our analysis aims to maximize.}
    \label{fig:combine_likelihood}
\end{figure}

\begin{figure}[ht]
    \centering
    \includegraphics[width= 0.85\textwidth]{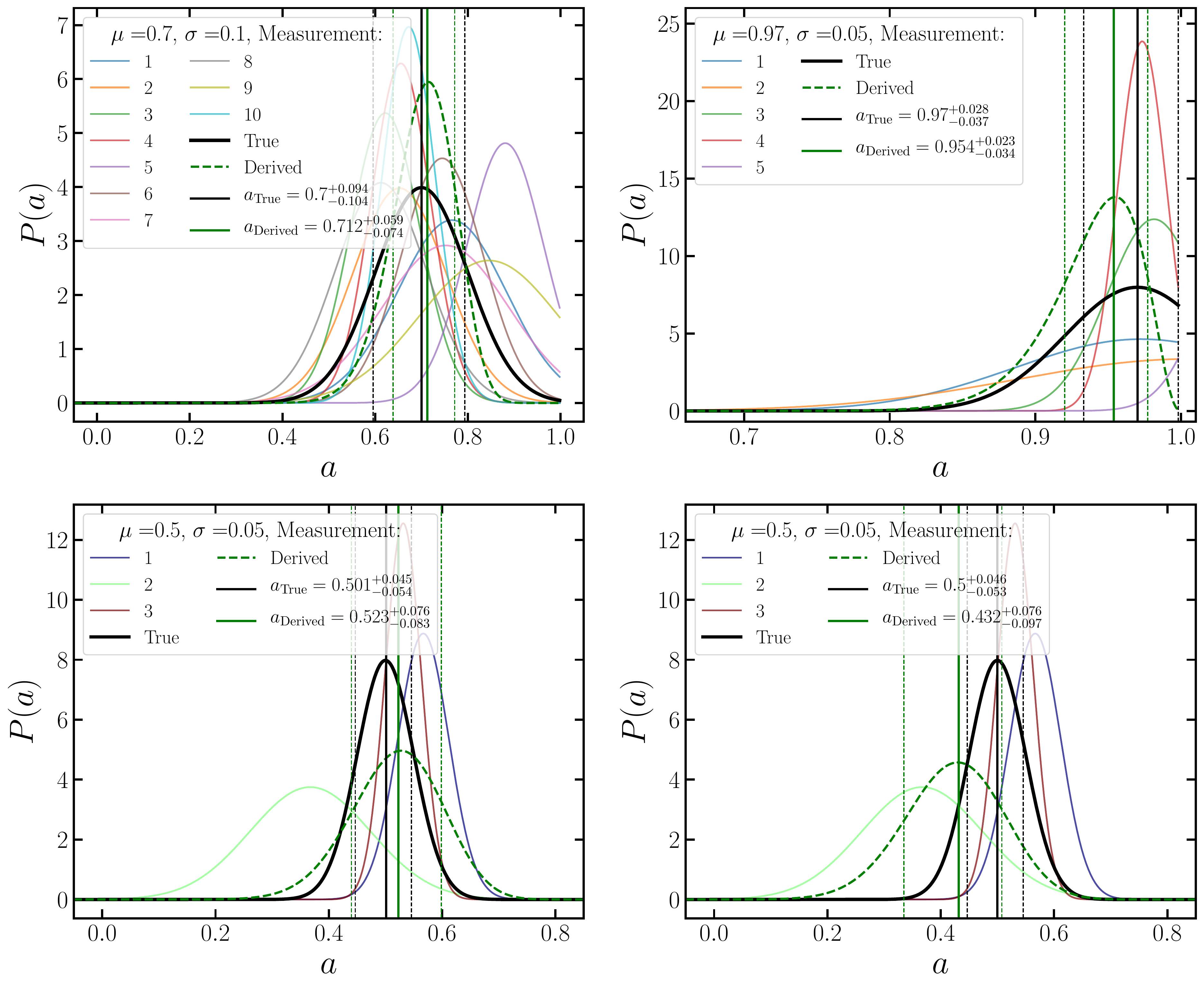}
    \caption{Top left: combining 10 Gaussian ``measurements" generated from a Gaussian distribution of median $\mu=0.7$ and $\sigma=0.1$ (shown through the solid black curve). Our combined measurement is shown through the dashed green curve, the solid vertical green and black lines represent the modes of the true and inferred distribution respectively, with the vertical dashed lines representing the $1\sigma$ credible intervals. Top right: combining 5 Gaussian ``measurements" generated from a Gaussian distribution of median $\mu=0.95$ and $\sigma=0.05$. The lines description is consistent with the previous panel. Note the different, reduced parameter range of the x axis. Bottom left: combining 3 Gaussian ``measurements" generated from a Gaussian distribution of median $\mu=0.5$ and $\sigma=0.05$. Bottom right: same as bottom left panel, but we weight the second ``measurement" eight times as much as the other two, influencing the inferred distribution toward that value.}
    \label{fig:tests_combining}
\end{figure}

\begin{rotatetable*}
\begin{deluxetable*}{l|c|c|c|c|ccc}
\centerwidetable
\tablecaption{Results of the MCMC analysis for IGR J17454-2919, GRS 1758-258, MAXI J1727-203, Swift J1753.5-0127, and MAXI J0637-430}
\label{tab:table1}
\tablewidth{\textwidth} 
\tabletypesize{\scriptsize}
\tablehead{
\colhead{} &
\colhead{IGR J17454-2919} &
\colhead{GRS 1758-258} &
\colhead{MAXI J1727-203} &
\colhead{Swift J1753.5-0127} &
\multicolumn{3}{c}{MAXI J0637-430}
}
\startdata\\
			ObsID & 80001046002 & 30401030002 & 90401329002 & 80001047002 & 80502324002 & 80502324004 & 80502324006 \\
			\hline
			$N_H\;[\times10^{22}\;cm^{-2}]$ & $8.7_{-0.3}^{+0.8}$ & $3.7_{-0.8}^{+1.2}$ & $6_{-2}^{+1}$ & $0.2_{-0.2}^{+1.0}$ & $0.006_{-0.005}^{+0.069}$ & $0.019_{-0.009}^{+0.092}$ & $0.03_{-0.02}^{+0.17}$ \\
			\hline
			$kT_{\rm in}\;[\rm keV]$ & $0.15_{-0.11}^{+0.03}$ & $0.56_{-0.05}^{+0.03}$ & $0.56_{-0.03}^{+0.02}$ & $0.49_{-0.01}^{+0.01}$ & $0.639_{-0.003}^{+0.002}$ & $0.599_{-0.003}^{+0.003}$ & $0.544_{-0.004}^{+0.003}$ \\
			$\rm norm_{d,A} $ & $8e+06_{-8e+06}^{+8e+07}$ & $168_{-74}^{+80}$ & $239_{-80}^{+70}$ & $585_{-140}^{+172}$ & $2e+03_{-6e+01}^{+6e+01}$ & $2e+03_{-8e+01}^{+6e+01}$ & $2e+03_{-9e+01}^{+1e+02}$ \\
			\hline
			$q_1$ & $9.8_{-3.6}^{+0.2}$ & $8.2_{-3.0}^{+0.8}$ & $9.8_{-2.7}^{+0.2}$ & $9.5_{-1.3}^{+0.5}$ & $9.9_{-1.0}^{+0.1}$ & $9.9_{-1.3}^{+0.1}$ & $9.9_{-2.1}^{+0.1}$ \\
			$q_2$ & $0.9_{-0.9}^{+0.9}$ & $0.1_{-0.1}^{+1.8}$ & $1e-07_{-1e-07}^{+1e-07}$ & $4_{-1}^{+4}$ & $1e-10_{-1e-10}^{+2e-03}$ & $2.5_{-1.5}^{+0.5}$ & $2.5_{-1.3}^{+0.5}$ \\
			$R_{\rm br} \;[r_g]$ & $8_{-3}^{+30}$ & $23_{-18}^{+40}$ & $25_{-18}^{+6}$ & $2.9_{-0.9}^{+29.8}$ & $95_{-41}^{+5}$ & $84_{-44}^{+16}$ & $95_{-55}^{+5}$ \\
			$a$ & $0.97_{-0.17}^{+0.03}$ & $0.991_{-0.019}^{+0.007}$ & $0.99_{-0.16}^{+0.01}$ & $0.997_{-0.003}^{+0.001}$ & $0.989_{-0.005}^{+0.004}$ & $0.977_{-0.013}^{+0.008}$ & $0.987_{-0.013}^{+0.006}$ \\
			$\theta\;[^\circ]$ & $60_{-14}^{+8}$ & $66_{-5}^{+8}$ & $64_{-7}^{+10}$ & $74_{-3}^{+3}$ & $67_{-3}^{+2}$ & $61_{-4}^{+2}$ & $67_{-10}^{+2}$ \\
			$\Gamma$ & $1.73_{-0.02}^{+0.03}$ & $1.63_{-0.01}^{+0.01}$ & $1.68_{-0.03}^{+0.03}$ & $2.00_{-0.03}^{+0.04}$ & $2.26_{-0.03}^{+0.03}$ & $2.22_{-0.01}^{+0.02}$ & $2.44_{-0.03}^{+0.26}$ \\
			$\log(\xi)$ & $2.7_{-0.4}^{+0.3}$ & $3.7_{-0.1}^{+0.2}$ & $4.0_{-0.2}^{+0.1}$ & $1.7_{-0.9}^{+0.6}$ & $3.7_{-0.3}^{+0.1}$ & $3.67_{-0.22}^{+0.05}$ & $3.9_{-2.5}^{+0.5}$ \\
			$A_{\rm Fe}$ & $3_{-1}^{+1}$ & $2.5_{-0.6}^{+0.7}$ & $1.0_{-0.5}^{+1.3}$ & $5_{-1}^{+2}$ & $9.9_{-1.3}^{+0.1}$ & $9.95_{-0.50}^{+0.05}$ & $9.8_{-1.9}^{+0.2}$ \\
			$\rm kT_e\;[\rm keV] $ & $29_{-28}^{+317}$ & $159_{-78}^{+126}$ & $409_{-194}^{+249}$ & $218_{-120}^{+119}$ & \nodata & $1e+03_{-2e+02}^{+2e+01}$ & $1e+03_{-4e+02}^{+2e+01}$ \\
			R & $0.28_{-0.07}^{+0.17}$ & $0.6_{-0.3}^{+0.3}$ & $0.7_{-0.3}^{+0.3}$ & $2.3_{-0.6}^{+0.8}$ & $1.1_{-0.3}^{+0.2}$ & $0.7_{-0.1}^{+0.1}$ & $0.8_{-0.3}^{+1.8}$ \\
			$\rm norm_{r,A} $ & $1e-03_{-7e-05}^{+4e-05}$ & $4e-03_{-2e-04}^{+3e-04}$ & $2e-03_{-3e-04}^{+2e-04}$ & $4e-04_{-1e-05}^{+1e-05}$ & $7e-04_{-8e-05}^{+9e-05}$ & $1e-03_{-5e-05}^{+6e-05}$ & $5e-04_{-7e-05}^{+7e-04}$ \\
			\hline
			$\rm norm_{d,B} $ & $2e+07_{-2e+07}^{+2e+08}$ & $202_{-89}^{+71}$ & $223_{-64}^{+79}$ & $550_{-137}^{+156}$ & $2e+03_{-5e+01}^{+6e+01}$ & $2e+03_{-7e+01}^{+6e+01}$ & $2e+03_{-8e+01}^{+1e+02}$ \\
			$\rm norm_{r,B} $ & $9e-04_{-5e-05}^{+6e-05}$ & $4e-03_{-2e-04}^{+3e-04}$ & $2e-03_{-3e-04}^{+2e-04}$ & $4e-04_{-1e-05}^{+1e-05}$ & $7e-04_{-8e-05}^{+8e-05}$ & $1e-03_{-4e-05}^{+6e-05}$ & $5e-04_{-7e-05}^{+7e-04}$ \\
			\hline
			$\chi^2/\nu$ & $457_{-5}^{+8}(445.48)/421$ & $452_{-3}^{+6}(442.56)/474$ & $453_{-5}^{+6}(440.59)/442$ & $398_{-4}^{+5}(385.18)/385$ & $401_{-4}^{+6}(387.88)/296$ & $392_{-4}^{+7}(382.78)/364$ & $286_{-5}^{+8}(271.90)/237$ \\
\enddata
\tablecomments{In this table, we report the mode of the posterior distributions in the MCMC analysis, along with the $1\sigma$ credible interval. For $\chi^2$, the number in parentheses indicates the best-fit $\chi^2$ value. For the normalization of the components, the subscripts d (from \texttt{diskbb}) and r (from \texttt{relxill}) represent the model that the values correspond to, while the subscripts A and B represent which NuSTAR FMP detector the spectrum was originating from.}
\end{deluxetable*}
\end{rotatetable*}
\newpage

\begin{rotatetable*}
\begin{deluxetable*}{l|ccc|cccc}
\centerwidetable
\tablecaption{Results of the MCMC analysis for V4641 Sgr and 4U 1543-47}
\label{tab:table2}
\tablewidth{\textwidth} 
\tabletypesize{\scriptsize}
\tablehead{
\colhead{} &
\multicolumn{3}{c}{V4641 Sgr} &
\multicolumn{4}{c}{4U 1543-47}
}
\startdata\\
			ObsID & 80002012002 & 90102011002 & 90601302002 & 90702326006 & 90702326008 & 90702326010 & 90702326012 \\
			\hline
			$N_H\;[\times10^{22}\;cm^{-2}]$ & $3.3_{-2.1}^{+0.2}$ & $4.9_{-1.1}^{+0.9}$ & $3.8_{-0.9}^{+0.9}$ & $1.8_{-0.3}^{+0.1}$ & $0.07_{-0.07}^{+0.18}$ & $1.3_{-0.3}^{+0.2}$ & $0_{-0}^{+0}$ \\
			\hline
			$\rm kT\;[\rm keV] $ & $1.98_{-0.05}^{+0.10}$ & $2.0_{-0.4}^{+0.1}$ & $1.8_{-0.2}^{+0.1}$ & \nodata & \nodata & \nodata & \nodata \\
			$A$ & $0.30_{-0.06}^{+0.09}$ & $0.04_{-0.03}^{+0.02}$ & $0.6_{-0.2}^{+0.2}$ & \nodata & \nodata & \nodata & \nodata \\
			$z$ & $-4e-02_{-9e-04}^{+2e-03}$ & $-0.03_{-0.01}^{+0.01}$ & $-0.033_{-0.002}^{+0.002}$ & \nodata & \nodata & \nodata & \nodata \\
			$\rm norm_{a,A} $ & $0.74_{-0.24}^{+0.08}$ & $0.5_{-0.1}^{+0.1}$ & $0.9_{-0.2}^{+0.4}$ & \nodata & \nodata & \nodata & \nodata \\
			\hline
			$N_H\;[\times10^{22}\;cm^{-2}]$ & $16_{-3}^{+7}$ & $52_{-17}^{+20}$ & $23_{-5}^{+6}$ & $52_{-2}^{+12}$ & $46_{-5}^{+4}$ & $65_{-8}^{+9}$ & $37_{-2}^{+2}$ \\
			$\log(\xi)$ & $0.3_{-0.1}^{+0.6}$ & $2.6_{-0.3}^{+0.2}$ & $0.3_{-0.3}^{+0.3}$ & $2.09_{-0.03}^{+0.12}$ & $1.92_{-0.30}^{+0.06}$ & $2.3_{-0.1}^{+0.1}$ & $1.89_{-0.19}^{+0.08}$ \\
			$f$ & $0.58_{-0.06}^{+0.12}$ & $0.94_{-0.17}^{+0.06}$ & $0.62_{-0.07}^{+0.05}$ & $0.511_{-0.055}^{+0.007}$ & $0.33_{-0.02}^{+0.02}$ & $0.39_{-0.02}^{+0.02}$ & $0.47_{-0.02}^{+0.01}$ \\
			\hline
			$kT_{\rm in}\;[\rm keV]$ & $1.16_{-0.02}^{+0.04}$ & $1.39_{-0.13}^{+0.03}$ & $1.19_{-0.04}^{+0.02}$ & $0.807_{-0.004}^{+0.008}$ & $0.864_{-0.004}^{+0.004}$ & $0.817_{-0.006}^{+0.006}$ & $0.813_{-0.004}^{+0.004}$ \\
			$\rm norm_{d,A} $ & $22_{-3}^{+4}$ & $2.8_{-0.8}^{+1.0}$ & $55_{-10}^{+17}$ & $2e+04_{-2e+03}^{+2e+03}$ & $1e+04_{-5e+02}^{+7e+02}$ & $2e+04_{-2e+03}^{+2e+03}$ & $2e+04_{-1e+03}^{+7e+02}$ \\
			\hline
			$q_1$ & $3.6_{-0.5}^{+0.7}$ & $4.2_{-0.6}^{+1.1}$ & $3.2_{-0.2}^{+2.2}$ & $7.0_{-0.4}^{+0.9}$ & $9.8_{-2.1}^{+0.2}$ & $8.0_{-1.0}^{+0.4}$ & $8_{-1}^{+2}$ \\
			$q_2$ & $0.09_{-0.09}^{+0.37}$ & $3e-08_{-3e-08}^{+2e-07}$ & $2e-08_{-2e-08}^{+1e-07}$ & $2e-07_{-2e-07}^{+1e-06}$ & $4e-07_{-4e-07}^{+2e-06}$ & $1e-07_{-1e-07}^{+8e-07}$ & $2e-07_{-2e-07}^{+1e-06}$ \\
			$R_{\rm br} \;[r_g]$ & $50_{-11}^{+16}$ & $26_{-6}^{+21}$ & $13_{-3}^{+9}$ & $14_{-2}^{+3}$ & $29.3_{-7.5}^{+0.7}$ & $10.3_{-0.3}^{+3.3}$ & $29.5_{-6.2}^{+0.5}$ \\
			$a$ & $0.66_{-0.09}^{+0.24}$ & $0.98_{-0.32}^{+0.02}$ & $0.7_{-0.8}^{+0.3}$ & $0.97_{-0.02}^{+0.01}$ & $0.90_{-0.48}^{+0.09}$ & $0.978_{-0.034}^{+0.007}$ & $0.95_{-0.32}^{+0.04}$ \\
			$\theta\;[^\circ]$ & $67.3_{-3.8}^{+0.3}$ & $65_{-2}^{+3}$ & $54_{-27}^{+16}$ & $68.3_{-1.0}^{+0.6}$ & $65_{-2}^{+3}$ & $69_{-1}^{+1}$ & $61_{-1}^{+2}$ \\
			$R_{\rm in} \;[r_{\rm ISCO}]$ & \nodata & \nodata & \nodata  & $1.01^{+0.13}_{-0.01}$ & $1.06^{+0.70}_{-0.06}$ & $1.01^{+0.17}_{-0.01}$  & $1.06^{+0.59}_{-0.06}$\\
			$\Gamma$ & $3.27_{-1.88}^{+0.04}$ & $2.13_{-0.09}^{+0.17}$ & $2.9_{-0.3}^{+0.2}$ & $2.67_{-0.06}^{+0.03}$ & $2.42_{-0.05}^{+0.05}$ & $2.77_{-0.03}^{+0.03}$ & $2.40_{-0.05}^{+0.07}$ \\
			$\log(\xi)$ & $4.66_{-1.37}^{+0.04}$ & $3.4_{-0.3}^{+0.2}$ & $4.62_{-0.37}^{+0.08}$ & $2.3_{-0.1}^{+0.2}$ & $2.3_{-0.4}^{+0.4}$ & $2.3_{-0.2}^{+0.2}$ & $1e-06_{-1e-06}^{+7e-06}$ \\
			$A_{\rm Fe}$ & $9_{-2}^{+1}$ & $3.6_{-0.6}^{+0.7}$ & $9.8_{-2.5}^{+0.2}$ & $9.93_{-0.75}^{+0.07}$ & $7_{-1}^{+3}$ & $9.9_{-1.3}^{+0.1}$ & $9.1_{-1.6}^{+0.9}$ \\
			$\rm kT_e\;[\rm keV] $ & $5.8_{-0.8}^{+59.8}$ & $35_{-12}^{+8}$ & $5.2_{-0.2}^{+1.5}$ & $87_{-13}^{+16}$ & $114_{-21}^{+24}$ & $1e+03_{-3e+02}^{+5e+01}$ & $109_{-22}^{+35}$ \\
			R & $9.8_{-4.0}^{+0.2}$ & $9_{-2}^{+3}$ & $8_{-3}^{+2}$ & $5.4_{-0.5}^{+0.5}$ & $0.47_{-0.07}^{+0.17}$ & $3.1_{-0.4}^{+0.4}$ & $1.5_{-0.2}^{+0.2}$ \\
			$\rm norm_{r,A} $ & $6e-05_{-3e-05}^{+3e-03}$ & $5e-05_{-9e-06}^{+2e-05}$ & $1e-03_{-6e-04}^{+1e-03}$ & $0.013_{-0.001}^{+0.002}$ & $0.029_{-0.003}^{+0.005}$ & $0.032_{-0.003}^{+0.004}$ & $0.010_{-0.002}^{+0.002}$ \\
			\hline
			$\rm norm_{a,B} $ & $0.6_{-0.1}^{+0.2}$ & $0.51_{-0.13}^{+0.08}$ & $1.0_{-0.4}^{+0.3}$ & \nodata & \nodata & \nodata & \nodata \\
			$\rm norm_{d,B} $ & $25_{-4}^{+3}$ & $2.8_{-0.6}^{+1.3}$ & $59_{-9}^{+19}$ & $2e+04_{-2e+03}^{+2e+03}$ & $1e+04_{-5e+02}^{+7e+02}$ & $2e+04_{-2e+03}^{+2e+03}$ & $2e+04_{-1e+03}^{+7e+02}$ \\
			$\rm norm_{r,B} $ & $6e-05_{-3e-05}^{+3e-03}$ & $5e-05_{-1e-05}^{+1e-05}$ & $4e-05_{-4e-05}^{+4e-04}$ & $0.012_{-0.001}^{+0.002}$ & $0.028_{-0.003}^{+0.005}$ & $0.031_{-0.003}^{+0.004}$ & $0.010_{-0.001}^{+0.002}$ \\
			\hline
			$\chi^2/\nu$ & $264_{-14}^{+19}(236.53)/141$ & $180_{-6}^{+8}(163.83)/141$ & $181_{-5}^{+9}(168.15)/94$ & $503_{-7}^{+9}(505.76)/393$ & $523_{-6}^{+6}(508.28)/418$ & $502_{-9}^{+12}(520.25)/403$ & $497_{-5}^{+6}(482.70)/399$ \\
\enddata
\tablecomments{In this table, we report the mode of the posterior distributions in the MCMC analysis, along with the $1\sigma$ credible interval. For $\chi^2$, the number in parentheses indicates the best-fit $\chi^2$ value. For the normalization of the components, the subscripts d (from \texttt{diskbb}), r (from \texttt{relxill}), and a (from \texttt{apec}) represent the model that the values correspond to, while the subscripts A and B represent which NuSTAR FMP detector the spectrum was originating from.}
\end{deluxetable*}
\end{rotatetable*}

\newpage

\movetabledown=3in
\begin{splitdeluxetable}{l|cccccBl|ccccc}
\rotate
\tablecaption{Results of the MCMC analysis for 4U 1957+11}
\label{tab:table3}
\tablehead{
\colhead{ObsID} &\colhead{30001015002} &\colhead{30402011002} &\colhead{30402011004} &\colhead{30402011006} &\colhead{30502007002}&\colhead{ObsID} &\colhead{30502007004} &\colhead{30502007006} &\colhead{30502007008} &\colhead{30502007010} &\colhead{30502007012}}
\startdata\\
			Model & relxillD$\_$19 & relxillD$\_$19 & relxill & relxill & relxill & Model & relxill & relxillCp & relxillCp & relxillCp & relxillCp \\
			\hline
			$N_H\;[\times10^{22}\;cm^{-2}]$ & $0.96_{-0.63}^{+0.02}$ & $0.97_{-0.38}^{+0.03}$ & $0.5_{-0.2}^{+0.2}$ & $0.8_{-0.2}^{+0.1}$ & $0.9_{-0.2}^{+0.1}$ & $N_H\;[\times10^{22}\;cm^{-2}]$ & $0.8_{-0.2}^{+0.1}$ & $1.9_{-0.4}^{+0.1}$ & $0.99_{-0.15}^{+0.01}$ & $0.5_{-0.2}^{+0.2}$ & $0.03_{-0.02}^{+0.26}$ \\
			\hline
			$kT_{\rm in}\;[\rm keV]$ & $1.460_{-0.007}^{+0.010}$ & $1.353_{-0.017}^{+0.008}$ & $1.58_{-0.01}^{+0.01}$ & $1.664_{-0.009}^{+0.006}$ & $1.401_{-0.008}^{+0.010}$ & $kT_{\rm in}\;[\rm keV]$ & $1.640_{-0.006}^{+0.010}$ & $1.77_{-0.02}^{+0.03}$ & $1.82_{-0.01}^{+0.02}$ & $1.607_{-0.007}^{+0.014}$ & $1.48_{-0.04}^{+0.02}$ \\
			$\rm norm_{d,A} $ & $10.0_{-0.4}^{+0.4}$ & $10.6_{-0.6}^{+0.6}$ & $7.2_{-0.3}^{+0.3}$ & $9.4_{-0.4}^{+0.3}$ & $12.1_{-0.4}^{+0.2}$ & $\rm norm_{d,A} $ & $10.7_{-0.2}^{+0.3}$ & $4.4_{-0.4}^{+0.4}$ & $5.6_{-0.4}^{+0.3}$ & $11.1_{-0.3}^{+0.2}$ & $9.8_{-1.4}^{+0.5}$ \\
			\hline
			$q_1$ & $9_{-3}^{+1}$ & $7_{-2}^{+1}$ & $9.6_{-3.9}^{+0.4}$ & $3.3_{-0.3}^{+4.0}$ & $5_{-1}^{+2}$ & $q_1$ & $3.3_{-0.3}^{+3.0}$ & $6.2_{-0.4}^{+2.9}$ & $3.3_{-0.3}^{+4.0}$ & $3.2_{-0.2}^{+2.7}$ & $3.2_{-0.2}^{+1.3}$ \\
			$q_2$ & $0.4_{-0.4}^{+1.2}$ & $0.09_{-0.09}^{+0.58}$ & $5e-07_{-5e-07}^{+3e-06}$ & $2.4_{-1.4}^{+0.6}$ & $6e-06_{-6e-06}^{+3e-05}$ & $q_2$ & $0.2_{-0.2}^{+1.3}$ & $3e-07_{-3e-07}^{+2e-06}$ & $1.9_{-0.8}^{+1.1}$ & $2.1_{-0.7}^{+0.3}$ & $0.3_{-0.3}^{+0.6}$ \\
			$R_{\rm br} \;[r_g]$ & $12_{-2}^{+16}$ & $10.8_{-0.8}^{+7.2}$ & $97_{-47}^{+3}$ & $9_{-3}^{+51}$ & $22_{-6}^{+22}$ & $R_{\rm br} \;[r_g]$ & $10_{-4}^{+32}$ & $18_{-9}^{+18}$ & $9_{-3}^{+37}$ & $8_{-2}^{+15}$ & $20_{-14}^{+29}$ \\
			$a$ & $0.8_{-0.5}^{+0.2}$ & $0.989_{-0.060}^{+0.009}$ & $0.93_{-0.17}^{+0.04}$ & $-0.8_{-0.2}^{+1.0}$ & $0.2_{-0.4}^{+0.4}$ & $a$ & $-0.4_{-0.6}^{+0.6}$ & $0.97_{-0.17}^{+0.02}$ & $-0.1_{-0.9}^{+0.4}$ & $-0.8_{-0.1}^{+0.8}$ & $0.3_{-1.0}^{+0.3}$ \\
			$\theta\;[^\circ]$ & $34_{-6}^{+14}$ & $55_{-12}^{+10}$ & $54_{-6}^{+4}$ & $43_{-5}^{+5}$ & $85_{-7}^{+1}$ & $\theta\;[^\circ]$ & $79_{-28}^{+8}$ & $48_{-8}^{+9}$ & $85_{-47}^{+2}$ & $80_{-13}^{+4}$ & $77_{-12}^{+7}$ \\
			$\Gamma$ & $3.10_{-0.13}^{+0.05}$ & $3.27_{-0.10}^{+0.07}$ & $2.49_{-0.04}^{+0.05}$ & $2.56_{-0.05}^{+0.07}$ & $2.6_{-0.1}^{+0.1}$ & $\Gamma$ & $2.67_{-0.08}^{+0.06}$ & $2.38_{-0.04}^{+0.05}$ & $2.28_{-0.03}^{+0.03}$ & $2.6_{-0.2}^{+0.2}$ & $1.9_{-0.2}^{+0.4}$ \\
			$\log(\xi)$ & $4.67_{-0.29}^{+0.03}$ & $4.67_{-0.34}^{+0.03}$ & $4.5_{-0.1}^{+0.2}$ & $4.6_{-0.3}^{+0.1}$ & $0.6_{-0.6}^{+0.9}$ & $\log(\xi)$ & $4e-06_{-4e-06}^{+3e-05}$ & $4.68_{-0.24}^{+0.02}$ & $0.2_{-0.2}^{+2.5}$ & $3e-06_{-3e-06}^{+2e-05}$ & $0.3_{-0.2}^{+1.5}$ \\
			$A_{\rm Fe}$ & $3.1_{-0.7}^{+2.2}$ & $4_{-2}^{+2}$ & $0.54_{-0.04}^{+0.35}$ & $0.7_{-0.2}^{+0.5}$ & $3.7_{-0.6}^{+5.1}$ & $A_{\rm Fe}$ & $2_{-2}^{+5}$ & $0.52_{-0.02}^{+0.18}$ & $0.8_{-0.3}^{+5.5}$ & $2_{-1}^{+4}$ & $8_{-3}^{+1}$ \\
			$\rm kT_e\;[\rm keV] $ & \nodata & \nodata & $106_{-6}^{+49}$ & $163_{-63}^{+206}$ & $119_{-19}^{+368}$ & $\rm kT_e\;[\rm keV] $ & $123_{-23}^{+230}$ & $18_{-1}^{+6}$ & $11.1_{-0.8}^{+1.0}$ & $3.9_{-0.9}^{+1.9}$ & $1.49_{-0.09}^{+0.15}$ \\
			R & $1.9_{-0.9}^{+0.9}$ & $3_{-2}^{+1}$ & $1.4_{-0.4}^{+0.4}$ & $0.5_{-0.2}^{+0.2}$ & $3_{-1}^{+2}$ & R & $0.4_{-0.2}^{+0.4}$ & $1.0_{-0.3}^{+0.5}$ & $0.04_{-0.04}^{+0.13}$ & $7_{-4}^{+4}$ & $3_{-2}^{+5}$ \\
			$\rm norm_{r,A} $ & $4e-04_{-1e-04}^{+2e-04}$ & $6e-04_{-2e-04}^{+2e-04}$ & $2e-03_{-3e-04}^{+4e-04}$ & $3e-03_{-6e-04}^{+6e-04}$ & $2e-03_{-6e-04}^{+2e-04}$ & $\rm norm_{r,A} $ & $4e-03_{-8e-04}^{+8e-04}$ & $4e-03_{-8e-04}^{+1e-03}$ & $6e-03_{-6e-04}^{+2e-04}$ & $1e-03_{-4e-04}^{+8e-04}$ & $7e-04_{-3e-04}^{+3e-04}$ \\
			\hline
			$\rm norm_{d,B} $ & $10.4_{-0.6}^{+0.4}$ & $10.3_{-0.6}^{+0.6}$ & $6.9_{-0.2}^{+0.4}$ & $9.3_{-0.4}^{+0.2}$ & $11.7_{-0.4}^{+0.3}$ & $\rm norm_{d,B} $ & $10.3_{-0.3}^{+0.2}$ & $4.3_{-0.5}^{+0.3}$ & $5.5_{-0.4}^{+0.3}$ & $10.8_{-0.2}^{+0.3}$ & $9.8_{-1.4}^{+0.5}$ \\
			$\rm norm_{r,B} $ & $5e-04_{-2e-04}^{+2e-04}$ & $5e-04_{-2e-04}^{+2e-04}$ & $2e-03_{-3e-04}^{+4e-04}$ & $3e-03_{-6e-04}^{+6e-04}$ & $1e-03_{-5e-04}^{+3e-04}$ & $\rm norm_{r,B} $ & $4e-03_{-7e-04}^{+8e-04}$ & $4e-03_{-1e-03}^{+9e-04}$ & $5e-03_{-6e-04}^{+3e-04}$ & $9e-04_{-3e-04}^{+8e-04}$ & $7e-04_{-3e-04}^{+3e-04}$ \\
			\hline
			$\chi^2/\nu$ & $161_{-4}^{+6}(152.37)/134$ & $135_{-6}^{+6}(131.43)/124$ & $379_{-4}^{+5}(371.22)/303$ & $338_{-5}^{+4}(332.34)/306$ & $274_{-5}^{+6}(264.80)/255$ & $\chi^2/\nu$ & $293_{-4}^{+5}(281.93)/286$ & $351_{-5}^{+4}(341.30)/343$ & $360_{-4}^{+5}(352.26)/331$ & $195_{-4}^{+5}(183.77)/161$ & $149_{-4}^{+5}(140.25)/122$ \\
\enddata
\tablecomments{\fontsize{6}{6}\selectfont In this table, we report the mode of the posterior distributions in the MCMC analysis, along with the $1\sigma$ credible interval. For $\chi^2$, the number in parentheses indicates the best-fit $\chi^2$ value. For the normalization of the components, the subscripts d (from \texttt{diskbb}) and r (from \texttt{relxill}) represent the model that the values correspond to, while the subscripts A and B represent which NuSTAR FMP detector the spectrum was originating from.}
\end{splitdeluxetable}

\newpage

\movetabledown=0.8in
\begin{rotatetable*}
\begin{deluxetable*}{l|cccccccc}
\centerwidetable
\tablecaption{Results of the MCMC analysis for H 1732-322}
\label{tab:table4}
\tablewidth{\textwidth} 
\tabletypesize{\scriptsize}
\tablehead{
\colhead{ObsID} &\colhead{80001044002} &\colhead{80001044004} &\colhead{80001044006} &\colhead{80002040002} &\colhead{80202012002} &\colhead{80202012004} &\colhead{80202012006} &\colhead{90401335002}}
\startdata\\
			Model & relxillD$\_$19 & relxillD$\_$19 & relxillCp & relxillD$\_$19 & relxillCp & relxillD$\_$19 & relxillCp & relxillCp \\
			\hline
			$N_H\;[\times10^{22}\;cm^{-2}]$ & $5.1_{-0.9}^{+0.4}$ & $6.8_{-0.5}^{+0.4}$ & $5.9_{-1.0}^{+0.6}$ & $6.2_{-2.0}^{+0.6}$ & $6.1_{-0.3}^{+0.5}$ & $5.3_{-0.5}^{+0.5}$ & $5.2_{-0.6}^{+0.4}$ & $5.5_{-0.6}^{+0.7}$ \\
			\hline
			$kT_{\rm in}\;[\rm keV]$ & $0.55_{-0.04}^{+0.02}$ & $0.504_{-0.012}^{+0.009}$ & $0.50_{-0.03}^{+0.02}$ & $0.47_{-0.04}^{+0.09}$ & $0.527_{-0.007}^{+0.010}$ & $0.525_{-0.016}^{+0.008}$ & $0.53_{-0.04}^{+0.01}$ & $0.53_{-0.02}^{+0.02}$ \\
			$\rm norm_{d,A} $ & $327_{-43}^{+212}$ & $2e+03_{-2e+02}^{+2e+02}$ & $915_{-318}^{+264}$ & $247_{-87}^{+629}$ & $1e+03_{-2e+02}^{+1e+02}$ & $1e+03_{-2e+02}^{+1e+02}$ & $355_{-86}^{+119}$ & $634_{-166}^{+112}$ \\
			\hline
			$q_1$ & $9.9_{-2.0}^{+0.1}$ & $9.1_{-2.3}^{+0.9}$ & $5.7_{-0.3}^{+3.6}$ & $9.8_{-2.9}^{+0.2}$ & $5_{-2}^{+2}$ & $8.3_{-0.9}^{+1.4}$ & $9_{-4}^{+1}$ & $7.3_{-0.6}^{+2.7}$ \\
			$q_2$ & $0.4_{-0.4}^{+1.6}$ & $0.1_{-0.1}^{+1.6}$ & $2.0_{-0.9}^{+1.0}$ & $0.6_{-0.3}^{+1.5}$ & $2.5_{-1.2}^{+0.5}$ & $0.1_{-0.1}^{+1.6}$ & $2.9_{-1.6}^{+0.1}$ & $0.6_{-0.5}^{+1.4}$ \\
			$R_{\rm br} \;[r_g]$ & $29_{-19}^{+36}$ & $17_{-11}^{+27}$ & $21_{-4}^{+54}$ & $82_{-48}^{+8}$ & $9_{-3}^{+35}$ & $21_{-14}^{+23}$ & $96_{-53}^{+4}$ & $33_{-23}^{+34}$ \\
			$a$ & $0.992_{-0.064}^{+0.006}$ & $0.984_{-0.048}^{+0.008}$ & $0.9_{-0.3}^{+0.1}$ & $0.996_{-0.016}^{+0.002}$ & $0.8_{-0.3}^{+0.1}$ & $0.987_{-0.050}^{+0.005}$ & $0.992_{-0.069}^{+0.006}$ & $0.99_{-0.22}^{+0.01}$ \\
			$\theta\;[^\circ]$ & $70_{-20}^{+2}$ & $54.8_{-0.9}^{+12.8}$ & $41_{-4}^{+8}$ & $62_{-7}^{+10}$ & $47_{-3}^{+5}$ & $61_{-8}^{+2}$ & $55_{-6}^{+17}$ & $47_{-6}^{+7}$ \\
			$\Gamma$ & $1.36_{-0.03}^{+0.02}$ & $1.45_{-0.01}^{+0.01}$ & $1.49_{-0.04}^{+0.02}$ & $1.43_{-0.08}^{+0.03}$ & $1.65_{-0.02}^{+0.02}$ & $1.59_{-0.02}^{+0.03}$ & $1.54_{-0.04}^{+0.03}$ & $1.53_{-0.03}^{+0.03}$ \\
			$\log(\xi)$ & $3.75_{-0.05}^{+0.12}$ & $3.69_{-0.06}^{+0.04}$ & $3.93_{-0.07}^{+0.06}$ & $3.99_{-0.67}^{+0.04}$ & $4.21_{-0.03}^{+0.13}$ & $3.91_{-0.12}^{+0.05}$ & $4.2_{-0.4}^{+0.1}$ & $3.95_{-0.13}^{+0.09}$ \\
			$A_{\rm Fe}$ & $3.2_{-0.5}^{+0.7}$ & $2.0_{-0.2}^{+0.3}$ & $1.8_{-0.3}^{+0.3}$ & $9.8_{-3.7}^{+0.2}$ & $3.9_{-0.8}^{+0.6}$ & $3.2_{-0.8}^{+0.3}$ & $2.7_{-0.5}^{+2.5}$ & $2.1_{-0.4}^{+0.3}$ \\
			$\rm kT_e\;[\rm keV] $ & \nodata & \nodata & $378_{-146}^{+22}$ & \nodata & $82_{-16}^{+60}$ & \nodata & $83_{-35}^{+39}$ & $183_{-58}^{+133}$ \\
			R & $0.5_{-0.1}^{+0.5}$ & $1.0_{-0.2}^{+0.5}$ & $1.2_{-0.2}^{+0.2}$ & $0.20_{-0.09}^{+0.18}$ & $1.2_{-0.2}^{+0.4}$ & $1.3_{-0.3}^{+0.4}$ & $0.27_{-0.08}^{+0.39}$ & $1.3_{-0.3}^{+0.4}$ \\
			$\rm norm_{r,A} $ & $8e-03_{-4e-04}^{+5e-04}$ & $8e-03_{-8e-04}^{+3e-04}$ & $5e-03_{-4e-04}^{+8e-04}$ & $7e-03_{-6e-04}^{+2e-04}$ & $4e-03_{-3e-04}^{+5e-04}$ & $5e-03_{-2e-04}^{+5e-04}$ & $5e-03_{-6e-04}^{+6e-04}$ & $5e-03_{-5e-04}^{+7e-04}$ \\
			\hline
			$\rm norm_{d,B} $ & $412_{-98}^{+174}$ & $2e+03_{-2e+02}^{+2e+02}$ & $1e+03_{-4e+02}^{+2e+02}$ & $258_{-107}^{+617}$ & $1e+03_{-2e+02}^{+1e+02}$ & $1e+03_{-2e+02}^{+2e+02}$ & $409_{-115}^{+104}$ & $638_{-133}^{+148}$ \\
			$\rm norm_{r,B} $ & $8e-03_{-4e-04}^{+5e-04}$ & $8e-03_{-8e-04}^{+3e-04}$ & $5e-03_{-4e-04}^{+8e-04}$ & $7e-03_{-6e-04}^{+2e-04}$ & $4e-03_{-3e-04}^{+5e-04}$ & $5e-03_{-2e-04}^{+5e-04}$ & $5e-03_{-6e-04}^{+6e-04}$ & $5e-03_{-5e-04}^{+7e-04}$ \\
			\hline
			$\chi^2/\nu$ & $596_{-6}^{+5}(581.18)/514$ & $668_{-5}^{+6}(653.13)/523$ & $507_{-5}^{+4}(499.54)/485$ & $542_{-7}^{+10}(535.72)/470$ & $618_{-4}^{+4}(604.41)/523$ & $595_{-6}^{+6}(582.15)/523$ & $566_{-5}^{+10}(555.77)/513$ & $557_{-6}^{+5}(545.11)/509$ \\
\enddata
\tablecomments{In this table, we report the mode of the posterior distributions in the MCMC analysis, along with the $1\sigma$ credible interval. For $\chi^2$, the number in parentheses indicates the best-fit $\chi^2$ value. For the normalization of the components, the subscripts d (from \texttt{diskbb}) and r (from \texttt{relxill}) represent the model that the values correspond to, while the subscripts A and B represent which NuSTAR FMP detector the spectrum was originating from.}
\end{deluxetable*}
\end{rotatetable*}

%\clearpage
%\onecolumngrid 

\begin{splitdeluxetable}{l|cccccccBl|ccccccBl|cccccc}
%\tablewidth{12in}
\movetableright=-3.7in
%\movetabledown=5in
%\rotate
\centerwidetable
\hskip-2in\tablecaption{Results of the MCMC analysis for MAXI J1820+070} 
\label{tab:table5}
\tabletypesize{\tiny}
\tablehead{\colhead{ObsID} & \colhead{90401309023} & \colhead{90401309025} & \colhead{90401309027} & \colhead{90401309033} & \colhead{90401309002} & \colhead{90401309004} & \colhead{90401309006} & \colhead{ObsID} & \colhead{90401309008} & \colhead{90401309010} & \colhead{90401309012}  & \colhead{90401309013} & \colhead{90401309014} & \colhead{90401309016} & \colhead{ObsID} & \colhead{90401309018} & \colhead{90401309019} & \colhead{90401309021} & \colhead{90401309026} & \colhead{90401309035} & \colhead{90401324002}}
\startdata\\
			Model & relxillD-19 & relxill & relxill & relxill & relxillCp & relxillCp & relxillCp & Model & relxillCp & relxillCp & relxillCp & relxillCp & relxillCp & relxillCp & Model & relxillCp & relxillCp & relxillCp & relxillCp & relxillCp & relxillCp \\
			\hline
			$N_H\;[\times10^{22}\;cm^{-2}]$ & $0.058_{-0.008}^{+0.085}$ & $0.053_{-0.003}^{+0.029}$ & $5e-02_{-7e-04}^{+7e-03}$ & $0.07_{-0.02}^{+0.31}$ & $0.07_{-0.02}^{+0.15}$ & $0.07_{-0.02}^{+0.20}$ & $0.053_{-0.003}^{+0.025}$ & $N_H\;[\times10^{22}\;cm^{-2}]$ & $0.054_{-0.004}^{+0.165}$ & $0.06_{-0.01}^{+0.12}$ & $0.057_{-0.007}^{+0.056}$ & $0.08_{-0.03}^{+0.36}$ & $0.057_{-0.007}^{+0.087}$ & $0.06_{-0.01}^{+0.12}$ & $N_H\;[\times10^{22}\;cm^{-2}]$ & $0.97_{-0.34}^{+0.03}$ & $0.07_{-0.02}^{+0.17}$ & $0.058_{-0.008}^{+0.073}$ & $0.07_{-0.02}^{+0.27}$ & $0.11_{-0.06}^{+0.32}$ & $0.2_{-0.2}^{+0.4}$ \\
			\hline
			$kT_{\rm in}\;[\rm keV]$ & $0.760_{-0.005}^{+0.004}$ & $0.762_{-0.002}^{+0.001}$ & $7e-01_{-8e-04}^{+7e-04}$ & $0.56_{-0.01}^{+0.01}$ & $0.73_{-0.07}^{+0.11}$ & $0.88_{-0.05}^{+0.05}$ & $1.47_{-0.04}^{+0.06}$ & $kT_{\rm in}\;[\rm keV]$ & $0.84_{-0.10}^{+-0.01}$ & $0.83_{-0.04}^{+0.04}$ & $0.86_{-0.02}^{+0.02}$ & $0.77_{-0.03}^{+0.06}$ & $0.80_{-0.02}^{+0.03}$ & $0.83_{-0.02}^{+0.02}$ & $kT_{\rm in}\;[\rm keV]$ & $0.74_{-0.04}^{+0.03}$ & $0.79_{-0.03}^{+0.03}$ & $0.83_{-0.02}^{+0.03}$ & $0.750_{-0.011}^{+0.006}$ & $0.65_{-0.07}^{+0.07}$ & $0.73_{-0.02}^{+0.04}$ \\
			$\rm norm_{d,A} $ & $1e+04_{-4e+02}^{+4e+02}$ & $1e+04_{-2e+02}^{+2e+02}$ & $1e+04_{-9e+01}^{+8e+01}$ & $2e+03_{-3e+02}^{+3e+02}$ & $56_{-33}^{+62}$ & $221_{-67}^{+102}$ & $31_{-6}^{+3}$ & $\rm norm_{d,A} $ & $381_{-380}^{+342}$ & $391_{-83}^{+136}$ & $327_{-40}^{+41}$ & $551_{-192}^{+278}$ & $510_{-64}^{+115}$ & $382_{-48}^{+82}$ & $\rm norm_{d,A} $ & $910_{-287}^{+326}$ & $533_{-129}^{+91}$ & $224_{-34}^{+34}$ & $1e+04_{-8e+02}^{+1e+03}$ & $68_{-46}^{+63}$ & $649_{-156}^{+377}$ \\
			\hline
			$q_1$ & $8.7_{-0.8}^{+0.5}$ & $9.98_{-0.20}^{+0.02}$ & $9.99_{-0.14}^{+0.01}$ & $9.6_{-1.7}^{+0.3}$ & $9.9_{-1.2}^{+0.1}$ & $9.8_{-1.3}^{+0.2}$ & $3.09_{-0.09}^{+0.82}$ & $q_1$ & $9.9_{-1.1}^{+0.1}$ & $9.93_{-0.48}^{+0.07}$ & $9.98_{-0.22}^{+0.02}$ & $8.6_{-2.1}^{+0.9}$ & $9.96_{-0.44}^{+0.04}$ & $9.97_{-0.25}^{+0.03}$ & $q_1$ & $9.92_{-0.99}^{+0.08}$ & $8.3_{-0.3}^{+0.9}$ & $9.4_{-0.6}^{+0.6}$ & $4_{-1}^{+2}$ & $5_{-1}^{+2}$ & $9.94_{-0.64}^{+0.06}$ \\
			$q_2$ & $0.08_{-0.08}^{+0.64}$ & $0.1_{-0.1}^{+1.1}$ & $2.6_{-1.5}^{+0.4}$ & $2.1_{-0.5}^{+0.9}$ & $1.0_{-0.8}^{+0.4}$ & $2.9_{-1.7}^{+0.1}$ & $1.77_{-0.07}^{+0.03}$ & $q_2$ & $2.6_{-1.5}^{+0.4}$ & $2.9_{-1.6}^{+0.1}$ & $2.9_{-1.6}^{+0.1}$ & $2.9_{-1.8}^{+0.1}$ & $4e-9_{-4e-09}^{+2e-03}$ & $2.8_{-1.6}^{+0.2}$ & $q_2$ & $0.2_{-0.2}^{+1.8}$ & $0.08_{-0.08}^{+0.50}$ & $2.9_{-1.5}^{+0.1}$ & $2.9_{-1.7}^{+0.1}$ & $2.7_{-1.4}^{+0.3}$ & $2.9_{-1.6}^{+0.1}$ \\
			$R_{\rm br} \;[r_g]$ & $11_{-1}^{+8}$ & $10.8_{-0.8}^{+9.1}$ & $46_{-8}^{+42}$ & $95_{-34}^{+5}$ & $14_{-2}^{+6}$ & $93_{-41}^{+7}$ & $10.3_{-0.3}^{+2.4}$ & $R_{\rm br} \;[r_g]$ & $98_{-46}^{+2}$ & $86_{-29}^{+14}$ & $92_{-31}^{+8}$ & $86_{-43}^{+14}$ & $10.6_{-0.6}^{+6.6}$ & $87_{-28}^{+13}$ & $R_{\rm br} \;[r_g]$ & $12_{-2}^{+50}$ & $11_{-1}^{+7}$ & $84_{-26}^{+16}$ & $96_{-52}^{+4}$ & $80_{-33}^{+20}$ & $96_{-38}^{+4}$ \\
			$a$ & $0.993_{-0.002}^{+0.001}$ & $0.990_{-0.002}^{+0.001}$ & $0.989_{-0.001}^{+0.001}$ & $0.978_{-0.009}^{+0.007}$ & $0.4_{-0.1}^{+0.1}$ & $0.88_{-0.04}^{+0.03}$ & $0.5_{-0.3}^{+0.1}$ & $a$ & $0.84_{-0.06}^{+0.04}$ & $0.90_{-0.02}^{+0.02}$ & $0.919_{-0.010}^{+0.007}$ & $0.93_{-0.06}^{+0.02}$ & $0.925_{-0.009}^{+0.039}$ & $0.961_{-0.004}^{+0.004}$ & $a$ & $0.977_{-0.005}^{+0.006}$ & $0.977_{-0.009}^{+0.004}$ & $0.956_{-0.012}^{+0.005}$ & $0.97_{-0.14}^{+0.02}$ & $0.87_{-0.14}^{+0.08}$ & $0.975_{-0.005}^{+0.005}$ \\
			$\theta\;[^\circ]$ & $72_{-2}^{+1}$ & $72.1_{-0.8}^{+0.3}$ & $71.0_{-0.5}^{+0.3}$ & $69_{-3}^{+2}$ & $39_{-3}^{+2}$ & $55_{-4}^{+2}$ & $77.8_{-6.3}^{+0.9}$ & $\theta\;[^\circ]$ & $52_{-3}^{+3}$ & $57_{-2}^{+1}$ & $58.4_{-1.0}^{+0.8}$ & $57_{-5}^{+4}$ & $59_{-2}^{+4}$ & $64.2_{-0.8}^{+0.7}$ & $\theta\;[^\circ]$ & $67_{-2}^{+1}$ & $65.4_{-1.6}^{+0.8}$ & $63_{-2}^{+1}$ & $54_{-4}^{+7}$ & $49_{-8}^{+8}$ & $67_{-2}^{+1}$ \\
			$\Gamma$ & $2.189_{-0.009}^{+0.005}$ & $2.11_{-0.01}^{+0.01}$ & $2.199_{-0.009}^{+0.007}$ & $1.841_{-0.008}^{+0.014}$ & $1.507_{-0.003}^{+0.005}$ & $1.561_{-0.003}^{+0.003}$ & $1.65_{-0.01}^{+0.02}$ & $\Gamma$ & $1.574_{-0.003}^{+0.002}$ & $1.571_{-0.003}^{+0.002}$ & $1.578_{-0.002}^{+0.002}$ & $1.597_{-0.005}^{+0.005}$ & $1.596_{-0.002}^{+0.005}$ & $1.630_{-0.002}^{+0.003}$ & $\Gamma$ & $1.652_{-0.004}^{+0.006}$ & $1.647_{-0.004}^{+0.003}$ & $1.616_{-0.003}^{+0.004}$ & $1.94_{-0.09}^{+0.08}$ & $1.649_{-0.005}^{+0.006}$ & $1.653_{-0.003}^{+0.006}$ \\
			$\log(\xi)$ & $3.47_{-0.07}^{+0.08}$ & $3.35_{-0.03}^{+0.04}$ & $3.55_{-0.03}^{+0.03}$ & $3.50_{-0.08}^{+0.07}$ & $3.08_{-0.02}^{+0.02}$ & $3.30_{-0.04}^{+0.04}$ & $1.77_{-0.08}^{+0.18}$ & $\log(\xi)$ & $3.30_{-0.03}^{+0.03}$ & $3.30_{-0.04}^{+0.02}$ & $3.29_{-0.03}^{+0.02}$ & $3.30_{-0.05}^{+0.08}$ & $3.30_{-0.07}^{+0.02}$ & $3.30_{-0.02}^{+0.03}$ & $\log(\xi)$ & $3.30_{-0.06}^{+0.06}$ & $3.30_{-0.03}^{+0.05}$ & $3.30_{-0.03}^{+0.06}$ & $3.7_{-0.4}^{+0.2}$ & $3.21_{-0.08}^{+0.11}$ & $3.34_{-0.05}^{+0.07}$ \\
			\hline
			$A_{\rm Fe}$ & $9.98_{-0.19}^{+0.02}$ & $9.97_{-0.18}^{+0.03}$ & $9.993_{-0.062}^{+0.007}$ & $7.0_{-1.1}^{+0.9}$ & $7_{-1}^{+1}$ & $5.0_{-0.2}^{+0.4}$ & $1.7_{-0.2}^{+0.4}$ & $A_{\rm Fe}$ & $5.0_{-0.1}^{+0.5}$ & $5.7_{-0.5}^{+0.7}$ & $6.5_{-0.4}^{+0.4}$ & $5.8_{-0.8}^{+1.4}$ & $6.4_{-0.4}^{+0.9}$ & $7.4_{-0.6}^{+0.6}$ & $A_{\rm Fe}$ & $9.8_{-1.2}^{+0.2}$ & $8.2_{-1.1}^{+0.9}$ & $9.0_{-0.6}^{+0.9}$ & $9.2_{-3.7}^{+0.7}$ & $9.6_{-1.2}^{+0.3}$ & $8.8_{-1.5}^{+0.8}$ \\
			$\rm kT_e\;[\rm keV] $ & \nodata & $1e+03_{-9e+01}^{+1e+01}$ & $1e+03_{-3e+01}^{+5e+00}$ & $1e+03_{-1e+02}^{+1e+01}$ & $45_{-4}^{+5}$ & $24.7_{-0.6}^{+0.6}$ & $37_{-4}^{+4}$ & $\rm kT_e\;[\rm keV] $ & $24.0_{-0.6}^{+0.4}$ & $23.1_{-0.6}^{+0.5}$ & $25.9_{-0.5}^{+0.4}$ & $27_{-1}^{+1}$ & $27.0_{-0.6}^{+0.9}$ & $30.8_{-0.8}^{+0.9}$ & $\rm kT_e\;[\rm keV] $ & $32_{-1}^{+3}$ & $33_{-1}^{+2}$ & $40_{-2}^{+2}$ & $9_{-1}^{+2}$ & $131_{-38}^{+57}$ & $37_{-2}^{+3}$ \\
			R & $1.7_{-0.2}^{+0.1}$ & $3.2_{-0.2}^{+0.2}$ & $6.1_{-0.3}^{+0.3}$ & $0.87_{-0.14}^{+0.09}$ & $0.18_{-0.02}^{+0.01}$ & $0.30_{-0.03}^{+0.03}$ & $1.4_{-0.5}^{+0.3}$ & R & $0.29_{-0.03}^{+0.02}$ & $0.32_{-0.03}^{+0.02}$ & $0.30_{-0.01}^{+0.02}$ & $0.26_{-0.02}^{+0.05}$ & $0.27_{-0.02}^{+0.03}$ & $0.34_{-0.01}^{+0.02}$ & R & $0.40_{-0.05}^{+0.05}$ & $0.33_{-0.02}^{+0.03}$ & $0.27_{-0.02}^{+0.02}$ & $3_{-1}^{+1}$ & $0.14_{-0.02}^{+0.05}$ & $0.41_{-0.04}^{+0.03}$ \\
			$\rm norm_{r,A} $ & $0.059_{-0.003}^{+0.002}$ & $1e-02_{-5e-04}^{+6e-04}$ & $4e-03_{-2e-04}^{+2e-04}$ & $3e-02_{-7e-04}^{+2e-04}$ & $6e-02_{-2e-03}^{+8e-04}$ & $0.152_{-0.001}^{+0.001}$ & $0.155_{-0.001}^{+0.002}$ & $\rm norm_{r,A} $ & $2e-01_{-7e-04}^{+1e-03}$ & $2e-01_{-1e-03}^{+9e-04}$ & $1e-01_{-5e-04}^{+6e-04}$ & $0.142_{-0.002}^{+0.002}$ & $1e-01_{-5e-04}^{+1e-03}$ & $1e-01_{-6e-04}^{+7e-04}$ & $\rm norm_{r,A} $ & $0.105_{-0.001}^{+0.001}$ & $1e-01_{-1e-03}^{+8e-04}$ & $7e-02_{-5e-04}^{+7e-04}$ & $0.003_{-0.001}^{+0.001}$ & $2e-02_{-8e-04}^{+4e-04}$ & $0.102_{-0.001}^{+0.001}$ \\
			\hline
			$\rm norm_{d,B} $ & $1e+04_{-4e+02}^{+4e+02}$ & $1e+04_{-1e+02}^{+2e+02}$ & $1e+04_{-8e+01}^{+8e+01}$ & $2e+03_{-3e+02}^{+3e+02}$ & $57_{-31}^{+68}$ & $247_{-76}^{+108}$ & $29_{-4}^{+5}$ & $\rm norm_{d,B} $ & $348_{-347}^{+311}$ & $384_{-83}^{+125}$ & $307_{-38}^{+38}$ & $538_{-204}^{+238}$ & $454_{-53}^{+108}$ & $342_{-44}^{+70}$ & $\rm norm_{d,B} $ & $874_{-273}^{+314}$ & $473_{-121}^{+69}$ & $232_{-41}^{+28}$ & $1e+04_{-7e+02}^{+1e+03}$ & $55_{-36}^{+55}$ & $604_{-178}^{+300}$ \\
			$\rm norm_{r,B} $ & $0.056_{-0.003}^{+0.002}$ & $1e-02_{-5e-04}^{+5e-04}$ & $4e-03_{-2e-04}^{+2e-04}$ & $3e-02_{-5e-04}^{+3e-04}$ & $6e-02_{-2e-03}^{+8e-04}$ & $0.143_{-0.001}^{+0.001}$ & $0.145_{-0.001}^{+0.002}$ & $\rm norm_{r,B} $ & $1e-01_{-5e-04}^{+1e-03}$ & $1e-01_{-8e-04}^{+1e-03}$ & $1e-01_{-4e-04}^{+6e-04}$ & $0.133_{-0.002}^{+0.001}$ & $1e-01_{-6e-04}^{+8e-04}$ & $1e-01_{-6e-04}^{+6e-04}$ & $\rm norm_{r,B} $ & $0.100_{-0.001}^{+0.001}$ & $1e-01_{-1e-03}^{+7e-04}$ & $7e-02_{-6e-04}^{+6e-04}$ & $3e-03_{-9e-04}^{+1e-03}$ & $2e-02_{-9e-04}^{+3e-04}$ & $1e-01_{-9e-04}^{+1e-03}$ \\
			\hline
			$\chi^2/\nu$ & $682_{-4}^{+18}(698.96)/482$ & $547_{-5}^{+6}(530.79)/451$ & $848_{-5}^{+6}(831.83)/455$ & $633_{-6}^{+4}(621.28)/523$ & $617_{-5}^{+6}(602.20)/525$ & $661_{-6}^{+6}(653.00)/520$ & $696_{-6}^{+8}(679.40)/534$ & $\chi^2/\nu$ & $685_{-8}^{+6}(668.34)/522$ & $615_{-5}^{+6}(600.40)/521$ & $890_{-5}^{+6}(875.93)/564$  & $546_{-5}^{+7}(544.73)/503$ & $809_{-7}^{+5}(798.27)/551$ & $737_{-5}^{+6}(730.36)/556$& $\chi^2/\nu$ & $558_{-6}^{+7}(549.13)/503$ & $684_{-5}^{+15}(674.27)/539$ & $893_{-4}^{+6}(884.07)/552$ & $410_{-7}^{+4}(396.64)/321$ & $613_{-5}^{+6}(600.87)/490$ & $631_{-4}^{+6}(621.92)/527$ \\
\enddata
\hskip-2in\tablecomments{\fontsize{6}{6}\selectfont In this table, we report the mode of the posterior distributions in the MCMC analysis, along with the $1\sigma$ credible interval. For $\chi^2$, the number in parentheses indicates the best-fit $\chi^2$ value. For the normalization of the components, the subscripts d (from \texttt{diskbb}) and r (from \texttt{relxill}) represent the model that the values correspond to, while the subscripts A and B represent which NuSTAR FMP detector the spectrum was originating from.}
\end{splitdeluxetable}

\begin{deluxetable*}{l|ccrr|ccr}
\tablecaption{All current spin measurements in X-ray binaries performed through relativistic reflection or continuum fitting}
\label{tab:all_spins}
\tablewidth{\textwidth} 
\tabletypesize{\scriptsize}
\tablehead{
\colhead{Source} & \colhead{Spin reflection} & \colhead{Conf.} & \colhead{Telescope} & \colhead{Reference}
 & \colhead{Spin continuum} & \colhead{Conf.} & \colhead{Reference}}
\startdata\\
IC 10 X-1  & \nodata   &   \nodata  &    \nodata & \nodata & $0.85_{-0.07}^{+0.04}$ & 90 & \cite{2016ApJ...817..154S} \\
M31 ULX-1 & \nodata   &   \nodata  &    \nodata & \nodata & $0.36_{-0.11}^{+0.10}$ & 90 & \cite{2012MNRAS.420.2969M} \\
M31 ULX-2  & \nodata   &   \nodata  &    \nodata & \nodata & $<-0.17$ & 97 & \cite{2014MNRAS.439.1740M} \\
M33 X-7  & \nodata   &   \nodata  &    \nodata & \nodata & $0.84_{-0.05}^{+0.05}$ & 68 & \cite{2008ApJ...679L..37L} \\
AT 2019wey & $0.97_{-0.03}^{+0.02}$ & 68 & NuSTAR & \cite{2021arXiv210907357F}  & \nodata   &   \nodata  &    \nodata \\
LMC X-3 & $0.24_{-0.05}^{+0.05}$ & 90 & NuSTAR & \cite{2021MNRAS.507.4779J} & $0.25_{-0.29}^{+0.2}$ & 90 & \cite{2014ApJ...793L..29S} \\
LMC X-1 & $0.9395_{-0.015}^{+0.015}$ & 90 & NuSTAR & \cite{2021MNRAS.507.4779J} & $0.92_{-0.07}^{+0.05}$ & 68 & \cite{2009ApJ...701.1076G} \\
1A 0620-00  & \nodata   &   \nodata  &    \nodata & \nodata & $0.12_{-0.19}^{+0.19}$ & 68 & \cite{2010ApJ...718L.122G} \\
MAXI J0637-430 & $0.97_{-0.02}^{+0.02}$ & 68 & NuSTAR & this paper & \nodata   &   \nodata  &    \nodata \\
GRS 1009-45  & \nodata   &   \nodata  &    \nodata & \nodata & $0.63_{-0.19}^{+0.16}$ & 68 & \cite{2016ApJ...825...45C} \\
GS 1124-683   & \nodata   &   \nodata  &    \nodata & \nodata & $-0.24_{-0.64}^{+0.05}$ & 90 & \cite{2014ApJ...784L..18M} \\
MAXI J1348-630 & $0.78_{-0.04}^{+0.04}$ & 90 & NuSTAR & \cite{2022MNRAS.511.3125J} & \nodata   &   \nodata  &    \nodata \\
GS 1354-645 & $>0.98$ & 68 & NuSTAR & \cite{2016ApJ...826L..12E}  & \nodata   &   \nodata  &    \nodata \\
MAXI J1535-571 & $0.985_{-0.004}^{+0.002}$ & 90 & NuSTAR & \cite{2022MNRAS.514.1422D}  & \nodata   &   \nodata  &    \nodata \\
4U 1543-47 & $0.98_{-0.02}^{+0.01}$ & 68 & NuSTAR & this paper & $0.8_{-0.05}^{+0.05}$ & 90 & \cite{2006ApJ...636L.113S} \\
XTE J1550-564 & $0.55_{-0.22}^{+0.15}$ & 90 & ASCA \& RXTE & \cite{2011MNRAS.416..941S} & $0.34_{-0.45}^{+0.37}$ & 90 &\cite{2011MNRAS.416..941S}\\
MAXI J1631-479 & $>0.94$ & 90 & NuSTAR & \cite{2020ApJ...893...30X}  & \nodata   &   \nodata  &    \nodata \\
4U 1630-472 & $0.985_{-0.014}^{+0.005}$ & 68 & NuSTAR & \cite{2014ApJ...784L...2K} & $0.92_{-0.04}^{+0.04}$ & 99.7 & \cite{2018ApJ...867...86P} \\
XTE J1650-500 & $0.79_{-0.01}^{+0.01}$ & 68 & XMM & \cite{2009ApJ...697..900M}  & \nodata   &   \nodata  &    \nodata \\
XTE J1652-453 & $0.45_{-0.02}^{+0.02}$ & 90 & XMM & \cite{2011MNRAS.411..137H}  & \nodata   &   \nodata  &    \nodata \\
GRO J1655-40 & $0.98_{-0.01}^{+0.01}$ & 68 & RXTE & \cite{2009ApJ...697..900M} & $0.7_{-0.05}^{+0.05}$ & 90 & \cite{2006ApJ...636L.113S} \\
Swift J1658.2-4242 & $>0.96$ & 90 & NuSTAR & \cite{2018ApJ...865...18X}  &  \nodata  &   \nodata  &    \nodata \\
MAXI J1659-152 & $<0$ & 95 & XMM \& RXTE & \cite{2020ApJ...888L..30R}  & $<0.44$   &    90 &    \cite{2021arXiv211203479F} \\
GX 339-4 & $0.95_{-0.08}^{+0.02}$ & 68 & NuSTAR & \cite{2016ApJ...821L...6P} & $<0.9$ & 90 & \cite{2010MNRAS.406.2206K} \\
IGR J17091-3624 & $0.07_{-0.2}^{+0.2}$ & 90 & NuSTAR & \cite{2018MNRAS.478.4837W} & $<0.2$ & 90 & \cite{2012ApJ...757L..12R} \\
SAX J1711.6-3808 & $0.6_{-0.2}^{+0.2}$ & 68 & BeppoSAX & \cite{2009ApJ...697..900M}  & \nodata   &   \nodata  &    \nodata \\
GRS 1716-249 & $>0.92$ & 90 & NuSTAR & \cite{2019ApJ...887..184T}  & \nodata   &   \nodata  &    \nodata \\
MAXI J1727-203 & $0.986_{-0.159}^{+0.012}$ & 68 & NuSTAR & this paper & \nodata   &   \nodata  &    \nodata \\
Swift J1728.0-3613 & $0.86_{-0.02}^{+0.02}$ & 68 & NuSTAR & Draghis et al. (2022 in prep) & \nodata   &   \nodata  &    \nodata \\
GRS 1739-278 & $0.8_{-0.2}^{+0.2}$ & 90 & NuSTAR & \cite{2015ApJ...799L...6M}  & \nodata   &   \nodata  &    \nodata \\
1E 1740.7-2942 & $>0.5$ & 68 & NuSTAR & \cite{2020MNRAS.493.2694S}  & \nodata   &   \nodata  &    \nodata \\
IGR J17454-2919 & $0.971_{-0.171}^{+0.027}$ & 68 & NuSTAR & this paper & \nodata   &   \nodata  &    \nodata \\
Swift J174540.2-290037 & $0.92_{-0.07}^{+0.05}$ & 90 & NuSTAR & \cite{2019ApJ...885..142M}  & \nodata   &   \nodata  &    \nodata \\
Swift J174540.7-290015 & $0.94_{-0.1}^{+0.03}$ & 90 & NuSTAR & \cite{2019ApJ...885..142M}  & \nodata   &   \nodata  &    \nodata \\
H 1743-322  & $0.98_{-0.02}^{+0.01}$   &   68  &    NuSTAR & this paper & $0.2_{-0.3}^{+0.3}$ & 68 & \cite{2012ApJ...745L...7S} \\
XTE J1752-223 & $0.92_{-0.06}^{+0.06}$ & 68 & RXTE & \cite{2018ApJ...864...25G}  & \nodata   &   \nodata  &    \nodata \\
Swift J1753.5-0127 & $0.997_{-0.002}^{+0.01}$ & 68 & NuSTAR & this paper  & \nodata   &   \nodata  &    \nodata \\
GRS 1758-258 & $0.991_{-0.019}^{+0.007}$ ($a>0.92$) & 68 (90) & NuSTAR & this paper (\cite{2022arXiv220801399J})  & \nodata   &   \nodata  &    \nodata \\
MAXI J1803-298 & $0.991_{-0.001}^{+0.001}$ & 68 & NuSTAR & \cite{2021arXiv211202794F}  & \nodata   &   \nodata  &    \nodata \\
MAXI J1813-095 & $>0.76$ & 90 & NuSTAR & \cite{2021RAA....21..125J}  & \nodata   &   \nodata  &    \nodata \\
V4641 Sgr & $0.86_{-0.02}^{+0.02}$ & 68 & NuSTAR & this paper  & \nodata   &   \nodata  &    \nodata \\
MAXI J1820+070  & $0.988_{-0.028}^{+0.006}$   &  68  &   NuSTAR & this paper & $0.14_{-0.09}^{+0.09}$ & 68 & \cite{2021ApJ...916..108Z} \\
MAXI J1836-194 & $0.88_{-0.03}^{+0.03}$ & 90 & Suzaku & \cite{2012ApJ...751...34R}  & \nodata   &   \nodata  &    \nodata \\
MAXI J1848-015 & $0.967_{-0.013}^{+0.013}$ & 90 & NuSTAR & \cite{2022ApJ...927..190P} & \nodata   &   \nodata  &    \nodata \\
EXO 1846-031 & $0.997_{-0.002}^{+0.001}$ & 68 & NuSTAR & \cite{2020ApJ...900...78D}  & \nodata   &   \nodata  &    \nodata \\
%Swift J1858.6-0814 & $0.0_{-0.0}^{+1.0}$ & 90 & NuSTAR & \cite{2020ApJ...890...57H}  & \nodata   &   \nodata  &    \nodata \\
XTE J1908+094 & $0.55_{-0.45}^{+0.29}$ & 68 & NuSTAR & \cite{2021ApJ...920...88D}  & \nodata   &   \nodata  &    \nodata \\
Swift J1910.2-0546 & $<-0.32$ & 90 & XMM & \cite{2013ApJ...778..155R}  & \nodata   &   \nodata  &    \nodata \\
GRS 1915+105 & $0.98_{-0.01}^{+0.01}$ & 68 & NuSTAR & \cite{2013ApJ...775L..45M} & $0.995_{-0.005}^{+0.002}$ & 90 & \cite{2020MNRAS.499.5891S} \\
Cyg X-1 & $>0.97$ & 68 & NuSTAR & \cite{2015ApJ...808....9P} & $>0.983$ & 99.7 & \cite{2014ApJ...790...29G} \\
4U 1957+11 & $0.95_{-0.04}^{+0.02}$ & 68 & NuSTAR & this paper & $>0.9$ & 90 & \cite{2012ApJ...744..107N} \\
V404 Cyg & $>0.92$ & 99 & NuSTAR & \cite{2017ApJ...839..110W}  & \nodata   &   \nodata  &    \nodata \\
\enddata
\tablecomments{Columns named ``Conf." show the confidence limit of the quoted measurement, in percentages.}
\end{deluxetable*}

\end{document}